\documentclass[aps,twocolumn,groupedaddress,nofootinbib,english,nobalancelastpage,floatfix,superscriptaddress,preprintnumbers]{revtex4-1}
\pdfoutput=1
\usepackage{flushend}
\DeclareMathAlphabet{\scr}{U}{rsfs}{m}{n}
\usepackage{graphicx}
\usepackage{amsmath,bm}
\usepackage{float}
\usepackage{soul}
\usepackage{hyperref}
\usepackage{color}

\definecolor{naviBlue}{RGB}{0,0,128}
 \hypersetup{
	colorlinks=true,        
	linkcolor=naviBlue,         
	citecolor=naviBlue,        
	filecolor=naviBlue,          
	urlcolor=naviBlue           
}           

\newcommand{\newc}{\newcommand}
\newc{\sigmav}{\langle\sigma v \rangle}
\renewcommand{\vec}[1]{\boldsymbol{#1}}
\newcommand{\D}{\mathrm{d}}
\newcommand{\E}{\mathrm{e}}
\newcommand{\I}{{\rm i}}
\newcommand{\closer}{\!\!\;}
\newcommand{\NNj}{{\closer N\closer N_{\closer j}}}
\newcommand{\slopeB}{\beta}
\newcommand{\eVdist}{\kern-0.06em}

\newcommand{\gev}{\:\text{Ge\eVdist V}}
\newcommand{\gv}{\:\text{G\eVdist V}}

\newcommand{\ie}{\emph{i.e.}}
\newcommand{\eg}{\emph{e.g.}}
\newcommand{\cf}{\emph{cf.}}
\newcommand{\oursubsubsection}[1]{\smallskip\emph{#1:}}

\begin{document}

\title{Dark matter or correlated errors: \\
Systematics of the AMS-02 antiproton excess}

\author{Jan Heisig}
\email{jan.heisig@uclouvain.be}
\affiliation{Centre for Cosmology, Particle Physics and Phenomenology (CP3), Universit\'e catholique de Louvain, Chemin du Cyclotron 2, B-1348 Louvain-la-Neuve, Belgium}

\author{Michael Korsmeier}
\email{michael.korsmeier@to.infn.it}
\affiliation{Dipartimento di Fisica, Universit\`a di Torino, Via P. Giuria 1, 10125 Torino, Italy}
\affiliation{Istituto Nazionale di Fisica Nucleare, Sezione di Torino, Via P. Giuria 1, 10125 Torino, Italy}
\affiliation{Institute for Theoretical Particle Physics and Cosmology, RWTH Aachen University, 52056 Aachen, Germany}

\author{Martin Wolfgang Winkler}
\email{martin.winkler@su.se}
\affiliation{The Oskar Klein Centre for Cosmoparticle Physics, Department of Physics, Stockholm University, Alba Nova, 10691 Stockholm, Sweden}
\preprint{CP3-20-19}
\preprint{TTK-20-14}

\begin{abstract}
Several studies have pointed out an excess in the AMS-02 antiproton spectrum at rigidities of 10--20\,GV. Its spectral properties were found to be consistent with a dark-matter particle of mass 50--100\,GeV which annihilates hadronically at roughly the thermal rate. In this work, we reinvestigate the antiproton excess including all relevant sources of systematic errors. Most importantly, we perform a realistic estimate of the correlations in the AMS-02 systematic error which could potentially ``fake'' a dark-matter signal. The  dominant systematics in the relevant rigidity range originate from uncertainties in the cross sections for absorption of cosmic rays within the detector material. For the first time, we calculate their correlations within the full Glauber-Gribov theory of inelastic scattering. The AMS-02 correlations enter our spectral search for dark matter in the form of covariance matrices which we make publicly available for the cosmic-ray community. We find that the global significance of the antiproton excess is reduced to below 1\,$\sigma$ once all systematics, including the derived AMS-02 error correlations, are taken into account. No significant preference for a dark-matter signal in the AMS-02 antiproton data is found in the mass range 10--10000\,GeV.
\end{abstract}

\maketitle

\section{Introduction}

Since their discovery about 40 years ago~\cite{Golden:1979bw,Buffington:1981zz} cosmic-ray antiprotons have been used as a sensitive probe of exotic cosmic-ray sources in our galaxy, such as dark-matter annihilation. As a matter of fact, their first measurement already exhibited an excess over the expected astrophysical background and has stimulated speculations about a dark-matter contribution~\cite{Silk:1984zy,Stecker:1985jc}. While significant theoretical and experimental progress has been made since, today the situation appears similar, although on an entirely different level of precision. By now several tens of thousands of antiproton events have been reported by the AMS-02 experiment on-board the International Space Station rendering statistical uncertainties subdominant over a large range of rigidities~\cite{Aguilar:2016kjl}. In this range systematic errors are at the level of a few percent and constitute the limiting factor in data analyses, which nonetheless allows us to search for a dark-matter contribution potentially as low as $\sim 10\%$ of the total antiproton flux. 

Recently, several groups have reported an
excess over the expected antiproton background in the rigidity range 10--20\,GV in the AMS-02 data which is compatible with a dark-matter annihilation signal~\cite{Cuoco:2016eej,Cui:2016ppb,Cuoco:2017rxb,Reinert:2017aga,Cui:2018klo,Cuoco:2019kuu,Cholis:2019ejx,Lin:2019ljc}. While the significance of the excess is highly controversial (ranging from $1\!-\!5\,\sigma$ in the aforementioned studies), a common picture of the preferred dark-matter properties has emerged. It hints at a particle of mass $m_\chi =50-\!100\:\text{GeV}$ which annihilates into hadronic final states with roughly a thermal cross section, $\langle \sigma v\rangle\sim 10^{-26}\:\text{cm}^2\text{s}^{-1}$. Intriguingly, dark matter with similar properties has been considered in the context of the galactic center gamma-ray excess~\cite{Goodenough:2009gk}. 

The key ingredient to test the dark-matter interpretation of the antiproton excess is a careful modeling of those systematic effects which could, alternatively, have caused the observed spectral feature. To this end, strong efforts have been made to improve the prediction and to quantify the uncertainties of antiproton production by cosmic-ray scattering~\cite{diMauro:2014zea,Kappl:2014hha,Kachelriess:2015wpa,Winkler:2017xor,Korsmeier:2018gcy}. The updated cross-section modeling entering the antiproton background was, indeed, found to somewhat reduce the antiproton excess~\cite{Reinert:2017aga,Cui:2018klo,Cuoco:2019kuu}.
However, one major piece was missing in all previous studies: the correlations of systematic errors in the AMS-02 data which have so far not been reported by the collaboration. Not surprisingly, these are of paramount importance given that correlated systematics can induce unwanted features in the data.
A proof of principle that correlation can have a potentially dramatic effect on the significance of the antiproton excess has been provided in Ref.~\cite{Cuoco:2019kuu}. In this case, correlations have been modeled by simple covariance functions characterized by a correlation length. 
A refined prescription in terms of covariance functions -- splitting the systematic uncertainty into several components -- has been introduced in Ref.~\cite{Derome:2019jfs} for the boron-to-carbon ratio and applied to cosmic-ray antiprotons in Ref.~\cite{Boudaud:2019efq}.

In this study we carefully derive estimates for the most relevant correlations in the AMS-02 (antiproton) data and investigate their implications for the tentative dark-matter signal. For this purpose, similar to Ref.~\cite{Boudaud:2019efq}, we collect all publicly available information to split the systematic error into its components which we then address individually.
In the rigidity range 10--20\,GV the dominant systematics in the antiproton flux  and $\bar p/p$ ratio 
arise from uncertainties in the cross sections for (anti)proton absorption in the AMS-02 detector for which the measured fluxes are corrected.

In the first part of this study, we undertake a detailed re-evaluation of the uncertainties of the involved nucleon-carbon absorption cross sections (the AMS-02 detector is dominantly composed of carbon). To this end, 
we perform a global fit of the absorption cross sections within the full Glauber-Gribov theory of inelastic scattering~\cite{Glauber1959,Sitenko:1959zh,Glauber:1970jm,Gribov:1968jf,Pumplin:1968bi}. 
It links the nuclear absorption cross section to the nucleon-nucleon scattering cross sections and nuclear density functions which are as well subject to experimental measurements.
We use this fit, for the first time in the literature, to reliably extract the correlations in the cross-section uncertainties which we then map to the systematic error in the antiproton flux. 
The second largest contribution to the correlated error stems from the effective acceptance, which the AMS-02 collaboration obtains from a comparison of their detector response between data and Monte Carlo simulation. We estimate the corresponding correlations from the shape (\ie~the ``wiggliness'') of the correction function employed in an AMS-02 analysis. Finally, for the subleading contributions we adopt the correlations estimated in Ref.~\cite{Boudaud:2019efq}.
The full covariance matrix of errors in the AMS-02 antiproton and $\bar{p}/p$ data, which we derive in this work, is made available in the ancillary files on arXiv.

In the second part of this study,
we perform a spectral search for dark matter in the AMS-02 antiproton data, where we fully include the systematic error correlation. The cosmic-ray fits are performed independently in two complementary cosmic-ray propagation setups (following~\cite{Reinert:2017aga} and~\cite{Cuoco:2019kuu}, respectively).
This allows us to draw solid conclusions eliminating further controversies in the assessment of the significance of the antiproton excess.

The remainder of this paper is organized as follows. In Sec.~\ref{sec:nucleonabsorption} we investigate the nucleon-carbon absorption cross sections and derive the corresponding correlation matrix. In Sec.~\ref{sec:correlations_pbar} we collect the various sources of systematic errors to build the overall covariance matrix for the AMS-02 antiproton flux and $\bar p/p$ flux ratio. Finally, in Sec.~\ref{sec:pbarexcess} we derive the implications for the antiproton excess following the two setups mentioned above before drawing our conclusions in Sec.~\ref{sec:concl}.
Appendices~\ref{app:nnxsfit} and \ref{sec:other_correlations}, respectively, provide additional details on the input cross sections of the Glauber-Gribov model and the derivation of error correlations for further cosmic-ray species used in the analyses.
Appendix~\ref{app:cr_prop} summarizes the best-fit values of all involved cosmic-ray propagation parameters in the two setups considered.

\section{Nucleon-nucleus absorption cross sections} \label{sec:nucleonabsorption}

In this section we describe the computation of the nucleon-nucleus 
absorption cross section for $\bar p$C and $p$C which is 
the key ingredient in the assessment of the AMS-02 systematic error.
We perform the computation in the framework of the Glauber model~\cite{Glauber1959,Sitenko:1959zh,Glauber:1970jm}. The theory is formulated based on the eikonal approximation and provides a successful theoretical description for the scattering of moderately relativistic particles off nuclei. 

In the following we first introduce the nuclear density function used in Sec.~\ref{sec:nucdens}. We then detail the computation within the Glauber model in Sec.~\ref{sec:Glauber} while shadowing effects due to inelastic screening are discussed in Sec.~\ref{sec:shad}. 
Parametrizations of the elementary nucleon-nucleon cross sections serving as the input for the Glauber-model computations are presented in Sec.~\ref{sec:nucnucparam}. Finally, we perform a global fit of all input parameters (from the nucleon-nucleon cross sections, the nuclear density function and the inelastic screening) 
to the respective data in Sec.~\ref{sec:global_fit} and derive a correlation matrix from the fit as detailed in Sec.~\ref{sec:correlation_xs_fit}.

\subsection{Parametrization of nuclear densities}
\label{sec:nucdens}

In this work we employ
the harmonic oscillator shell model density~\cite{elton1961nuclear,DeJager:1987qc,Pi:1992ug} which provides a good description for the light nuclei, $3\le A\le 16$~\cite{Rybczynski:2013yba}. It reads
\begin{equation}
\label{eq:HOSdens}
\rho (r) =\frac{4}{\pi^{4/3} C^3}\left[ 1 + \frac{A-4}{6}\left(\frac{r}{C}\right)^2\right] \E^{-r^2/C^2}\,,
\end{equation}
with
\begin{equation}
C = \sqrt{\frac{\langle r_\text{ch}^2\rangle_A - \langle r_\text{ch}^2\rangle_p}{5/2-4/A}}\,,
\end{equation} 
where  $\langle r_\text{ch}^2\rangle_p$ and $\langle r_\text{ch}^2\rangle_A$ are the mean square charge radii
of the proton and nucleus, respectively. We take  
$\langle r_\text{ch}^2\rangle_p=0.7714\,\text{fm}^2$ and $\langle r_\text{ch}^2\rangle_{{}^{12}\text{C}}=6.1\,\text{fm}^2$ (for carbon) as nominal values~\cite{Rybczynski:2013yba,Angeli:2013epw}.
We estimate the uncertainty on $\langle r_\text{ch}^2\rangle_{{}^{12}\text{C}}$ 
to be $0.44\,\text{fm}^2$ by comparing the above value to
the one obtained when taking into account nucleon-nucleon repulsion with an 
expulsion radius of $d=0.9\,$fm (see~\cite{Rybczynski:2013yba}).

\subsection{Computations within the Glauber model}
\label{sec:Glauber}

The absorption cross section of a nucleon $N$ on a nucleus $A$ is obtained by subtracting the respective elastic and quasielastic ($pA\to pA^*)$ 
part from the total cross section (see \emph{e.g.}~\cite{Kopeliovich:1989iy}):
\begin{equation}
\sigma_\text{abs} = \sigma_\text{tot}-\sigma_\text{el}- \sigma_\text{qel}\,.
\end{equation}
Within the Glauber model~\cite{Glauber1959,Sitenko:1959zh,Glauber:1970jm}, neglecting Coulomb effects and spin-orbit interactions~\cite{Faldt:1977qs,Glauber:1978fw}, it is described by (see \emph{e.g.}~\cite{Larionov:2016xeb} for a recent account on the subject):
\begin{equation}
\begin{split}
\label{eq:sigmaabsGlaub}
\sigma_\text{abs}^\text{GM} 
&=\int \D^2 b \left[1 - \left(1- \frac{2\,\text{Im}\chi_N(\vec b)}{A}\right)^{\!A\,} \right]\\
&\simeq \int \D^2 b \left(1-\E^{-2\,\text{Im}\chi_N(\vec b)}\right)\,,
\end{split}
\end{equation}
where $\vec b$ is the impact parameter and $\chi_N(\vec b)$
the nuclear phase-shift function. The last expression in Eq.~\eqref{eq:sigmaabsGlaub} corresponds to the optical approximation valid in the limit of large $A$ (see \emph{e.g.}~the discussion in Ref.~\cite{Kopeliovich:2005us}) which is, however, not used in the numerical analysis.\footnote{The relative difference between the exponential and non-exponential form is found to be up to a few percent for the case under consideration.} The phase-shift function reads
\begin{equation}
\begin{split}
\label{eq:phasesh}
\chi_N(\vec b) = \;&\;\frac\I2 \sum_{j=1}^A\sigma_\NNj (1 - \I \alpha_{\NNj})
\\
&\!\!\times\int\!\frac{\D^2q}{(2\pi)^2}\,\E^{\I \vec b\cdot\vec q} \,\E^{-\slopeB_\NNj q^2/2} \int \!\D^3 r\, \rho_j (\vec r)\, \E^{-\I \vec q\cdot{\vec r}_\text{T}}\,,
\end{split}
\end{equation}
where $\sigma_{N\closer N_j}$ is the total nucleon-nucleon cross section, $\alpha_\NNj$ is the ratio of the real-to-imaginary part of the forward scattering amplitude and $\slopeB_\NNj$ is the slope of the differential inelastic cross section in the forward direction, $\slopeB= \frac{\D}{\D t}\left[\ln\frac{\D\sigma_\text{el}}{\D t}(s,t)\right]_{t=0}$, with $t$ being the four-momentum transfer squared. In general, these three quantities depend on the momentum of the incoming nucleon, $p_\text{lab}$, in the laboratory-frame.
Furthermore, ${\vec r}_\text{T}$ is the transverse part of $\vec r$. 

For a spherical symmetric nuclear density function, $\rho(r)$, the integrals in Eq.~\eqref{eq:phasesh} can be rewritten to
\begin{equation}
\begin{split}
\label{eq:chiNsph}
\chi_N(b) = &\;\frac\I2 \sum_{j=1}^A\sigma_\NNj (1 - \I \alpha_\NNj)\\
& \!\!\!\!\times\int_0^\infty\!\!\!\D q\, q\,J_0(b q) \,\E^{-\slopeB_\NNj q^2/2} 
\int_0^\infty \!\!\!\D r\, r^2  J_0(r q)\,\rho_j(r)\,,
\end{split}
\end{equation}
where $J_n$ denotes the Bessel function of the first kind. Note that the two integrals in
Eq.~\eqref{eq:chiNsph} are real such that the term proportional to $\alpha_\NNj$ does not
contribute to $\text{Im}\chi_N$. 

For the harmonic oscillator shell model density, introduced in Sec.~\ref{sec:nucdens} the integrals in Eq.~\eqref{eq:chiNsph} can be solved analytically leading to:
\begin{equation}
\begin{split}
\text{Im}\chi_N(b) = &\;\sum_{j=1}^A \frac{2}{A}\,\sigma_\NNj \;\exp\!\left({-\frac{b^2}{2 \slopeB_\NNj +C^2}}\right)\\
& \!\times \frac{  4 b^2 C^2+36 \slopeB_\NNj ^2+5 C^4+28 \slopeB_\NNj  C^2}{3 \pi  \left(2 \slopeB_\NNj +C^2\right)^3} \,,
\end{split}
\end{equation}
which we use to solve Eq.~\eqref{eq:sigmaabsGlaub} numerically.

\subsection{Shadowing corrections}
\label{sec:shad}

The Glauber model as described above accounts for elastic screening, \emph{i.e.}~multiple nucleon scattering where the intermediate particle is the nucleon. While providing a good description to the experimental data at low momenta, $p_\text{lab}\lesssim10\,$GeV~\cite{Larionov:2016xeb}, for larger beam momenta it is expected to gradually lose its applicability. This is due to the increasing importance of inelastic screening effects, \emph{i.e.}~rescatterings with excited intermediate states~\cite{Gribov:1968jf,Pumplin:1968bi,Kaidalov:1973fz,VonBochmann:1970xx,Karmanov:1973va} (see also~\cite{Kopeliovich:2003tz,Frankfurt:2011cs} and references therein for more recent accounts on the subject). These effects increase the shadowing correction leading to a reduction of the total cross section.

Here we follow Ref.~\cite{RamanaMurthy:1975vfu}
where inelastic screening has been considered for the case of neutron
projectiles using the expression~\cite{Karmanov:1973va}:
\begin{equation}
\begin{split}
\label{eq:sigmashad}
\Delta \sigma_\text{inel} = 
&
\,- 4\pi \int \D^2 b \;\E^{- \frac12 \sum_{j=1}^A   \sigma_\NNj  \int \D z \, \rho(b,z) }
\\ &\;\times 
\int_{M_0^2}^\infty \D M^2\,\frac{\D^2\sigma(t=0)}{\D M^2 \D t}\, \left| F(q_\text{L},b)\right|^2 \,,
\end{split}
\end{equation}
where $ \D^2\sigma/(\D M^2 \D t)(t\!=\!0) $ is the differential diffraction cross section for the process $N+N_j\to N_j+X$ evaluated at $t=0$,
$q_\text{L} = (M^2-m_N^2) m_N/s$ is the longitudinal momentum transfer in the production of the (exited intermediate) state $X$ with mass $M$, $M_0^2 = (m_N + m_\pi)^2\simeq 1.17\,\text{GeV}^2$, and $F$ is the form factor:
\begin{equation}
F(q_\text{L},b) =  \int \D z \,\rho\!\left(\!\sqrt{b^2+z^2}\right) \E^{\I q_\text{L} z}\,.
\end{equation}
For the purpose of solving Eq.~\eqref{eq:sigmashad} we parametrize the differential diffraction cross section by
\begin{equation}
\begin{split}
\frac{\D^2\sigma(t=0)}{\D M^2 \D t} & = \,\bigg[
a_1  \;\delta \!\left(M^2 - 2.54\,\text{GeV}^2\right)\\
&\;\,+
 \frac{a_2}{M^2-m_N^2} \;\Theta \!\left(M^2 - 5\,\text{GeV}^2\right)\bigg] \,\frac{\text{mb\,}}{\text{GeV}^2}\,,
\end{split}
\end{equation}
where the first and second term provide an effective description of the resonant and continuum contribution, respectively. Note that the structure of the resonant part is in general more complex~\cite{RamanaMurthy:1975vfu,Bartenev:1974gz,Jenkovszky:2012hf}.\footnote{In Ref.~\cite{RamanaMurthy:1975vfu} the resonant part is fitted by a fifth order polynomial.} However, $F$ is a slowly varying function of $M^2$ for $M^2<5\,\text{GeV}^2$, in particular for $p_\text{lab}\gtrsim10\,$GeV where inelastic screening effects are relevant. Hence, the exact shape of the resonant part is not relevant for the computation of the integral in Eq.~\eqref{eq:sigmashad}. We treat $a_1$  and $a_2$ as effective free parameters choosing $a_1=10.6,\,a_2=4.05$ as nominal values and assigning a relative uncertainty of $20\%$.
These values are consistent with the results of~\cite{RamanaMurthy:1975vfu} obtained from a fit to the Fermilab data~\cite{Bartenev:1974gz} on $p +p \to p +X$ and $p +d \to d +X$, where the latter was corrected for binding of the deuteron in an approximate way.\footnote{Note that we choose the continuum part to be proportional to $1/(M^2-m_N^2)$ (instead of $1/M^2$~\cite{Karmanov:1973va,RamanaMurthy:1975vfu}) for $M^2 \ge5 \,\text{GeV}^2$. While the numerical differences are irrelevant considering the involved uncertainties, our choice allows for an analytical calculation of the $M^2$ integral in Eq.~\eqref{eq:sigmashad} and hence significantly improves the numerical performance of the computation.} 

The final absorption cross section is obtained from
\begin{equation}
\sigma_\text{abs} = \sigma_\text{abs}^\text{GM} + \Delta \sigma_\text{inel}\,.
\end{equation}

\subsection{Nucleon-nucleon cross-sections parametrizations}
\label{sec:nucnucparam}

For the nucleon-nucleon cross sections used in the computations within the Glauber model we consider the parametrizations
\begin{equation}
\hspace{-2.5ex}\sigma_{\bar pp}  \!=\sigma_\text{asymp}  
\left(1+ \frac{c_1}{\Delta_s^{1/4}}+\frac{{c_2}}{\Delta_s^{1/2}}+\frac{{c_3}}{\Delta_s}+\frac{{c_4}}{\Delta_s^2}+\frac{{c_5}}{\Delta_s^3}\right),\label{eq:ppcross}
\end{equation}
\vspace{-4.15ex}
\begin{equation}
\hspace{-4.1ex}\sigma_{\bar pn}  \!=\sigma_\text{asymp}  
 \left(1+\frac{{c_6}}{\Delta_s^{1/4}}+\frac{{c_7}}{\Delta_s^{1/2}}+\frac{{c_8}}{\Delta_s}+\frac{{c_9}}{\Delta_s^2}\right),
\end{equation}
\vspace{-2.75ex}
\begin{equation}
\sigma_{pp}  \!
=\sigma_\text{asymp}  
\left(1\!+\!\frac{c_{10}}{\Delta_s^{\!1/4}}\!+\!\frac{c_{11}}{\Delta_s^{\!1/2}}\!+\!\frac{c_{12}}{\Delta_s}\!+\!\frac{c_{13}}{\Delta_s^{\!3/2}}
\!+\!\frac{c_{14}}{\Delta_s^2}\!+\!\frac{c_{15}}{\Delta_s^3}\right),
\end{equation}
\vspace{-3.2ex}
\begin{equation}
\sigma_{pn}  \!=\sigma_\text{asymp}  
 \left(1+\frac{c_{16}}{\Delta_s^{1/4}}+\frac{c_{17}}{\Delta_s^{1/2}}+\frac{c_{18}}{\Delta_s}+\frac{c_{19}}{\Delta_s^{3/2}}+\frac{c_{20}}{\Delta_s^2}\right),\label{eq:pncross}
\end{equation}
where $\Delta_s(p_\text{lab}) =\left(s(p_\text{lab})-s(0)\right)/\text{GeV}^2$ with
\begin{equation}
    s  =  2 m_N \!\left( \sqrt{m_N^2+p_\text{lab}^2}+m_N \right)\,,
\end{equation}
and $\sigma_\text{asymp}$ taken from Ref.~\cite{Ishida:2009fp}:
\begin{equation}
\sigma_\text{asymp}  = \left[ 36.04 + 0.304 \,\ln^2\!\left(\frac{s}{33.1\,\text{GeV}^2}\right)\right] \text{mb}\,.
\end{equation}
The expressions in Eqs.~\eqref{eq:ppcross}--\eqref{eq:pncross} are valid at $p_\text{lab}\gtrsim 1\gev$. Furthermore, we parametrize the slope of the differential inelastic cross section in the forward direction by:
\begin{align}
\slopeB_{\bar p p} & =\left[d_1+\frac{d_2}{p_\text{lab}/\text{GeV}}+d_3 \ln\!\left(\frac{p_\text{lab}}{\text{GeV}}\right)\right] \text{GeV}^{-2}\,,
\\
\slopeB_{p p}& =\left[d_4+\frac{d_5}{p_\text{lab}/\text{GeV}}+d_6 \ln\!\left(\frac{p_\text{lab}}{\text{GeV}}\right)\right] \text{GeV}^{-2}\,,
\end{align}
and assume $\slopeB_{\bar p n} = \slopeB_{\bar p p}$ and $\slopeB_{p n} = \slopeB_{p p}$.
We determine the best-fit values and covariance matrices for the involved parameters $c_1,\, \dots, \,c_{20},\, d_1,\, \dots, \,d_6$  by performing fits to the cross-section data collected in Ref.~\cite{Tanabashi:2018oca}\footnote{We observe that the cross-section data~\cite{Tanabashi:2018oca} contain underestimated systematic errors. This becomes obvious from the fact that, with errors taken at face value, the cross-section data are poorly fitted by any smooth function. We, therefore, introduce an additional systematic error of $5\%$ for each data point, which we assumed to be fully correlated within one experiment. In the case of $\bar{p}p$-scattering, where inconsistencies are larger, we furthermore had to multiply the errors by a factor of 2. With the described procedure, we arrived at a realistic goodness of fit of $\sim 1/ \text{dof}$.} and the data on the slope parameter~\cite{Okorokov:2015bha}\footnote{While the slope parameter is defined at momentum-transfer $t=0$, the experimental data~\cite{Okorokov:2015bha} were taken at small, but non-vanishing $t$. This causes small systematic differences in the normalization of data taken at the same laboratory momentum. To evade a degradation of the fit, we, therefore, bin the data with a bin size of $0.2$ in units of $\log_{10}p_{\text{lab}}$ ($p_{\text{lab}}$ in GeV). For each bin we, furthermore, add a systematic uncertainty of $5\%$ to account for the error caused by the non-vanishing $t$ extraction.}.
The best-fit values and the covariance matrices are given in Appendix~\ref{app:nnxsfit}. The corresponding uncertainty band for the nucleon-nucleon cross sections and slopes (minimal and maximal uncertainty over the full momentum range) are provided in Table~\ref{tab:uncertaintysigmabeta} in the Appendix.

\subsection{Global fit}\label{sec:global_fit}

To compute the correlations in the cross-section uncertainties we perform global fits of the $\bar p$C and $p$C absorption cross section within the Glauber model (including inelastic screening) varying the input parameters $c_i,\,d_i$ as well as $\langle r_\text{ch}^2\rangle_{{}^{12}\text{C}},\,a_1,\,a_2$ according to their nominal values and (correlated) uncertainties (using covariance matrices shown in Appendix~\ref{app:nnxsfit}). Note that the parameters $\langle r_\text{ch}^2\rangle_{{}^{12}\text{C}},\,a_1,\,a_2$ are varied independently for $\bar p$C and $p$C, considering the respective parametrization to be an effective description that might account for additional effects not explicitly considered in the calculation.

For the experimental measurements of the $\bar p$C and $p$C absorption cross sections, we use data from Refs.~\cite{Allaby:1970pv,Abrams:1972ab,Denisov:1973zv,Carroll:1978hc,Nakamura:1984xw}\footnote{
The data set~\cite{Abrams:1972ab} originally contains a large number of bins in the momentum range $1.6\!-\!3.25\gev$. Since systematic errors of $2\!-\! 5$\% dominate over statistical errors, their correlations would be important. Due to their unavailability, we combine the data set into three bins, and conservatively take the error to be 5\% in each bin (statistical errors are negligible after the combination). By reducing the number of bins, we minimize the impact of the (unknown) error correlations.} and~\cite{Denisov:1973zv,Carroll:1978hc,Chen:1955nkq,Moskalev:1956,Booth:1957,Bowen:1958,Batty:1958,Low:1958,Longo:1962zz,Grigorov:1964,Basilova:1966,Bellettini:1966zz,Grigorov:1970yt,Alakoz:1971,Renberg:1972jf,McGill:1974zz,Lindstrom:1975xr,Baros:1978,Heckman:1978ib,Akhababyan:1979,Bobchenko:1979hp,Afanasev:1984,Grchurin:1985,Abgrall:2011ae}, respectively. Since we observe (mild) inconsistencies in the data, in addition to the reported experimental errors, we included an additional normalization error of 10\% for each data point which we took to be fully correlated within the same experiment (but uncorrelated between different experiments). This normalization error also accounts for the fact that information on the inclusion of the quasi-inelastic part of the cross section is not available for a large fraction of the data (see \eg~the discussion in Refs.~\cite{Carroll:1978hc,Abgrall:2011ae}).

\begin{figure*}[t]
\centering
\setlength{\unitlength}{1\textwidth}
\begin{picture}(1,0.35)
 \put(-0.0055,-0.02){\includegraphics[width=0.53\textwidth, trim= {3.3cm 2.2cm 3cm 2cm}, clip]{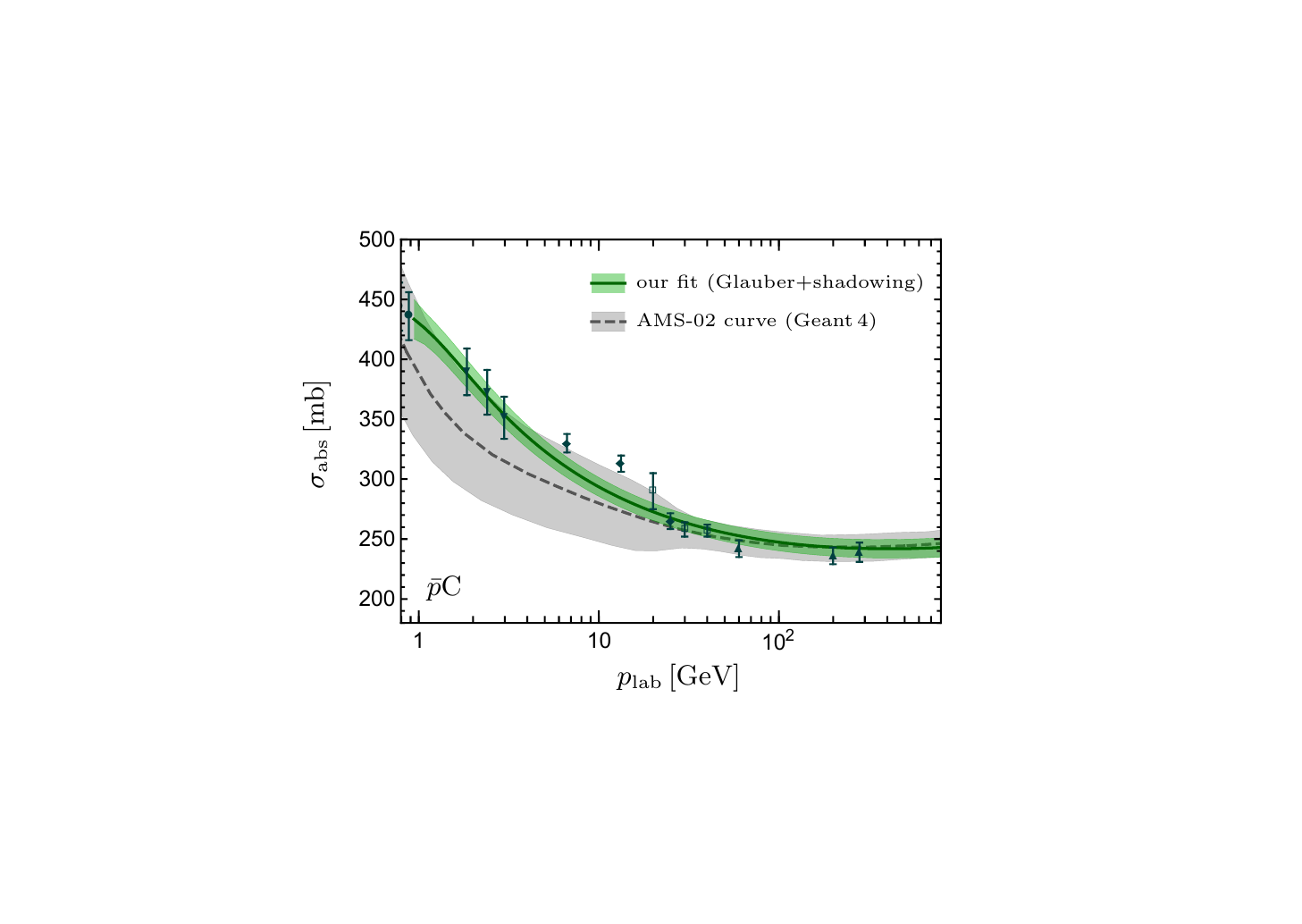}}
 \put(0.506,-0.02){\includegraphics[width=0.53\textwidth, trim= {3.3cm 2.2cm 3cm 2cm}, clip]{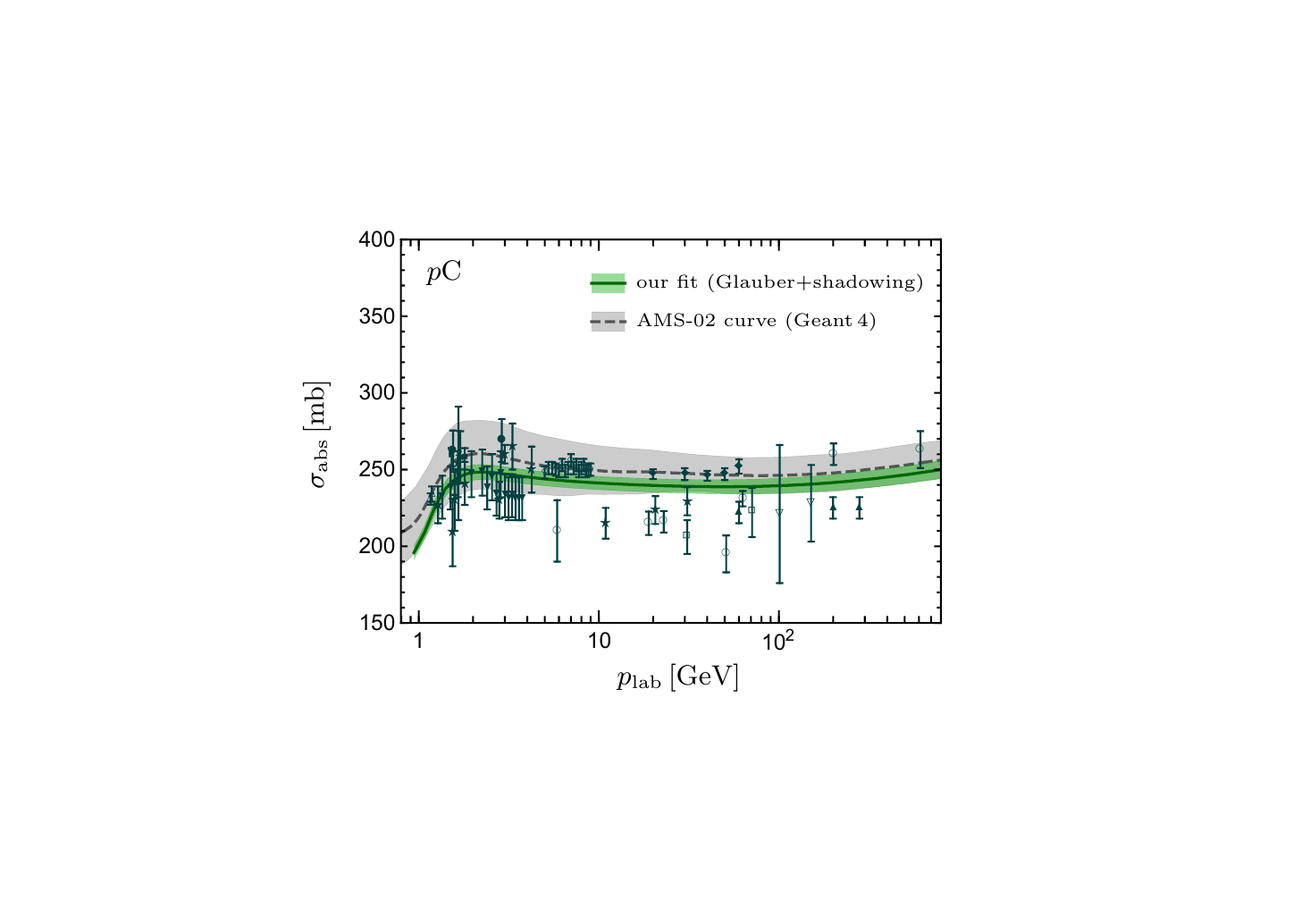}}
\end{picture}
\caption{
Absorption cross section for $\bar p$C (left panel) and $p$C (right panel) as a function of the projectile momentum $p_\text{lab}$. The solid dark green curve and green shaded band denote our best-fit cross section and its $1\sigma$ uncertainty, respectively. The data points (containing $1\sigma$ error bars) of different experiments are denoted by individual symbols except for the star, which represents a collection of 15 experiments each of which, however, only provides one data point. Note that the 10\% normalization error of each experiment is not included in the error bars. For comparison, we also show the corresponding cross sections used in the AMS-02 analyses stemming from an implementation in Geant 4 (dashed curve and gray shaded band).
\label{fig:pbCpCxs}  
}
\end{figure*}

We sample the 15- (17-)dimensional input parameter space for $\bar p$C ($p$C) with the multimodal nested sampling algorithm \textsc{MultiNest}~\cite{Feroz:2008xx}.\footnote{We use 5000 live points, a sampling efficiency of 0.65 and an evidence tolerance of $10^{-18}$.} The corresponding best fits are shown in Fig.~\ref{fig:pbCpCxs} as the solid (dark green) curves. The green band denotes the $1\sigma$ uncertainty band. For comparison we display the respective results used in the AMS-02 analyses taken from~\cite{Zuccon:2019} (gray dashed curve and gray shaded error band). The best-fit values for the input parameters in the global fits are displayed in Table~\ref{tab:globalfits}. It is interesting to note that they are very close to their nominal values (not involving $\bar p$C and $p$C data). A similar observation is made by performing the fit without the $\chi^2$-contribution from $\bar p$C and $p$C data (not shown here) which  provides very similar results for the $\bar p$C and $p$C absorption cross section as in Fig.~\ref{fig:pbCpCxs}. We, hence, observe that our computation within the Glauber model including inelastic screening effects provides an excellent description of the measured $\bar p$C and $p$C cross section over the whole considered range of momenta.\footnote{This holds strictly for $\bar p$C. However, in the case of $p$C we observe a significant deviation of $a_2$ from its nominal value towards a reduction of inelastic screening effects, \ie~a slightly larger cross section for $p_\text{lab}\gtrsim100\,$GeV as compared to a fit without $p$C data.
While such a reduction can arise from multi-nucleon correlations (see \eg~\cite{Alvioli:2009iw}) we notice that 
it is mainly driven by the cosmic-ray data reported in Ref.~\cite{Grigorov:1970yt} (open circles in Fig.~\ref{fig:pbCpCxs}) which are subject to further systematics, as discussed in Ref.~\cite{Ganguli:1973rja}. In fact, omitting this data fully reconciles the discrepancy resembling the same situation as for $\bar p $C. However, the inclusion or omission of this data set 
does not have a significant effect on
the correlation matrix extracted from our fit. 
}

\begin{table}[t]
\begin{center}
\renewcommand{\arraystretch}{1.5}
\begin{tabular}{|c | c  c  | c c |} 
\hline
 & \multicolumn{2}{c |}{$\bar p$C} & \multicolumn{2}{c|}{$p$C}\\
 \hline
               &   $c_1$  &  0.0712   &  $c_{10}$ &  $-0.404$ \\    
               &   $c_2$  &  1.42     &  $c_{11}$ &  2.43   \\
$\sigma_{Np}$  &   $c_3$  &  1.96     &  $c_{12}$ &  $-9.31$  \\  
               &   $c_4$  &  $-2.23$    &  $c_{13}$ &  20.5   \\ 
               &   $c_5$  &  0.814    &  $c_{14}$ &  $-16.4$  \\  
               &          &           &  $c_{15}$ &  3.06   \\
 \hline
               &   $c_6$  &  $-0.0194$  &  $c_{16}$ &  0.102  \\  
               &   $c_7$  &  1.69     &  $c_{17}$ &  0.261  \\  
$\sigma_{Nn}$  &   $c_8$  &  1.42     &  $c_{18}$ &  $-0.120$ \\   
               &   $c_9$  &  $-0.998$   &  $c_{19}$ &  $-0.805$ \\  
               &          &           &  $c_{20}$ &  0.497  \\ 
 \hline
               &   $d_1$  &  10.7     &  $d_4$    &  7.65   \\
 $\beta_{Np}$  &   $d_2$  &  5.77     &  $d_5$    &  $-5.14$  \\
               &   $d_3$  &  0.307    &  $d_6$    &  0.827  \\
 \hline
$a_1$    & \multicolumn{2}{c |}{11.2} &\multicolumn{2}{c|}{10.6} \\
$a_2$    & \multicolumn{2}{c |}{4.40} &\multicolumn{2}{c|}{1.44} \\
$\; \langle r_\text{ch}^2\rangle_{{}^{12}\text{C}}\,/\,\text{fm}^2\;$ & \multicolumn{2}{c |}{6.17} &\multicolumn{2}{c|}{6.06} \\
\hline
$\chi^2_{\sigma_{Np}}$     &  \multicolumn{2}{c |}{0.033} &\multicolumn{2}{c|}{0.086} \\
$\chi^2_{\sigma_{Nn}}$     &  \multicolumn{2}{c |}{0.12} &\multicolumn{2}{c|}{0.22} \\
$\chi^2_{\beta_{Np}}$      &  \multicolumn{2}{c |}{0.009} &\multicolumn{2}{c|}{0.13} \\
$\chi^2_{a_1}$             &  \multicolumn{2}{c |}{0.040} &\multicolumn{2}{c|}{0.0005} \\
$\chi^2_{a_2}$             &  \multicolumn{2}{c |}{0.080} &\multicolumn{2}{c|}{4.62} \\
$\chi^2_{\langle r_\text{ch}^2\rangle_{{}^{12}\text{C}}}$  &  \multicolumn{2}{c |}{0.028} &\multicolumn{2}{c|}{0.0077} \\
$\chi^2_{\sigma_\text{abs}}/n_\text{dof}$ &  \multicolumn{2}{c |}{14.0/13} &\multicolumn{2}{c|}{56.8/72} \\
\hline
$\chi^2_{\text{tot}}$ &    \multicolumn{2}{c |}{14.3} &\multicolumn{2}{c|}{61.9} \\
\hline
\end{tabular}
\renewcommand{\arraystretch}{1}
\end{center}
\caption{
Best-fit values of the input parameters and the respective $\chi^2$ contributions for the global fits of the absorption cross section for $\bar p$C and $p$C.
}
\label{tab:globalfits}
\end{table}

\subsection{Correlation matrices for AMS-02 cross sections}\label{sec:correlation_xs_fit}

From the global fit described above, we can infer 
a covariance matrix for $\sigma_\text{abs}$ evaluated at the values of $p_\text{lab}$ that
correspond to the rigidity bins chosen by AMS-02. In this way, we can incorporate correlations
in the AMS-02 flux measurements arising from $\sigma_\text{abs}$ uncertainties. 

Generally, the covariance matrix of a quantity $\vec\theta$ can be computed using
\begin{equation}
\label{eq:covmatdev}
\mathcal{V}_{ij} = \sum_k w^k \left( \theta_i^k - \bar\theta_i \right) \left( \theta_j^k - \bar\theta_j \right),
\end{equation}
where $w^k$ is the statistical weight (as given by \textsc{MultiNest}) of the $k$th point in the statistical 
ensemble (the \textsc{MultiNest} chain) and $\bar\theta_i = \sum_k w^k \theta_i^k$.
For the absorption cross-section uncertainties $\theta_i^k = \sigma^k_\text{abs}(p_{\text{lab},i})$, where the $p_{\text{lab},i}$ correspond to the bins of AMS-02. The corresponding correlation matrices read
\begin{equation}
\rho_{ij}=\frac{\mathcal{V}_{ij}}{\sqrt{\mathcal{V}_{ii}\mathcal{V}_{jj}}}\,.
\end{equation}
We expect the obtained correlations to exhibit a dependence on the theoretical framework (Glauber-Gribov) and our parametrizations of input functions. While our choices are thoroughly motivated, quantifying this dependence goes beyond the scope of this work.
The correlation matrices of the antiproton-carbon and proton-carbon absorption cross-section uncertainties are provided in the ancillary files on arXiv.
Note that in the following we make use of these correlation matrices only. The absolute cross-section uncertainties are taken as reported by the AMS-02 Collaboration (corresponding to the gray bands in Fig.~\ref{fig:pbCpCxs}). This is a conservative approach since the uncertainties employed by AMS-02 are larger compared to those in our fit. Details of our approach will be described in Sec.~\ref{sec:covariance}.

\section{Error correlations in the AMS-02 antiproton data}\label{sec:correlations_pbar}

To determine the error correlations in the AMS-02 $\bar{p}$ and $\bar{p}/p$ data, we follow a two-step procedure: We first split the systematic errors into individual contributions (as described below). Then, in Sec.~\ref{sec:covariance}, we derive the correlations for each sub-error and build up the full AMS-02 covariance matrices. Our cosmic-ray fits will also require the AMS-02 covariance matrices for the proton and helium fluxes as well as the B/C ratio as an input. Their calculation (which proceeds analogous to the antiproton case) is described in Appendix~\ref{sec:other_correlations}.
 
\begin{figure*}[t]
\centering
\setlength{\unitlength}{1\textwidth}
\begin{picture}(0.99,0.44)
 \put(-0.0055,-0.02){\includegraphics[width=0.53\textwidth, trim= {3.3cm 2.2cm 3cm 0.8cm}, clip]{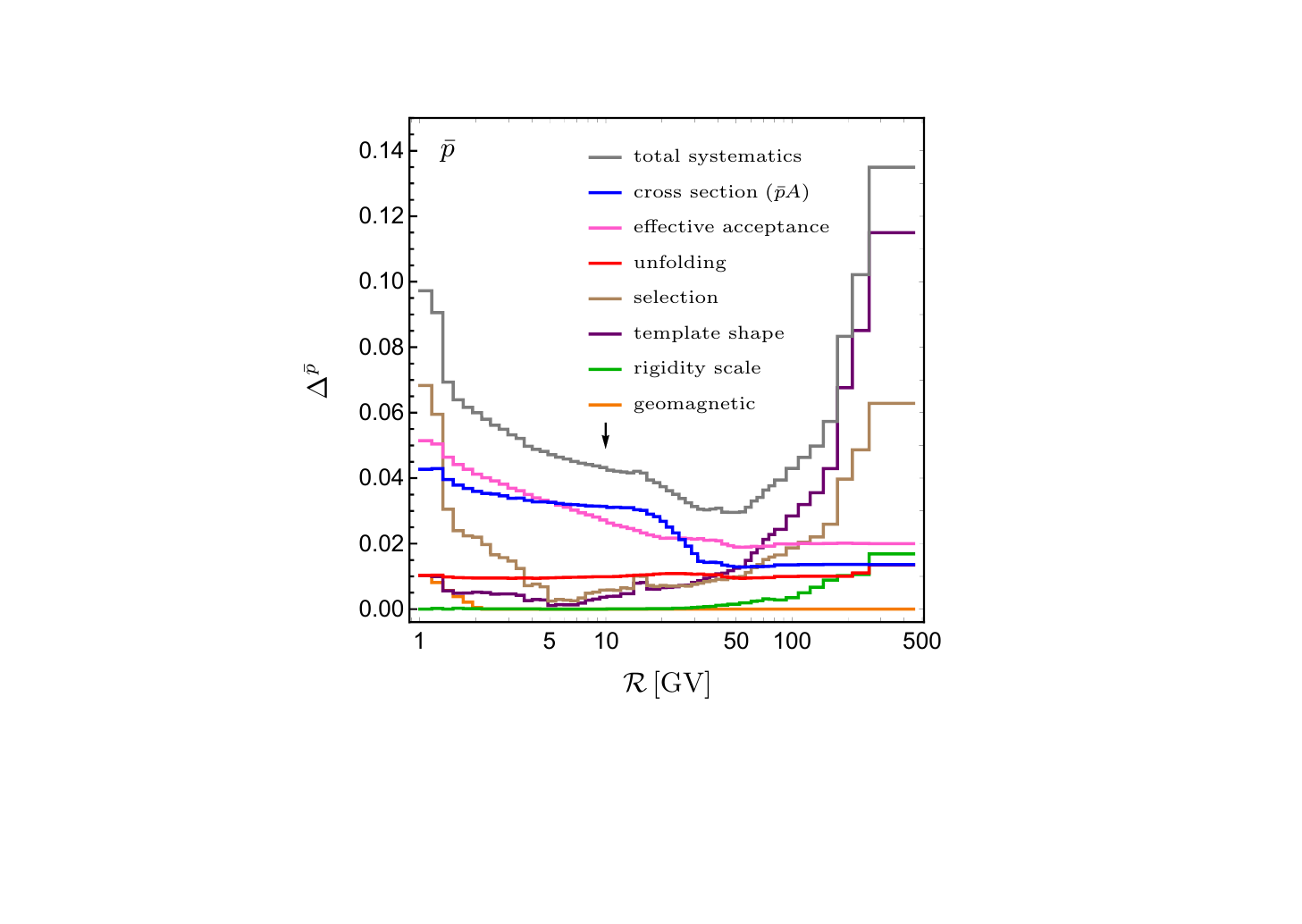}}
 \put(0.506,-0.02){\includegraphics[width=0.53\textwidth, trim= {3.3cm 2.2cm 3cm 0.8cm}, clip]{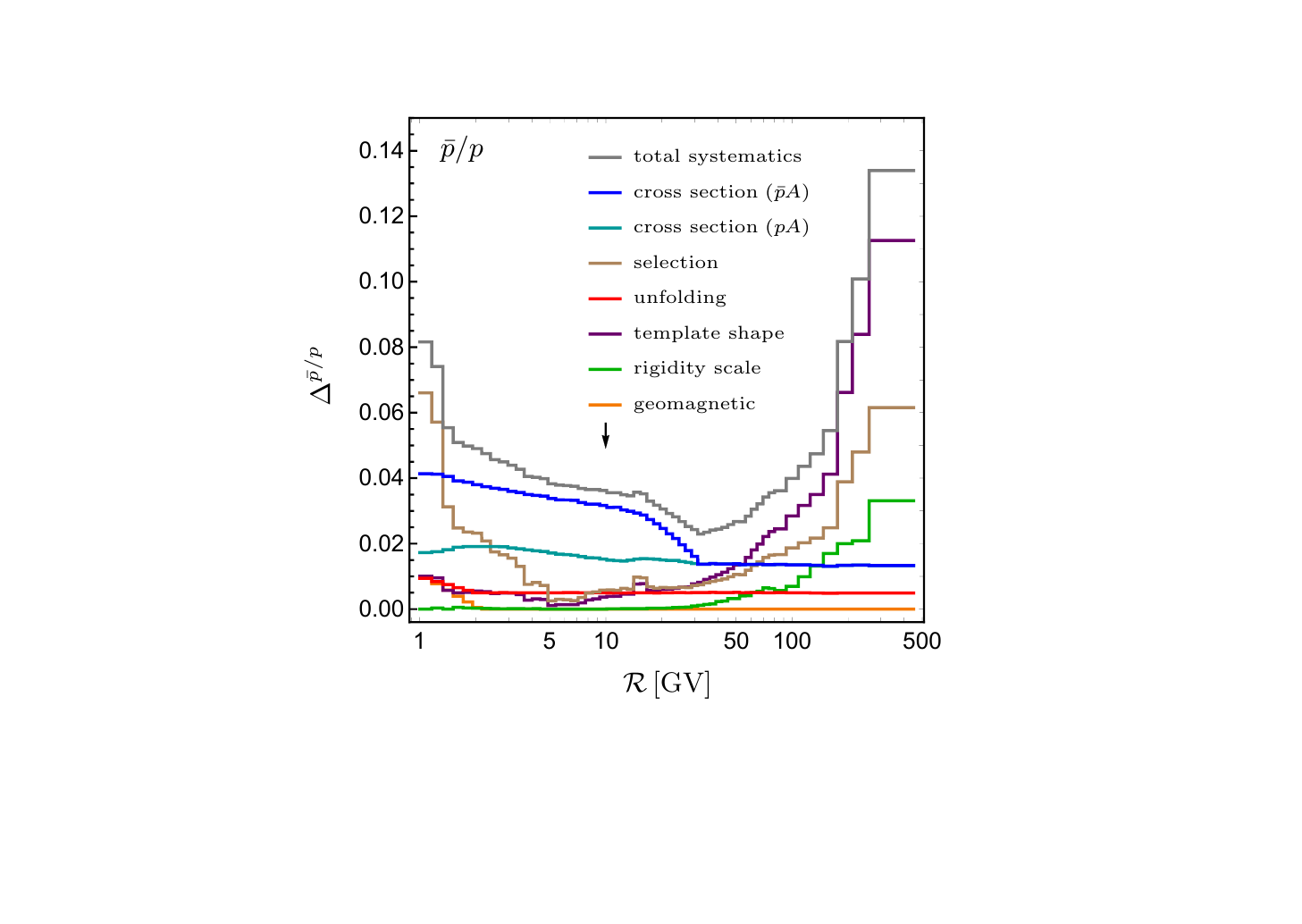}}
\end{picture}
\caption{
Reconstructed relative systematic errors in the AMS-02 antiproton (left) and $\bar{p}/p$ data (right). The contributions in the legend are ordered according to their size at 10\,GV as indicated by the arrow.
\label{fig:systematicerrors}
}
\end{figure*}

\subsection{Systematic errors}

In the following, we will denote relative systematic errors in the antiproton flux by $\Delta^{\bar{p}}$ and in the $\bar{p}/p$ ratio by $\Delta^{\bar{p}/p}$. 

\oursubsubsection{Unfolding error}
Detector resolution effects cause the migration of events into neighboring rigidity bins. This must be corrected for through the unfolding procedure. The choice of the migration matrices (characterizing the migration probabilities) is associated with a systematic error. 
This unfolding error is $\Delta^{\bar{p}}_{\text{unf}}=1\%$ at $\mathcal{R}<200\gv$ and $1.5\%$ at $\mathcal{R}=450\gv$ for the antiproton flux. The error partially cancels in the $\bar{p}/p$ ratio for which it becomes $\Delta^{\bar{p}/p}_{\text{unf}}=1\%$ at $\mathcal{R}=1\gv$ and $0.5\%$ at $\mathcal{R}>2\gv$~\cite{Aguilar:2016kjl}. Between the stated rigidity intervals we interpolate logarithmically. 

\oursubsubsection{Cross-section error}
The (rigidity-dependent) AMS-02 acceptance is sensitive to the fraction of cosmic rays which are absorbed in the detector. The survival probability $P_N$ of the incoming particle $N$ ($N=\bar{p},\,p$) with momentum $p$ is estimated as
\begin{equation}\label{eq:survival}
  P_N = \exp\left({-\sum\limits_A n_{A}(p) \,\sigma_\text{abs}^{NA}(p)}\right)\,,
\end{equation}
where $n_{A}(p)$ accounts for the amount of detector material with mass number $A$ which has to be traversed by the incoming cosmic ray, while $\sigma_\text{abs}^{NA}$ is the corresponding absorption cross section. We note that the material thickness acquires an effective momentum-dependence due to cuts on track length performed in the AMS-02 analysis. For simplicity, we will neglect subdominant material admixtures and assume that the AMS-02 detector is entirely comprised of carbon, as the corresponding cross-section error correlations are expected to be very similar. 

We can extract the cross-section error $\Delta^{\bar{p}/p}_{\text{xs}}$ in the $\bar{p}/p$ data by (quadratically) subtracting the unfolding error (as derived above) from the acceptance error as given in Ref.~\cite{Chen:2017}. Notice that $\Delta^{\bar{p}/p}_{\text{xs}}$ is the quadratic sum of the proton and antiproton contribution to the cross-section uncertainties, i.e.
\begin{equation}\label{eq:sum_xs_errors}
  \Delta^{\bar{p}/p}_{\text{xs}}=\sqrt{(\Delta^{p}_{\text{xs}})^2+(\Delta^{\bar{p}}_{\text{xs}})^2}\,.
\end{equation}
From Eq.~\eqref{eq:survival}, it, furthermore, follows that
\begin{equation}\label{eq:relative_xs_errors}
  \frac{\Delta^{\bar{p}}_{\text{xs}}}{\Delta^p_{\text{xs}}}=\frac{\Delta{\sigma_\text{abs}^{\bar{p}\text{C}}}}{\Delta{\sigma_\text{abs}^{p\text{C}}}}\,,
\end{equation}
at linear order. Here, $\Delta\sigma_\text{abs}^{\bar{p}\text{C}}$ and $\Delta\sigma_\text{abs}^{p\text{C}}$ denote the (absolute) uncertainties in the antiproton and proton absorption cross sections on carbon, respectively, which we extract from~\cite{Zuccon:2019}. By combining Eqs.~\eqref{eq:sum_xs_errors} and~\eqref{eq:relative_xs_errors}, we then extract $\Delta^{p}_{\text{xs}}$ and $\Delta^{\bar{p}}_{\text{xs}}$. Of course, only $\Delta^{\bar{p}}_{\text{xs}}$ contributes to the systematic error in the antiproton flux. 

\oursubsubsection{Scale error}
The absolute rigidity scale of measured events can be affected by misalignment of the tracker planes and small uncertainties in the magnetic field map of the inner tracker. The AMS-02 collaboration estimated the corresponding systematic uncertainty by comparing electron and positron rigidity measurements in tracker and electromagnetic calorimeter. The scale error $\Delta^{\bar{p}/p}_{\text{scale}}$ is directly given in Ref.~\cite{Chen:2017}. Since the rigidity scale error affects protons and antiprotons in the opposite way, one finds $\Delta^{\bar{p}}_{\text{scale}}=0.5\,\Delta^{\bar{p}/p}_{\text{scale}}$.

\oursubsubsection{Effective acceptance error}
A residual systematic error in the effective folded acceptance is estimated by comparing efficiencies in several detector parts as extracted from Monte Carlo simulation with direct measurements. The effective acceptance error amounts to $\Delta^{\bar{p}}_{\text{eff.\,acc.\!}}=5\%$ at $\mathcal{R}=1\gv$ and 2\% at $\mathcal{R}>20\gv$~\cite{Aguilar:2016kjl}. Between $1$ and $20\gv$, we perform a logarithmic interpolation. The effective acceptance error affects antiprotons and protons in the same way, it cancels in the $\bar{p}/p$ ratio.

\oursubsubsection{Geomagnetic error}
To reject indirect cosmic rays produced in the earth's atmosphere, AMS-02 applies a rigidity cut above the geomagnetic cutoff. The measured cosmic-ray fluxes (at low rigidity) exhibit a small residual dependence on the exact numerical choice of the cutoff factor. The corresponding systematic error is estimated by varying the cutoff factor in the event selection. The geomagnetic error is $\Delta^{\bar{p}}_{\text{geo}}=1\%$ at $\mathcal{R}=1\gv$ and vanishes at $\mathcal{R}>2\gv$~\cite{Aguilar:2016kjl}, in between we interpolate logarithmically. Since the geomagnetic error is significantly smaller for protons~\cite{Aguilar:2016kjl}, we assume $\Delta^{\bar{p}/p}_{\text{geo}}=\Delta^{\bar{p}}_{\text{geo}}$.

\oursubsubsection{Template shape and selection error}
AMS-02 uses templates to separate signal from background events. Systematic uncertainties arise from the choice of the template shape. The template error mostly affects antiprotons. It is dominated by the contribution related to the charge confusion of incoming protons. In addition, the event selection is affected by a systematic error related to the cuts on the track shape which are used to identify a certain cosmic-ray species. The selection error again mostly affects antiprotons. From~\cite{Chen:2017}, we can extract the systematic error on the event number which corresponds to the quadratic sum $\sqrt{\big(\Delta^{\bar{p}/p}_{\text{geo}}\big)^2+\big(\Delta^{\bar{p}/p}_{\text{template}}\big)^2+\big(\Delta^{\bar{p}/p}_{\text{selection}}\big)^2}$. To derive the individual errors, we use the geomagnetic error from above and assume the following relative size of template and selection errors
$  \Delta^{\bar{p}/p}_{\text{selection}}/\Delta^{\bar{p}/p}_{\text{template}} = 0.48 + 6.5/\mathcal{R}^{0.78} $. This function was chosen to reproduce the ratio of the two errors at several rigidities as given in Ref.~\cite{Aguilar:2016kjl}. Since both errors are mainly relevant for antiprotons, we take $\Delta^{\bar{p}}_{\text{selection}}=\Delta^{\bar{p}/p}_{\text{selection}}$ and $\Delta^{\bar{p}}_{\text{template}}=\Delta^{\bar{p}/p}_{\text{template}}$.

\medskip

When (quadratically) summing the individual systematic errors in the antiproton flux, a tiny mismatch with the published overall systematic error is observed. We eliminate the mismatch by rescaling all individual errors with a correction factor which varies at most by a few percent from unity over the full rigidity range. Figure~\ref{fig:systematicerrors} summarizes the resulting systematic errors in the antiproton flux and the $\bar{p}/p$ ratio as a function of rigidity.\footnote{The reconstructed systematic errors obtained here differ slightly from those in Ref.~\cite{Boudaud:2019efq} as we choose to include additional information provided in Ref.~\cite{Chen:2017}.}

\subsection{Covariance matrices for AMS-02 errors}\label{sec:covariance}

After splitting the AMS-02 systematic error into its various components, we will now assign correlation matrices $\rho^{\bar{p}}_a$, $\rho^{\bar{p}/p}_a$ ($a=\text{unf},\,\text{xs},\,\text{scale},\,\dots$) to each of the sub-errors. The leading uncertainty (in the regime where the systematic error dominates over the statistical error) derives from the absorption cross sections. The reported, \ie~absorption-corrected (anti)proton flux scales inversely with the (anti)proton survival probability which was defined in Eq.~\eqref{eq:survival}. Therefore, at linear order in the cross-section error, the correlation matrices $\rho^{\bar{p}}_\text{xs}$ and $\rho^{p}_\text{xs}$
are identical to the correlation matrices of uncertainties in the absorption cross sections $\sigma_\text{abs}^{\bar{p}\text{C}}$ and $\sigma_\text{abs}^{p\text{C}}$, respectively, (assuming that the same rigidity bins are chosen) which were derived in Sec.~\ref{sec:correlation_xs_fit}.
Note that the correlation matrix $\rho^{\bar{p}/p}_\text{xs}$ contains the contributions from $\rho^{\bar{p}}_\text{xs}$ and $\rho^{p}_\text{xs}$ (weighted by the relative magnitude of the antiproton and proton cross-section uncertainties).

For the remaining uncertainties, we follow the approach of~\cite{Derome:2019jfs,Boudaud:2019efq} and define the following correlation matrix\footnote{In Ref.~\cite{Boudaud:2019efq} the covariance function approach has also been employed to model antiproton absorption cross-section uncertainties and a correlation length $\ell_{\text{xs}}=1$ was chosen. In contrast, the correlations we employ in this work -- derived from our fit within the Glauber-Gribov theory -- do not reduce to simple correlation function. If we \eg~separate our antiproton correlation matrix into five sub-blocks of equal size and fit each subblock to the form Eq.~\eqref{eq:correlation_parameterization}, we find that the correlation length varies in the range $\ell_{\text{xs}}\sim 0.5\! -\! 3$ (with $\ell_{\text{xs}}$ increasing towards high rigidity).}
\begin{equation}\label{eq:correlation_parameterization}
   \left(\rho^{\bar{p}}_a\right)_{ij} = \exp\left[  -\frac{1}{2} \left(\frac{\log_{10}(\mathcal{R}_i/\mathcal{R}_j)}{\ell_a}\right)^2        \right]\,,
\end{equation}
for each systematic uncertainty in the antiproton flux. Here, $\mathcal{R}_i$ denotes the (mean) rigidity of the $i$-th bin. The correlation lengths (in units of energy decade) depend on the error under consideration. The correlation matrices $\big(\rho^{\bar{p}/p}_a\big)_{ij}$ of uncertainties in the $\bar{p}/p$ ratio are defined analogously.

\begin{figure}[t]
\centering
\setlength{\unitlength}{1\textwidth}
\begin{picture}(0.53,0.335)
 \put(-0.0055,-0.02){\includegraphics[width=0.53\textwidth, trim= {3.3cm 2.2cm 3cm 2cm}, clip]{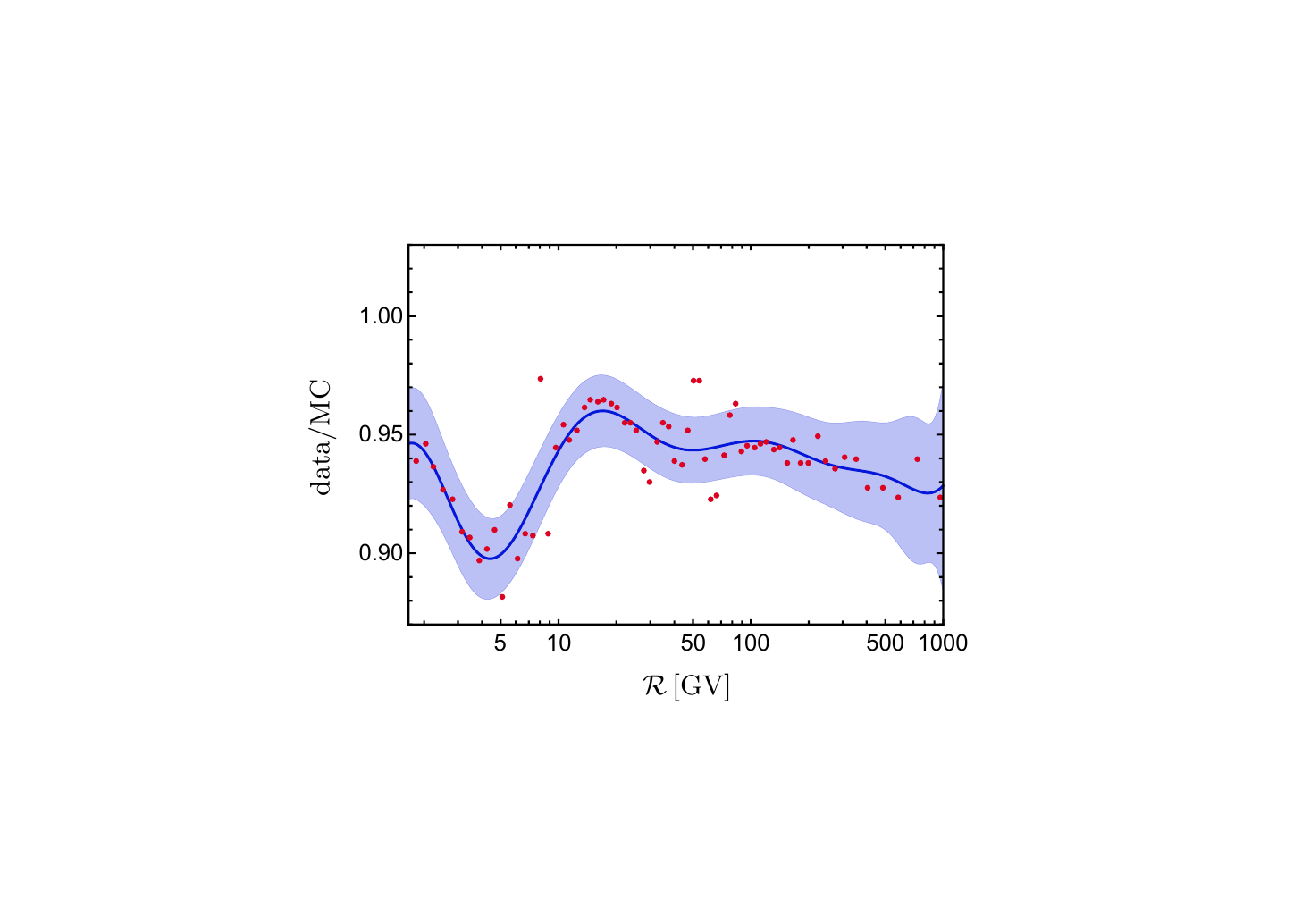}}
\end{picture}
\caption{Effective acceptance correction function extracted from~\cite{Konak:2019}. The `wiggliness' of the function provides a measure for correlation length of the effective acceptance error in the AMS-02 antiproton data.}
\label{fig:effective_acceptance}
\end{figure}

Apart from the cross-section error, the effective acceptance error plays a significant role. It is derived from a data versus Monte Carlo comparison and may receive contributions from mismodeling of efficiencies in various detector parts or from small errors in the detector composition model. Since it amounts to a collection of different residual errors, it is difficult to gain any intuitive insights into the corresponding error correlations. However, a realistic estimate of the correlation length can be obtained by analyzing the `wiggliness' of the data/MC correction function employed by AMS-02\@. In the AMS-02 analysis, the latter is determined from proton data and then assumed to be identical for antiprotons (this is why the effective acceptance correction and the corresponding error cancel in the $\bar{p}/p$ ratio). We extract the data/MC correction function from the proton flux analysis in the Ph.D. thesis~\cite{Konak:2019} by taking the ratio of effective and geometric efficiency.\footnote{We note that the systematic errors derived in Ref.~\cite{Konak:2019} differ somewhat from the official AMS-02 proton analysis~\cite{Aguilar:2015ooa}. However, we only use~\cite{Konak:2019} to extract the correlation length of the effective acceptance error which should be hardly affected by small analysis differences compared to~\cite{Aguilar:2015ooa}. We validated that our determination of $\ell_{\text{eff.\,acc.}}$ is not particularly sensitive to the analysis choices of~\cite{Konak:2019} by a comparison with the (inofficial) AMS-02 helium analysis presented in Ref.~\cite{Alaoui:2016}. The effective acceptance error correlation length obtained from Fig.\ 3.31 in Ref.~\cite{Alaoui:2016} differs by less than 20\% from the value obtained from Fig.~\ref{fig:effective_acceptance} in the main text.} In Fig.~\ref{fig:effective_acceptance}, we fit a polynomial of 12th degree in $\log_{10}\mathcal{R}$ to the error function which well reproduces its overall shape.\footnote{We have tested polynomial fits of lower and higher degree, but found that they do either not reproduce the shape of the data/MC correction function well or induce unphysical wiggles.}

From this fit, we can directly extract the correlations in the effective acceptance error. By a subsequent fit of the correlations to the form Eq.~\eqref{eq:correlation_parameterization}, we finally obtain\footnote{To perform a $\chi^2$-fit, we have to assign an error in each rigidity bin shown in Fig.~\ref{fig:effective_acceptance}. For definiteness we have chosen this error to be be $5\%$. We note, however, that this error cancels out in the calculation of the correlations. Therefore, the choice of error does not affect our determination of the correlation length $\ell_{\text{eff.\,acc.}}$.}
\begin{equation}\label{eq:effacceptancelength}
  \ell_{\text{eff.\,acc.}}=0.15\,.
\end{equation}

\begin{figure*}[t]
\centering
\setlength{\unitlength}{1\textwidth}
\begin{picture}(1,0.37)
 \put(-0.0054,-0.02){\includegraphics[width=0.49\textwidth, trim= {2.74cm 1.4cm 2.9cm 0.8cm}, clip]{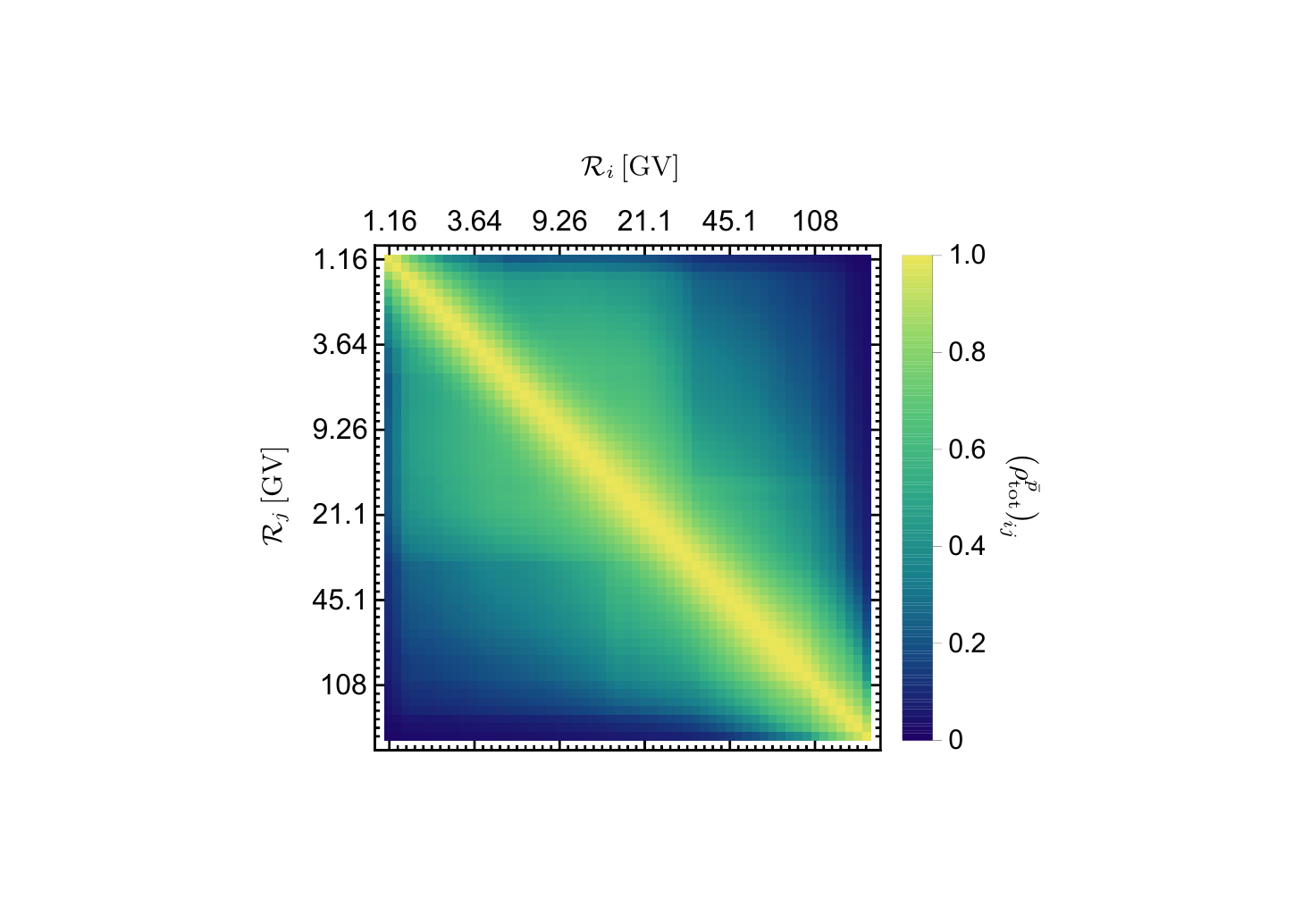}}
 \put(0.5045,-0.02){\includegraphics[width=0.49\textwidth, trim= {2.74cm 1.4cm 2.9cm 0.8cm}, clip]{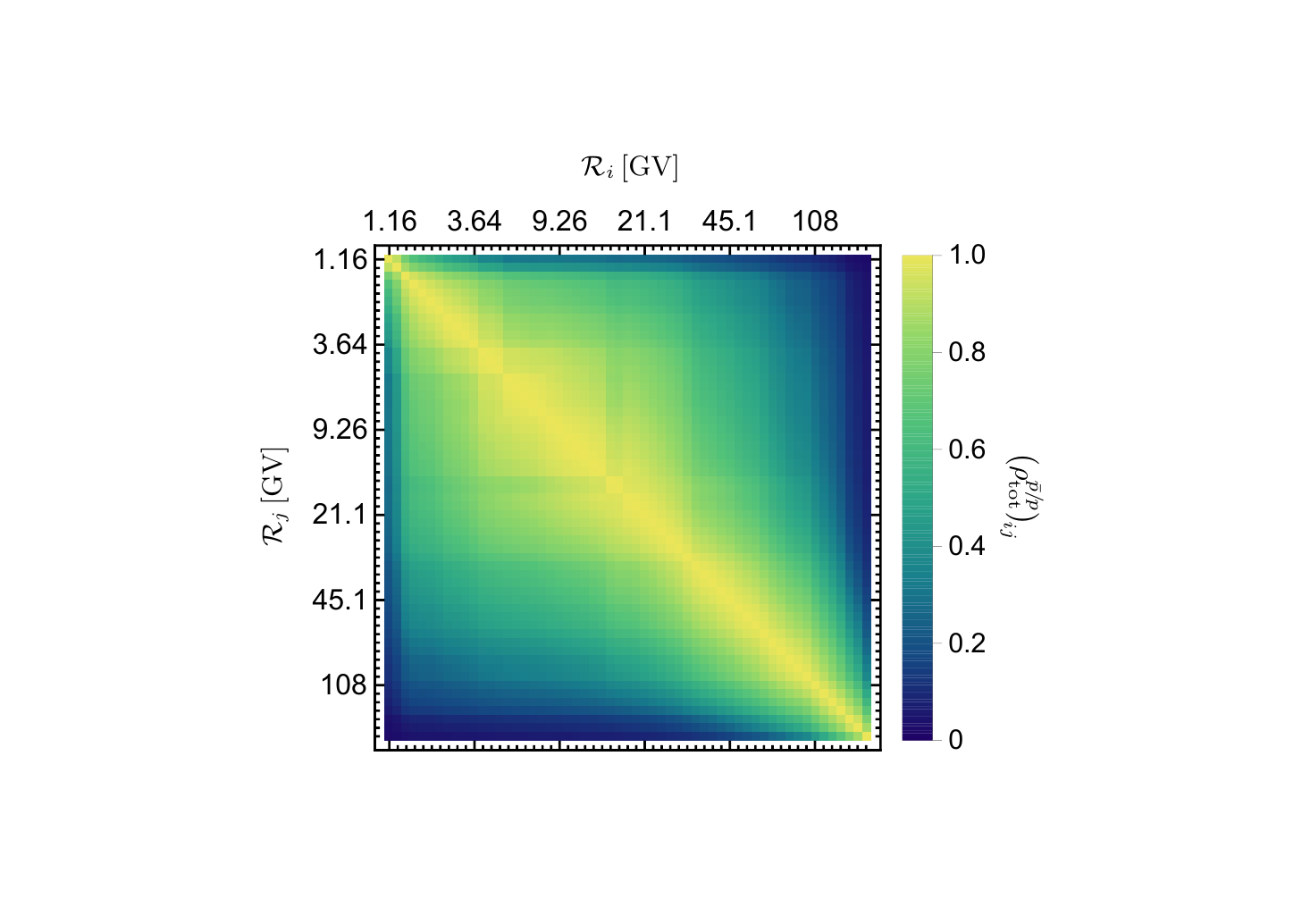}}
\end{picture}
\caption{
Correlations of the AMS-02 systematic errors in the antiproton flux (left panel) and the $\bar{p}/p$ ratio (right panel).
For better orientation we display the rapidity of every tenth bin.
\label{fig:ams_correlations}
}
\end{figure*}

As an alternative approach to derive the correlation length of the effective acceptance, we have considered the correction function shown in Fig.~\ref{fig:effective_acceptance} to be an estimate for the systematic uncertainty of the effective acceptance itself. To this end, we have defined a likelihood for the parameter $\ell_{\text{eff.\,acc.}}$ from Eq.~\eqref{eq:correlation_parameterization} and two nuisance parameters describing the overall size of the systematic uncertainty and a pure re-normalization uncertainty. A similar strategy was used in Ref.~\cite{Cuoco:2019kuu}. Profiling over the three parameters we obtain $\ell_{\text{eff.\,acc.}}\sim 0.1$ which is comparable with the result above. While we are thus confident that we obtained a reasonable estimate of the effective acceptance correlations from the available public information, their precise form can only be provided by the AMS-02 collaboration. At the precision level, we also expect small differences in $\ell_{\text{eff.\,acc.}}$ between different cosmic-ray species due to specific analysis cuts. We neglect such differences in this work.

The remaining uncertainties play a subleading role. They are always subdominant to either the two previously discussed systematic errors or the statistical error. For those errors, we refrain from a detailed analysis and adopt the correlation lengths estimated in Ref.~\cite{Boudaud:2019efq}:
\begin{equation}\label{eq:length}
\begin{split}
  \ell_{\text{scale}}=4\,,\qquad \ell_{\text{unf}}=1\,,\qquad  \ell_{\text{geo}}=1\,,
  \\ \ell_{\text{selection}}=0.5\,,\qquad    \ell_{\text{template}}=0.5\,.
\end{split}
\end{equation}

In the next step, we build the covariance matrix for each sub-error by multiplying the entries of the correlation matrix by the AMS-02 errors (as displayed in Fig.~\ref{fig:systematicerrors}) in the corresponding bins. The covariance matrices for each sub-error are then added to build the AMS-02 covariance matrix for the full systematic error. Figure~\ref{fig:ams_correlations} illustrates the overall correlations in the AMS-02 antiproton and $\bar{p}/p$ systematic errors as derived by our method. It can be seen that the systematic error in the antiproton flux is correlated on a shorter length scale compared to the error in the $\bar{p}/p$ ratio. This is because the effective acceptance error, which has a relatively short correlation length, affects the antiproton flux, but not the $\bar{p}/p$ ratio. The full AMS-02 covariance also containing the statistical error is provided in the ancillary files.

\section{Implications for the AMS-02 antiproton excess}
\label{sec:pbarexcess}

The AMS-02 error correlations can now be used to gain insights into cosmic-ray spectra. Of particular interest is the question how the correlations affect the interpretation of the antiproton excess at $\mathcal{R}= 10\!-\!20\;\text{GV}$. The latter has been considered as a possible dark-matter signal in a number of studies~\cite{Cuoco:2016eej,Cui:2016ppb,Cuoco:2017rxb,Reinert:2017aga,Cui:2018klo,Cuoco:2019kuu,Cholis:2019ejx,Lin:2019ljc}. At the same time, the significance of the excess is rather controversial (ranging from $1$ to $5\,\sigma$ within the mentioned references). The previous studies took the systematic errors in the AMS-02 antiproton flux to be uncorrelated (or modeled correlations in a simplistic way). 

In the following, we will reinvestigate the antiproton excess in Sec.~\ref{sec:results}, fully including the derived correlations in the AMS-02 systematic errors. We decided to perform two complementary likelihood analyses on the AMS-02 data. The two analyses differ substantially in the modeling of cosmic-ray propagation and in the considered species. Hence, we can directly verify the robustness of the conclusions we draw on the correlations with respect to the propagation model. Before we describe our analysis methods in Sec.~\ref{sec:crmde} we will briefly review the production and propagation of cosmic rays in Sec.~\ref{sec:crprop}.

\subsection{Cosmic-ray production and propagation}
\label{sec:crprop}

Cosmic rays are mainly composed of galactic matter which has been energized by supernova shock acceleration. This so-called primary component includes protons, helium and heavier nuclei like carbon and oxygen. Primaries, when they propagate through the galaxy, induce secondary cosmic rays by scattering processes in the interstellar disk. The source term for a secondary $a$, which denotes its differential production rate per volume, time and energy, takes the form
\begin{equation}
  q^\text{sec}_a = \sum\limits_{i,j} 4\pi \int \D T^\prime \left(\frac{\D\sigma_{ij\rightarrow a}}{\D T}\right) \,n_j \;\Phi_i(T^\prime)\,,
\end{equation}
where $i$ runs over the relevant primary cosmic-ray species ($\Phi_i$ and $T^\prime$ denote their flux and kinetic energy, respectively) and $j$ over the target nuclei in the galactic disk ($n_j$ denotes their number density). Secondary production is the main source of certain nuclear cosmic rays like boron, but also plays a major role in the generation of antimatter. 

For antiprotons, which are the main concern of this work, we will employ the differential production cross sections $\D\sigma_{ij\rightarrow \bar{p}}/\D T$ (with $i,j=p,\text{He}$) derived in Refs.~\cite{Kappl:2014hha,Winkler:2017xor}. Since production cross sections are only known to a few percent precision, they comprise an important source of systematic errors in the modeling of antiproton fluxes. These will be fully included in our cosmic-ray analysis. One of our fits will also include the boron-to-carbon ratio B/C. 
The boron production cross sections with uncertainties are taken from~\cite{Reinert:2017aga} (see also~\cite{Genolini:2018ekk,Evoli:2019wwu}).

In addition to the secondary background, primary antiprotons can be induced by dark-matter annihilation. The primary source term reads
\begin{equation}
  q^\text{prim}_{\bar{p}} = \frac{\rho_\chi^2}{m_\chi^2} \frac{\langle \sigma v \rangle}{2} \frac{\D N}{\D T}\,,
\end{equation}
where $\rho_\chi$, $m_\chi$ and $\langle \sigma v \rangle$ stand for the dark-matter energy density, mass, and annihilation cross section. The antiproton energy spectrum per annihilation is denoted by $\D N/\D T$. In this work we will consider 
\begin{equation}
  \chi\chi\rightarrow \bar{b}b
\end{equation}
as an exemplary annihilation channel and extract $\D N/\D T$ from~\cite{Cirelli:2010xx}. The scan range for the dark-matter mass is taken as $m_\chi=$10--10000$\,\text{GeV}$.

Independent of their origin, cosmic rays follow complicated trajectories on their passage through the galaxy. Scattering on magnetic field inhomogeneities induces a random walk which can be described as spatial diffusion. Convective winds may blow charged particles away from the galactic disk. In addition, cosmic-ray interactions with matter lead to energy losses and annihilation, while Alfv\'en waves induce reacceleration. The collection of processes is encoded in the diffusion equation. In this work, we will consider two approaches to solve the diffusion equation following our previous studies~\cite{Reinert:2017aga,Cuoco:2019kuu}: the semi-analytic solution in the two-zone diffusion model~\cite{Maurin:2001sj,Donato:2001ms,Maurin:2002ua} and a fully numerical solution based on the GALPROP code~\cite{Moskalenko:1997gh,Strong:1998pw,Strong:2001gh}. In both schemes, diffusion is assumed to occur homogeneously and isotropically in the galactic halo. Magnetohydrodynamics considerations suggest a power-law form of the diffusion coefficient $K$ (in the GALPROP code, $K$ is denoted as $D_{xx}$),
\begin{equation}\label{eq:eta}
K \propto \beta^{\eta} \mathcal{R}^\delta\,,
\end{equation}
where $\mathcal{R}$ is the rigidity and $\beta$ the velocity of the cosmic-ray particle. While $\eta=1$ in the standard implementation (as well as in our previous works~\cite{Reinert:2017aga,Cuoco:2019kuu}), here, we will include $\eta$ as a free fit parameter. This change is partly motivated by recent studies~\cite{DiBernardo:2009ku,Maurin:2010zp,Genolini:2019ewc,Weinrich:2020cmw} which observed a substantial improvement in the fit to secondary nuclear cosmic rays within the diffusion model with free $\eta$. In addition, the freedom of $\eta$ can be viewed as a conservative choice since it tends to slightly reduce the significance of a potential dark-matter signal (we will return to this point later). From a physical perspective, an increase of the diffusion coefficient (negative $\eta$) towards low rigidity can be motivated by wave damping on cosmic rays~\cite{Ptuskin:2005ax}. Following our previous work~\cite{Reinert:2017aga,Cuoco:2019kuu} we also include a break in the power law index $\delta$ at $\mathcal{R}\sim 300\:\text{GV}$ as required to fit nuclear primary and secondary cosmic rays~\cite{Genolini:2017dfb, Aguilar:2018njt}. The high-energy break does, however, virtually not affect our dark-matter analysis.

Both, the two-zone diffusion model and the numerical implementation of GALPROP, allow for convective winds perpendicular to the galactic plane. Both include energy losses, annihilation and reacceleration processes, although slight differences in the implementation exist (concerning e.g. the modeling of the interstellar material and the spatial extension of the reacceleration zone). For details we refer to the original references~\cite{Moskalenko:1997gh,Strong:1998pw,Strong:2001gh,Maurin:2001sj,Donato:2001ms,Maurin:2002ua}. Some custom modifications to the default setups have been described in our previous works~\cite{Reinert:2017aga,Cuoco:2019kuu}. 

\subsection{Methodology}
\label{sec:crmde}
To investigate the significance a possible dark-matter signal in the AMS-02 antiproton data, we consider two complementary setups which we describe in the following. Within both setups, we include the derived AMS-02 covariance matrices of systematic errors for all species in the fit.

\oursubsubsection{Setup 1}
The first setup implements the approach described in Ref.~\cite{Reinert:2017aga}. It is based on the two-zone diffusion model of cosmic-ray propagation. The primary fluxes of $p$, He, C, N, O, Ne, Mg, Si are taken as an input to predict the secondary fluxes of antiprotons and boron. In addition, a primary antiproton component from dark-matter annihilation is included. Solar modulation is taken into account through an improved force-field approximation~\cite{Cholis:2015gna}, which -- to describe charge-breaking effects -- contains one parameter in addition to the Fisk potential.
 
In setup 1, a simultaneous fit to the AMS-02 B/C ratio~\cite{Aguilar:2016vqr}, the AMS-02 antiproton flux~\cite{Aguilar:2016kjl} and the antiproton flux ratio between AMS-02 and PAMELA~\cite{Adriani:2012paa} is performed. The combination of B/C and $\bar{p}$ is needed to determine the propagation parameters and the significance of a dark-matter excess, while the AMS/PAMELA flux ratio fixes the charge breaking parameter in the solar modulation (see~\cite{Reinert:2017aga} for details). The total goodness of fit is determined by $\chi^2=\chi^2_{\bar{p}} +\chi^2_{\text{B/C}} + \chi^2_{\text{AMS/PAM}}$ which is evaluated with and without a dark-matter component. Uncertainties in the production cross sections of antiprotons and boron are fully included via covariance matrices~\cite{Reinert:2017aga}.

\oursubsubsection{Setup 2}
The second setup follows the approach of~\cite{Cuoco:2019kuu} and employs the GALPROP code for cosmic-ray propagation. The primary source terms of proton and helium are modeled as broken power laws. In addition, a primary antiproton source term from dark-matter annihilation is included. From this input, the full network of cosmic-ray propagation, scattering and propagation of secondaries is employed to determine the proton, helium and secondary antiproton flux. A major difference compared to setup~1 is that the propagation parameters are constrained by primary fluxes and the $\bar p/p$ ratio instead of the B/C ratio. Solar modulation is implemented through the standard force-field approximation (with individual Fisk potentials for both charge signs). A more refined treatment can be evaded as the AMS-02 data is cut at rigidity $\mathcal{R}=5\:\text{GV}$. The propagation and solar modulation parameters as well as the significance of a dark-matter excess is determined by a simultaneous fit to the AMS-02~\cite{Aguilar:2015ooa,Aguilar:2015ctt} and Voyager~\cite{Stone150} data on protons and helium as well as the AMS-02 antiproton-over-proton~\cite{Aguilar:2016kjl} data. Uncertainties in the production cross sections of antiprotons are again taken into account through the covariance matrix matrix from~\cite{Reinert:2017aga}.\footnote{We have checked that a slight difference of the covariance matrix~\cite{Reinert:2017aga} compared to the one used in Ref.~\cite{Cuoco:2019kuu} does virtually not affect our results.}

\subsection{Results}
\label{sec:results}
The results of our fits are summarized in Table~\ref{tab:res}, which provides the goodness-of-fit, dark-matter parameters and significances for the best-fit points of the setups~1 and~2. The $\chi^2$ values with and without a dark-matter component in the antiproton flux are included in each column, the latter is displayed in parentheses. To highlight the impact of the derived AMS-02 correlation matrices, we compare results with and without the error correlations. The corresponding cosmic-ray spectra and residuals are depicted in Figs.~\ref{fig:fit_mw} and~\ref{fig:fit_mk}. The best-fit propagation parameters within the two setups are listed in Appendix~\ref{app:cr_prop}. The stated local significance of the dark-matter signal refers to a $\Delta\chi^2$-test for one degree of freedom. For the global significance we take into account the look-elsewhere effect by evaluating the probability distribution among mock-data sets created under the background-only hypothesis.\footnote{Note that in general the global probability distribution of the considered $\Delta\chi^2$ deviates from the standard $\chi^2$-distribution with two degrees of freedom as Wilks' theorem does not apply to cases, where one parameter (in this case the dark-matter mass) is only defined under the alternative hypothesis~\cite{Davies:1987zz,Gross:2010qma}. The difference is found to be sizable for the low significances observed here.}

\begin{table}[th]
\begin{center}
\renewcommand{\arraystretch}{1.3}
\begin{tabular}{ |l|cc| }
\multicolumn{3}{c}{--- Setup 1 ---}                                                              \\ \hline
\,                                                             & w/o corr.     & with corr.      \\ \hline
$\chi^2_{\text{B/C}}$  (63\,bins)                              & 40.4\,(41.4)  & 67.3\,(67.3)    \\
$\chi^2_{\bar{p}}$ (57\,bins)                                  & 26.2\,(28.8)  & 58.8\,(58.9)    \\
$\chi^2_{\text{AMS/PAM}}$\,(17\,bins)                          & 11.5\,(11.4)  & 11.2\,(11.3)    \\
\,                                                             &               &                 \\
$\chi^2_\text{tot}$                                            & 78.2\,(81.7)  & 137.3\,(137.6)  \\
No.~of fit param.                                                & 7(9)         & 7(9)           \\
$m_{\text{DM}} \,[\text{GeV}]$                                 & 76            & 88              \\
$\langle\sigma v\rangle\,[10^{-26}\,\text{cm}^3\!/\text{s}]$   & 0.73          & 0.4             \\
$\Delta\chi^2_\text{tot}$                                      & 3.5           & 0.3             \\
local sig.                                                     & $1.9\,\sigma$   & $0.5\,\sigma$ \\
global sig.                                                    & $0.8\,\sigma$   & $-$           \\ \hline
\multicolumn{3}{c}{}     \\
\multicolumn{3}{c}{--- Setup 2 ---}                                                               \\ \hline
\,                                                             & w/o corr.      & with corr.      \\ \hline
$\chi^2_{\bar p/p}$\,(42\,bins)                                & $11.4\,(16.2)$ & 45.6\,(46.4)    \\
$\chi^2_{p}$\,(50\,bins)                                       & $ 1.8\,( 3.2)$ & 104.5\,(104.9)  \\
$\chi^2_{\text{He}}$\,(50\,bins)                               & $ 4.8\,( 4.5)$ & 78.4\,(77.6)    \\
$\chi^2_{p,\text{Voy}}$\,(9\,bins)                             & $ 1.8\,( 1.7)$ & 2.9\,(4.3)      \\
$\chi^2_{\text{He},\text{Voy}}$\,(5\,bins)                     & $ 0.3\,( 1.0)$ & 1.8\,(2.0)      \\
\,                                                             &                &                 \\
$\chi^2_\text{tot}$                                            & $20.3\,(27.2)$ & 233.1\,(236.3)  \\
No.~of fit param.                                                & 16(18)         & 16(18)          \\
$m_{\text{DM}} \,[\text{GeV}]$                                 & 76             & 66              \\
$\langle\sigma v\rangle\,[10^{-26}\,\text{cm}^3\!/\text{s}]$   & 0.91           & 0.74            \\
$\Delta\chi^2_\text{tot}$                                      & 6.9            & 3.2             \\
local sig.                                                     & $2.6\,\sigma$      &  $1.8\,\sigma$             \\ 
global sig.                                                    & $1.8\,\sigma$      &  $0.5\,\sigma$             \\ \hline
\end{tabular}
\renewcommand{\arraystretch}{1}
\end{center}
\caption{
          $\chi^2$ values, number of free fit parameters, dark-matter parameters and significances for the best fits within setup~1 (top) and 2 (bottom). For each setup the first and second column corresponds to the fit without and with correlations in the AMS-02 errors, respectively. The absolute $\chi^2$ values given in each column refer to the fits with (without) dark matter. 
        }
\label{tab:res}
\end{table}

\begin{figure*}[t!h]
\centering
\setlength{\unitlength}{1\textwidth}
\begin{picture}(1,0.95)
 \put(-0.0054,-0.02){\includegraphics[width=1\textwidth, trim= {1.55cm 5.5cm 1.7cm 4.5cm}, clip]{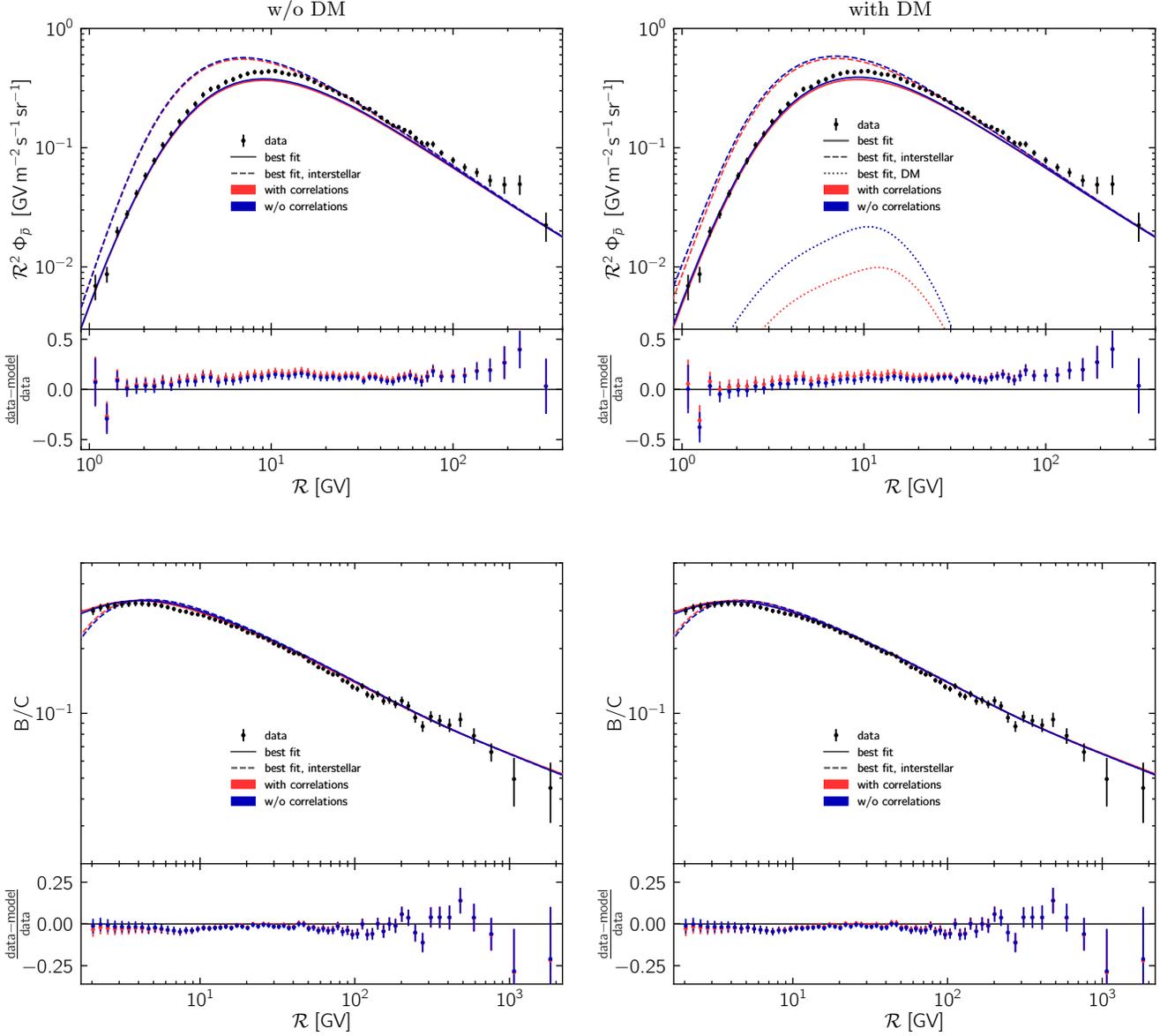}}
\end{picture}
\caption{Antiproton flux (top) and B/C flux ratio (bottom) of the fit without (left) and with dark matter (right) within setup~1. The solid red and blue curves (light and dark gray in the print gray-scale version) denote the best-fit spectra at the top of the atmosphere with and without correlations in the AMS-02 errors, respectively. 
The dashed curves denote the corresponding interstellar fluxes. The dotted curves in the upper right plot show the respective best-fit contributions from dark matter. Error bars denote the statistical and systematic uncertainties (according to the diagonal entries of the total experimental covariance matrix). The red (blue) data points in the lower panels show the residuals of the fit with (without) correlation. For the red points, we remark that error bars only depict the diagonal entries of the covariance matrix, namely they do not show the impact of correlations.
\label{fig:fit_mw}
}
\end{figure*}

\begin{figure*}[t!h]
\centering
\setlength{\unitlength}{1\textwidth}
\begin{picture}(1,0.95)
 \put(-0.0054,-0.02){\includegraphics[width=1\textwidth, trim= {1.55cm 5.5cm 1.7cm 4.5cm}, clip]{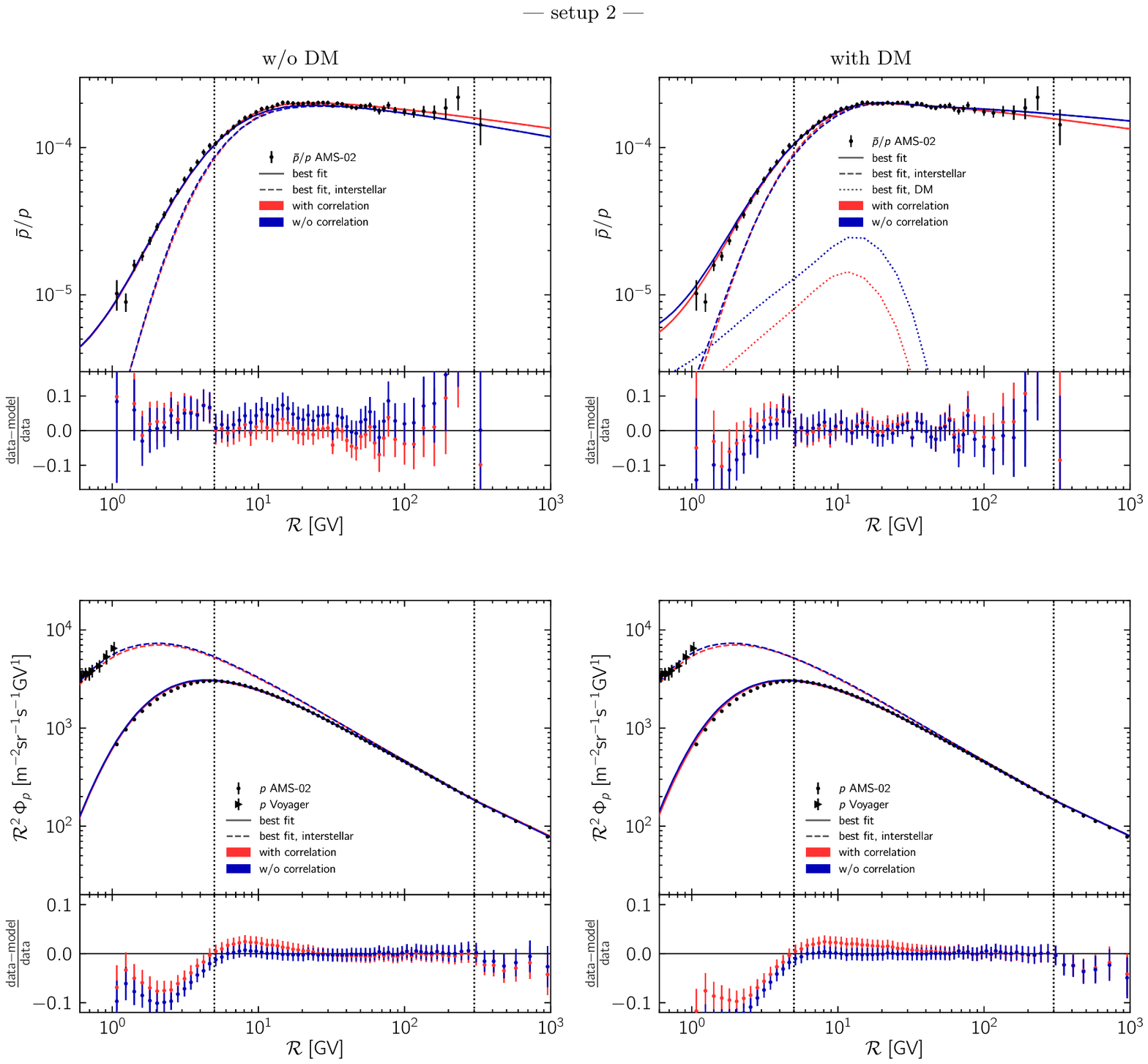}}
\end{picture}
\caption{ 
          Antiproton-to-proton ratio (top) and proton flux (bottom) of the fit without (left) and with dark matter (right) within setup~2. 
          The solid red and blue curves (light and dark gray in the print gray-scale version) denote the best-fit spectra at the top of the atmosphere with and without correlations in the AMS-02 errors, respectively. The dashed curves denote the corresponding interstellar fluxes. We display the cosmic-ray measurements of AMS-02 (proton and antiproton-to-proton ratio) and Voyager (proton). The cosmic-ray fit of the AMS-02 data is restricted to rigidities between the dotted black lines. Residuals are shown only for the AMS-02 data points. 
          Error bars denote the statistical and systematic uncertainties (according to the diagonal entries of the total experimental covariance matrix).
          The red (blue) data points in the lower panels show the residuals of the fit with (without) correlation.           \label{fig:fit_mk}
        }
\end{figure*}

To make the connection to previous studies, let us first turn to the fit results \emph{without} including error correlations. Up to the extra propagation parameter $\eta$ (\cf~Eq.~\eqref{eq:eta}), this case corresponds to the default configuration in Ref.~\cite{Reinert:2017aga} (for setup~1)\footnote{Compared to~\cite{Reinert:2017aga}, we also updated the boron absorption cross section in the interstellar disk by scaling the proton-carbon cross section derived in Sec.~\ref{sec:nucleonabsorption} to boron. This led to a slight reduction in $\chi^2_{\text{B/C}}$, but did otherwise not affect our analysis.} and to the configuration ``XS uncertainty by covariance matrix'' in Ref.~\cite{Cuoco:2019kuu} (for setup~2). Not surprisingly, we qualitatively reproduce the results given in these references. An antiproton excess is observed at $\mathcal{R}= 10$--$20\:\text{GV}$. The latter is compatible with a dark-matter particle of mass $m_\chi\sim 80\:\text{GeV}$ and annihilation cross section $\langle \sigma v \rangle \sim 10^{-26}\:\text{cm}^3/s$ into bottom quarks. However, in setup~1, the global significance is only $\sim 1\,\sigma$, while it reaches $\sim 2\,\sigma$ in setup~2. In both setups, the significance is slightly smaller compared to Refs.~\cite{Reinert:2017aga,Cuoco:2019kuu}, which is due to the additional freedom in the propagation. The extra diffusion parameter $\eta$ allows for a stronger ``bending'' of cosmic-ray fluxes towards low energy and, hence, absorbs a small fraction of the excess.

When we turn to the fits including the correlations in the AMS-02 systematic errors, we observe an overall increase in the $\chi^2$ values. This is to be expected, since -- with a realistic modeling of systematic errors -- one expects a total $\chi^2$ comparable to the number of degrees of freedom (dof). Setup~1 nicely fulfills this criterion, which gives us confidence that the derived AMS-02 error correlations are indeed realistic. In setup~2, a somewhat higher $\chi^2/\text{dof}\sim 1.5$ occurs. This could possibly stem from a mild underestimation of systematic errors in the proton and helium data of AMS-02. We wish to point out that the unofficial AMS-02 helium analysis performed in the Ph.D. thesis~\cite{Alaoui:2016} indeed derived larger uncertainties compared to the published data. Alternatively, it could indicate that we slightly overestimated the correlation length in the proton and helium systematic errors. Even if this is the case, it would not affect our conclusions on the dark-matter excess as we have explicitly verified.\footnote{Technically, we checked that the significance of the dark-matter excess is only marginally affected when we set the correlation length of the effective acceptance error to a smaller value of $\ell_{\text{eff.\,acc.}}\sim 0.1$. Note that we chose to alter the correlations of the effective acceptance error as other systematic error sources do not support correlations on short rigidity scales.}

Our key result is that the significance of the dark-matter excess decreases substantially, once we include error correlations in the AMS-02 data. In setup~1, the preference for a dark-matter signal disappears completely, but even in setup~2 the global significance drops below $1\,\sigma$. Correspondingly, the best-fit dark-matter signal is reduced by about a factor of 2 in both setups. It is thus obvious that the systematic errors of AMS-02 provide sufficient freedom to absorb the antiproton excess, once their correlations are properly taken into account.

In both setups, the correlations in the absorption cross-section uncertainties, which we derived in Sec.~\ref{sec:nucleonabsorption}, play an important role. However, there is also an interesting distinction: In setup~1, the effective acceptance error in the antiproton flux contributes to reducing the significance of the antiproton excess. In setup~2, the $\bar{p}/p$ ratio is employed in the fit (instead of the antiproton flux). Since the effective acceptance error cancels in $\bar{p}/p$ one might think that it is irrelevant in this case. However, we observe, that the fit then takes advantage of the effective acceptance error in the proton flux. This can be seen in Fig.~\ref{fig:fit_mk}, where the excess in $\bar{p}/p$ partly shifts into the proton channel, once correlations are included.

We finally wish to emphasize that our conclusion, namely that the antiproton excess is explainable by systematic effects, is not built on the systematic error correlations alone. Rather, the full interplay between all uncertainties matters -- those in the cosmic-ray propagation, those in the antiproton production cross sections and those in the AMS-02 systematic errors. To emphasize this point, let us turn to an example: We have tested the significance of the antiproton excess in setup 2 without including production cross-section uncertainties and in a more restrictive propagation setup (without the diffusion parameter $\eta$). In this case, the correlations in the AMS-02 systematic errors had a contrary effect: Compared to uncorrelated systematic errors, they strongly increased the significance of the antiproton excess to $\sim 5\,\sigma$. The same observation has already been made in Ref.~\cite{Cuoco:2019kuu}. We should thus refine our previous statement: The AMS-02 systematic errors alone might not absorb the observed spectral feature in the antiproton flux. However, in combination with the other main uncertainties, the correlated systematic errors can account for the previously seen excess such that there is no longer a preference for a dark-matter signal.

Finally, we remark that our results should not be understood as to exclude the dark-matter interpretation of the antiproton feature at $\mathcal{R}=10$--$20\gv$. Any improvement in the description of cosmic-ray propagation and the modeling of antiproton production cross sections will reduce the uncertainties and help to determine whether an antiproton excess exists. Both, the dark-matter interpretation as well as the interpretation as a combination of systematic effects will thus undergo further scrutiny in the next years. Obviously, additional information on systematic errors, provided by the AMS-02 collaboration, would be extremely useful in this regard.

\section{Conclusions}
\label{sec:concl}

Four years ago, the AMS-02 collaboration published their first measurement of the cosmic-ray antiproton flux. After background subtraction, it seemingly revealed a small residual component in the spectrum at rigidity $\mathcal{R}= 10\!-\!20\:\text{GV}$. Since then, the antiproton excess has been subject to major controversy in the community. On the one hand, its possible explanation in terms of annihilating dark matter has caused a wave of excitement. In particular, since the favored mass $m_\chi=50\!-\!100\gev$ and (hadronic) annihilation cross section $\langle\sigma v \rangle = 0.5\!-\!3 \times 10^{-26}\:\text{cm}^3\text{s}^{-1}$ could also tentatively be linked to the galactic center gamma ray excess. On the other hand, systematic effects have been a matter of concern.

Since the antiproton excess was first observed, major improvements in the modeling of the antiproton background and the description of cosmic-ray propagation have been made. But one decisive piece has been missing so far: the full covariance matrix of systematic errors in the AMS-02 antiproton data. Correlated systematic errors are so important because they can cause features in the residuals and potentially `fake' a dark-matter excess. Therefore, in this work, we comprehensively derived the correlations in the AMS-02 data with the purpose of further scrutinizing the excess.

In the rigidity range $\mathcal{R}= 10\!-\!20\:\text{GV}$, the systematic error dominantly descends from cosmic-ray absorption in the AMS-02 detector. The relevant nuclear cross sections which enter the AMS-02 simulation carry major uncertainties. These directly translate to the measured fluxes of antiprotons, protons and heavier cosmic rays. We carefully computed the mentioned absorption cross sections within the Glauber model and performed a global fit to the data from nucleon-nucleus and nucleon-nucleon scattering experiments, the latter of which serve as input to the theoretical model. Including inelastic shadowing effects, we obtained an excellent fit to the data. This computation also allowed us to reliably infer the correlations of the absorption cross-section error in the AMS-02 systematics. Furthermore, we found a data-driven estimate for correlations in the effective acceptance error and used covariance functions for subdominant contributions related to the AMS trigger and unfolding procedure. From the correlations of individual errors, we constructed the covariance matrix for the total experimental error in the antiproton and proton flux as well as the $\bar p/p$ ratio. Similarly, we also derived the corresponding covariances for He and B/C which also enter our cosmic-ray analysis. The final AMS-02 covariance matrices are made available for public use in the ancillary files on arXiv.\footnote{The diagonal components of the covariance matrices correspond to the errors published by AMS-02. The off-diagonal components have been derived in this work.}

In the final step, we turned to a reevaluation of the antiproton excess. To this end we performed a spectral search for dark matter in the AMS-02 antiproton data, where we fully took into account the derived correlations in the AMS-02 systematic errors. Our fits also incorporate a detailed modeling of the uncertainties in the antiproton production cross sections which affect the prediction of the astrophysical antiproton background. The results were obtained in two complementary propagation setups with a conservative choice of propagation parameters. In the first setup, we observe that the antiproton excess disappears, once the AMS-02 systematic error correlations are taken into account -- without correlations, a local (global) excess of $1.9\,\sigma$ ($0.8\,\sigma$) had been found. In the second setup, the systematic error correlations reduce the significance of the excess from $2.6\,\sigma$ ($1.8\,\sigma$) locally (globally) to $1.8\,\sigma$ ($0.5\,\sigma$). We conclude, that with all uncertainties properly taken into account, the AMS-02 data do not prefer a dark-matter interpretation at this stage. The fact that we obtain consistent results in both propagation setups gives us further confidence in the robustness of this conclusion. 

While our findings emphasize the importance of error correlations in the AMS-02 data, we remark that including the systematic errors in the background model turns out to be equally crucial. We found that ignoring uncertainties in the antiproton production cross sections can even lead to the (wrong) conclusion that AMS-02 systematic error correlations increase the significance of the antiproton excess. Only the interplay between correlated errors in the data, uncertainties in the antiproton production and sufficient freedom in the cosmic-ray propagation allows the fit to fully absorb the antiproton excess (without a dark-matter component).

Our result should not be understood as to exclude a dark-matter candidate with mass $m_\chi=50\!-\!100\gev$ and thermal annihilation cross section. Rather, with the full account of systematic errors, our cosmic-ray fit is only weakly sensitive to dark matter of this type. It can simply not distinguish between a dark-matter excess at the $10\%$-level and a correlated fluctuation in the systematics. Further reducing these systematics must be a main objective in the next years. It is of paramount importance to strengthen the efforts to precisely measure cosmic-ray cross sections at accelerators. At the same time, existing and upcoming nuclear cosmic-ray data will help to further advance the physics of cosmic-ray propagation. In the precision era, which cosmic-ray physics has entered, it is no longer the statistics, but the control of systematic errors which decides on the prospects of indirect dark-matter detection.

\section*{Acknowledgments}

We thank Henning Gast, Boris Kopeliovich, Michael Kr\"amer and Paolo Zuccon for very helpful discussions and comments. 
The authors would like to express special thanks to the Mainz Institute for Theoretical Physics (MITP) of the Cluster of Excellence PRISMA+ (Project ID 39083149) for its hospitality and support.

J.H.~acknowledges support from the F.R.S.-FNRS, of which he is a postdoctoral researcher. M.W.W acknowledges support by the Vetenskapsr\r{a}det (Swedish Research Council) through Contract No.~638-2013-8993 and the Oskar Klein Centre for Cosmoparticle Physics. 

The work of M.K. is supported by the  `Departments of Excellence 2018-2022' grant awarded by the Italian Ministry of Education, University and Research (MIUR) L.\ 232/2016
as well as by Istituto Nazionale di Fisica Nucleare (INFN) and by the Italian Space Agency through the ASI INFN Agreement No.\ 2018-28-HH.0: "Partecipazione italiana al GAPS -- General AntiParticle Spectrometer". Furthermore, M.K. acknowledges support from research grants: TAsP (Theoretical Astroparticle Physics) funded by INFN, `The Dark Universe: A Synergic Multimessenger Approach' No.\ 2017X7X85K, PRIN 2017, funded by MIUR, and `The Anisotropic Dark Universe' No.\ CSTO161409, funded by Compagnia di Sanpaolo and University of Turin.
Simulations were performed with computing resources
granted by RWTH Aachen University under
the Project No. rwth0085.

\vspace{2.3ex}
J.H.~dedicates this work to his father, Klaus Heisig, who passed away during the early stage of J.H.'s work on the project.

\appendix

\section{Nucleon-nucleon cross-section fit}\label{app:nnxsfit}

In this Appendix, we display the best-fit values and covariance matrices for the parametrizations of the nucleon-nucleon cross sections and the slopes of the differential inelastic cross sections as described in Sec.~\ref{sec:nucnucparam}. The best-fit values and the corresponding covariance matrices read:
\begin{widetext}
\begin{align}
\left(
\begin{array}{c}
c_1 \\ c_2 \\ c_3 \\ c_4 \\ c_5
\end{array}
\right)
= 
\left(
\begin{array}{c}
0.0995 \\1.36 \\ 2.01 \\ -2.25 \\  0.816
\end{array}
\right),
\quad
&\mathcal{V}_{\sigma_{\bar p p}}  =  
\left(
\begin{array}{ccccc}
0.0322 &  -0.0802 &  0.079 &  -0.048 &  0.0161 \\ 
-0.0802 &  0.216 &  -0.231 &  0.153 &  -0.0542 \\ 
0.079 &  -0.231 &  0.284 &  -0.216 &  0.0806 \\ 
-0.048 &  0.153 &  -0.216 &  0.183 &  -0.0723 \\ 
0.0161 &  -0.0542 &  0.0806 &  -0.0723 &  0.0296 \\
\end{array}
\right),
\\
\left(
\begin{array}{c}
c_6 \\ c_7 \\ c_8 \\ c_9
\end{array}
\right)
 = 
\left(
\begin{array}{c}
 0.0505 \\ 1.56 \\ 1.42 \\ -0.933
\end{array}
\right)\,,\quad
&\mathcal{V}_{\sigma_{\bar p n}}  =  
\left(
\begin{array}{cccc}
0.0872 &  -0.224 &  0.241 &  -0.103 \\ 
-0.224 &  0.659 &  -0.838 &  0.4 \\ 
0.241 &  -0.838 &  1.37 &  -0.773 \\ 
-0.103 &  0.4 &  -0.773 &  0.549 \\
\end{array}
\right)\,,
\\
\left(
\begin{array}{c}
c_{10} \\ c_{11} \\ c_{12} \\ c_{13} \\ c_{14} \\ c_{15} 
\end{array}
\right)
= 
\left(
\begin{array}{c}
 -0.424 \\ 2.51 \\ -9.48 \\ 20.8 \\ -16.5 \\ 3.08
\end{array}
\right)\,,\quad
&\mathcal{V}_{\sigma_{p p}}  =  
\left(
\begin{array}{cccccc}
0.00942 &  -0.0303 &  0.0719 &  -0.0996 &  0.0564 &  -0.00744 \\ 
-0.0303 &  0.11 &  -0.293 &  0.434 &  -0.257 &  0.036 \\ 
0.0719 &  -0.293 &  0.875 &  -1.38 &  0.853 &  -0.125 \\ 
-0.0996 &  0.434 &  -1.38 &  2.25 &  -1.43 &  0.215 \\ 
0.0564 &  -0.257 &  0.853 &  -1.43 &  0.919 &  -0.142 \\ 
-0.00744 &  0.036 &  -0.125 &  0.215 &  -0.142 &  0.0227 \\
\end{array}
\right),
\\
\left(
\begin{array}{c}
c_{16} \\ c_{17} \\ c_{18} \\ c_{19} \\ c_{20}
\end{array}
\right)
= 
\left(
\begin{array}{c}
 0.0480 \\ 0.417 \\ -0.345 \\ -0.626 \\ 0.443
\end{array}
\right)\,,\quad
&\mathcal{V}_{\sigma_{p n}}  =  
\left(
\begin{array}{ccccc}
0.021 &  -0.0539 &  0.0738 &  -0.0588 &  0.0179 \\ 
-0.0539 &  0.147 &  -0.218 &  0.183 &  -0.058 \\ 
0.0738 &  -0.218 &  0.372 &  -0.344 &  0.116 \\ 
-0.0588 &  0.183 &  -0.344 &  0.336 &  -0.118 \\ 
0.0179 &  -0.058 &  0.116 &  -0.118 &  0.0424 \\
\end{array}
\right)\,,
\\
\left(
\begin{array}{c}
d_1 \\ d_2 \\ d_3 \end{array}
\right)
 = 
\left(
\begin{array}{c}
10.7 \\ 5.81 \\ 0.320
\end{array}
\right)\,,\quad
&\mathcal{V}_{\slopeB_{\bar p p}}  =  
\left(
\begin{array}{ccc}
0.386 &  -0.371 &  -0.0793 \\ 
-0.371 &  0.533 &  0.075 \\ 
-0.0793 &  0.075 &  0.019 \\
\end{array}
\right),
\\
\left(
\begin{array}{c}
d_4 \\ d_5 \\ d_6 
\end{array}
\right)
 = 
\left(
\begin{array}{c}
 7.76 \\ -5.24 \\ 0.794
 \end{array}
\right)\,,\quad
&\mathcal{V}_{\slopeB_{p p}}  =  
\left(
\begin{array}{ccc}
0.111 &  -0.0961 &  -0.0272 \\ 
-0.0961 &  0.0864 &  0.0237 \\ 
-0.0272 &  0.0237 &  0.00827 \\
\end{array}
\right)\,.
\end{align}
\end{widetext}

For the corresponding uncertainty bands in the nucleon-nucleon cross sections and slopes, see Table~\ref{tab:uncertaintysigmabeta}.
\begin{table}[h!]
    \centering
    \begin{tabular}{|c|c|c|c|c|c|}
    \hline
        $\sigma_{\bar p p}$ & $\sigma_{\bar p n}$ & $\sigma_{p p}$ & $ \sigma_{pn}$  & $ \slopeB_{\bar p p} $ & $\slopeB_{p p}$  \\
        $\; 1-2\% \;$ & $\; 2-14\%\;$ & $\;\sim 1\% \;$ & $\; 1-2\% \;$ & $\; 4-7\%\;$ & $\; 3-6\%\;$\\[1mm]
        \hline
    \end{tabular}
    \caption{Uncertainties on the nucleon-nucleon cross sections and slopes.}
    \label{tab:uncertaintysigmabeta}
\end{table}

\section{Error correlations in the AMS-02 proton, helium and B/C data}\label{sec:other_correlations}

The calculation of the covariance matrices of AMS proton, helium and B/C errors proceeds analogously to those for antiprotons and $\bar{p}/p$ which was described in Sec.~\ref{sec:correlations_pbar}. We first split the systematic errors into their components.

\oursubsubsection{Proton flux}
In the proton case, the error related to absorption cross sections is $1\%$ at $\mathcal{R}=1\gv$, $0.6\%$ from $\mathcal{R}=10\!-\!300\gv$ and $0.8\%$ at $\mathcal{R}=1800\gv$~\cite{Aguilar:2015ooa}. Between the given rigidities we interpolate logarithmically. The effective acceptance error is obtained by (quadratically) subtracting the cross-section error from the acceptance error given in the Supplementary Material of Ref.~\cite{Aguilar:2015ooa}. The unfolding, scale and trigger uncertainties can directly be taken from the same reference.

\oursubsubsection{Helium flux}
Cross-section errors are taken to be $1\%$ for helium over the full rigidity range~\cite{Aguilar:2015ctt}. Effective acceptance, unfolding, scale and trigger error are extracted in the same way as for the proton flux from the Supplementary Material of Ref.~\cite{Aguilar:2015ctt}.

\oursubsubsection{B/C flux ratio}
For B/C we employ the effective acceptance error of $1\%$ over the full rigidity range~\cite{Aguilar:2016vqr}. The cross-section error is obtained by (quadratically) subtracting the effective acceptance error from the total acceptance error given in the Supplementary Material of Ref.~\cite{Aguilar:2016vqr}. The systematic unfolding, scale, trigger and background subtraction errors are taken from the same reference. The background subtraction error (which was absent in the proton and helium fluxes) is related to the spallation of heavier cosmic-ray species within the AMS detector.\\[5mm]
After characterizing the individual error components, we assign a correlation matrix to each of them (following our approach in Sec.~\ref{sec:covariance}). The cross-section error correlations for the proton flux can directly be taken from our fit (see Sec.~\ref{sec:correlation_xs_fit}). The cross-section error correlations for helium and B/C are expected to be of similar shape. Therefore, we refrain from performing cross-section fits for the absorption of these species and rather adopt the correlations from the $\sigma_{pC}$-fit. However, the natural unit for correlations is the momentum (per nucleon) which (approximately) corresponds to half the rigidity in the case of nuclei. Since the AMS data are provided in terms of the rigidity, we thus have to stretch the correlations from the proton case by a factor of two in order to apply them to helium and B/C.\footnote{Note that very recently measurements of nucleus-carbon interaction cross sections with the AMS-02 detector have been reported~\cite{2020NuPhA.99621712Y}.}

The remaining correlations matrices are again taken to be of the form Eq.~\eqref{eq:correlation_parameterization}. We consistently choose the scale and effective acceptance correlation lengths as in Eqs.~\eqref{eq:length} and~\eqref{eq:effacceptancelength}. For the trigger error which contains the geomagnetic error, we take $\ell_{\text{trigger}}=1$ and for the unfolding error $\ell_{\text{unf}}=0.5$~\cite{Derome:2019jfs}. Similarly, for the background subtraction error which only affects B/C we assume $\ell_{\text{background}}=1$ (since the contamination should be strongly correlated between a few neighboring bins).

The main difference compared to the covariance matrices derived by Ref.~\cite{Derome:2019jfs} consists of the modeling of the acceptance error correlations (including those related to absorption cross sections and effective acceptances). These correlations have been obtained by educated guesses in Ref.~\cite{Derome:2019jfs}, while we extracted the cross-section correlations from our fits in the Glauber-Gribov theory and effective acceptance correlations through the procedure described in Sec.~\ref{sec:covariance}.

\section{Cosmic-ray propagation parameters}\label{app:cr_prop}

Table~\ref{tab:propa} summarizes the best-fit propagation parameters for the two setups considered. For the definition of all parameters [except for $\eta$ which was defined in Eq.~\eqref{eq:eta}] and the respective details of the propagation model we refer to Refs.~\cite{Reinert:2017aga} (for setup~1) and~\cite{Cuoco:2019kuu} (for setup~2). In setup~1, a degeneracy between $K_0$ and $L$ arises and, therefore, $L$ was fixed to the lowest value $L=4\,\text{kpc}$ suggested by positron data~\cite{Reinert:2017aga} (the value $L=4\,\text{kpc}$ is also motivated from a recent Be/B analysis in the same propagation setup~\cite{Weinrich:2020ftb}, see also Ref.~\cite{Evoli:2019iih}).
In addition to the parameters given in the table, in both setups a rigidity break in the diffusion coefficient is taken into account in the same way as in Ref.~\cite{Reinert:2017aga} and \cite{Cuoco:2019kuu}, \ie~$\Delta\delta=0.157$, ${\cal R}_b=275\,\text{GV}$, $s=0.074$ and $\delta-\delta_2=0.12$, $R_1=300\,\text{GV}$, respectively. Note that although some parameters in the two setups are equivalent, we choose the very same notation as in the respective references to avoid ambiguities. 

\begin{table}[b]
\begin{center}
\begin{tabular}{ |l|cc| }
\multicolumn{3}{c}{--- Setup 1 ---}                                                               \\ \hline
\,                                                             & w/o corr.     & with corr.       \\ \hline
$K_0\;[\text{kpc}^2/\text{Gyr}]$                               & 38.9\,(37.8)  & 40.3\,(39.8)     \\ 
$\delta$                                                       & 0.468\,(0.472)  & 0.462\,(0.464) \\
$\eta$                                                         & $-1.05$\,($-1.12$)  & $-0.85$\,($-0.86$)    \\
$V_a\;[\text{km}/\text{s}]$                                    & 43.7\,(44.2)  & 45.5\,(45.1)     \\
$V_c\;[\text{km}/\text{s}]$                                    & 0\,(0)        & 0\,(0)           \\
$L\,[\text{kpc}]$                                              & 4\,(4)        & 4\,(4)           \\
$\phi_0\;[\text{GV}]$                                          & 0.72\,(0.72)  & 0.72\,(0.72)     \\
$\phi_1\;[\text{GV}]$                                          & 0.81\,(0.77)  & 0.75\,(0.75)     \\
\hline
$m_{\text{DM}} \,[\text{GeV}]$                                 & 76            & 88               \\
$\langle\sigma v\rangle\,[10^{-26}\,\text{cm}^3\!/\text{s}]$   & 0.73          & 0.4              \\ 
\hline
\multicolumn{3}{c}{}  \\
\multicolumn{3}{c}{--- Setup 2 ---}                                                                     \\ \hline
\,                                                             & w/o corr.          & with corr.        \\ \hline
$D_0\;\mathrm{[10^{28}\,cm^2/s]}$                              & 6.98\,(2.77)       & 3.89\,(2.26)      \\ 
$\delta$                                                       & 0.338 \,(0.421)    & 0.385\,(0.383)    \\
$\eta$                                                         & $-0.03$\,($-1.01$) & $-0.57$\,($-0.61$)\\
$v_\mathrm{A}\;[\text{km}/\text{s}]$                           & 17.4\,(15.5)       & 16.5\,(17.2)      \\
$v_{0,\mathrm{c}}\;[\text{km}/\text{s}]$                       & 12.4\,(2.23)       & 5.45\,(5.85)      \\
$z_\mathrm{h}\;\mathrm{[kpc]}$                                 & 6.87\,(3.27)       & 4.08\,(2.39)      \\
$\varphi_{\mathrm{SM,AMS},p,\mathrm{He}} \,\mathrm{[GV]}$      & 0.59\,(0.62)       & 0.60\,(0.61)      \\ 
$\varphi_{\mathrm{SM,AMS},\bar p} \,\mathrm{[GV]}$             & 0.53\,(0.56)       & 0.69\,(0.51)      \\ 
$\gamma_{1,p}$                                                 & 1.93\,(2.16)       & 2.05\,(2.10)      \\
$\gamma_1$                                                     & 1.94\,(2.24)       & 2.16\,(2.21)      \\
$\gamma_{2,p}$                                                 & 2.47\,(2.41)       & 2.43\,(2.43)      \\
$\gamma_2$                                                     & 2.43\,(2.37)       & 2.39\,(2.39)      \\
$R_{0}\;\mathrm{[GV]}$                                         & 7.58\,(1.28)       & 9.19\,(1.38)      \\
$s$                                                            & 0.46\,(0.30)        & 0.50\,(0.43)     \\
\hline
$m_{\text{DM}} \,[\text{GeV}]$                                 & 76             & 66                    \\
$\langle\sigma v\rangle\,[10^{-26}\,\text{cm}^3\!/\text{s}]$   & 0.91           & 0.74                  \\
\hline
\end{tabular}
\renewcommand{\arraystretch}{1}
\end{center}
\caption{
          Propagation, solar-modulation and dark-matter parameters yielding the best fit within the setup 1 (left) and 2 (right) with and without including correlations in the AMS-02 systematic errors. The values given in each column refer to the fits with (without) dark matter.
        }
\label{tab:propa}
\end{table}

\addcontentsline{toc}{section}{References}
\bibliography{article_ref}

\begin{thebibliography}{108}%
\makeatletter
\providecommand \@ifxundefined [1]{%
 \@ifx{#1\undefined}
}%
\providecommand \@ifnum [1]{%
 \ifnum #1\expandafter \@firstoftwo
 \else \expandafter \@secondoftwo
 \fi
}%
\providecommand \@ifx [1]{%
 \ifx #1\expandafter \@firstoftwo
 \else \expandafter \@secondoftwo
 \fi
}%
\providecommand \natexlab [1]{#1}%
\providecommand \enquote  [1]{``#1''}%
\providecommand \bibnamefont  [1]{#1}%
\providecommand \bibfnamefont [1]{#1}%
\providecommand \citenamefont [1]{#1}%
\providecommand \href@noop [0]{\@secondoftwo}%
\providecommand \href [0]{\begingroup \@sanitize@url \@href}%
\providecommand \@href[1]{\@@startlink{#1}\@@href}%
\providecommand \@@href[1]{\endgroup#1\@@endlink}%
\providecommand \@sanitize@url [0]{\catcode `\\12\catcode `\$12\catcode
  `\&12\catcode `\#12\catcode `\^12\catcode `\_12\catcode `\%12\relax}%
\providecommand \@@startlink[1]{}%
\providecommand \@@endlink[0]{}%
\providecommand \url  [0]{\begingroup\@sanitize@url \@url }%
\providecommand \@url [1]{\endgroup\@href {#1}{\urlprefix }}%
\providecommand \urlprefix  [0]{URL }%
\providecommand \Eprint [0]{\href }%
\providecommand \doibase [0]{http://dx.doi.org/}%
\providecommand \selectlanguage [0]{\@gobble}%
\providecommand \bibinfo  [0]{\@secondoftwo}%
\providecommand \bibfield  [0]{\@secondoftwo}%
\providecommand \translation [1]{[#1]}%
\providecommand \BibitemOpen [0]{}%
\providecommand \bibitemStop [0]{}%
\providecommand \bibitemNoStop [0]{.\EOS\space}%
\providecommand \EOS [0]{\spacefactor3000\relax}%
\providecommand \BibitemShut  [1]{\csname bibitem#1\endcsname}%
\let\auto@bib@innerbib\@empty
\bibitem [{\citenamefont {Golden}\ \emph {et~al.}(1979)\citenamefont {Golden},
  \citenamefont {Horan}, \citenamefont {Mauger}, \citenamefont {Badhwar},
  \citenamefont {Lacy}, \citenamefont {Stephens}, \citenamefont {Daniel},\ and\
  \citenamefont {Zipse}}]{Golden:1979bw}%
  \BibitemOpen
  \bibfield  {author} {\bibinfo {author} {\bibfnamefont {R.~L.}\ \bibnamefont
  {Golden}}, \bibinfo {author} {\bibfnamefont {S.}~\bibnamefont {Horan}},
  \bibinfo {author} {\bibfnamefont {B.~G.}\ \bibnamefont {Mauger}}, \bibinfo
  {author} {\bibfnamefont {G.~D.}\ \bibnamefont {Badhwar}}, \bibinfo {author}
  {\bibfnamefont {J.~L.}\ \bibnamefont {Lacy}}, \bibinfo {author}
  {\bibfnamefont {S.~A.}\ \bibnamefont {Stephens}}, \bibinfo {author}
  {\bibfnamefont {R.~R.}\ \bibnamefont {Daniel}}, \ and\ \bibinfo {author}
  {\bibfnamefont {J.~E.}\ \bibnamefont {Zipse}},\ }\href {\doibase
  10.1103/PhysRevLett.43.1196} {\bibfield  {journal} {\bibinfo  {journal}
  {Phys. Rev. Lett.}\ }\textbf {\bibinfo {volume} {43}},\ \bibinfo {pages}
  {1196} (\bibinfo {year} {1979})}\BibitemShut {NoStop}%
\bibitem [{\citenamefont {Buffington}\ \emph {et~al.}(1981)\citenamefont
  {Buffington}, \citenamefont {Schindler},\ and\ \citenamefont
  {Pennypacker}}]{Buffington:1981zz}%
  \BibitemOpen
  \bibfield  {author} {\bibinfo {author} {\bibfnamefont {A.}~\bibnamefont
  {Buffington}}, \bibinfo {author} {\bibfnamefont {S.~M.}\ \bibnamefont
  {Schindler}}, \ and\ \bibinfo {author} {\bibfnamefont {C.~R.}\ \bibnamefont
  {Pennypacker}},\ }\href {\doibase 10.1086/159247} {\bibfield  {journal}
  {\bibinfo  {journal} {Astrophys. J.}\ }\textbf {\bibinfo {volume} {248}},\
  \bibinfo {pages} {1179} (\bibinfo {year} {1981})}\BibitemShut {NoStop}%
\bibitem [{\citenamefont {Silk}\ and\ \citenamefont
  {Srednicki}(1984)}]{Silk:1984zy}%
  \BibitemOpen
  \bibfield  {author} {\bibinfo {author} {\bibfnamefont {J.}~\bibnamefont
  {Silk}}\ and\ \bibinfo {author} {\bibfnamefont {M.}~\bibnamefont
  {Srednicki}},\ }\href {\doibase 10.1103/PhysRevLett.53.624} {\bibfield
  {journal} {\bibinfo  {journal} {Phys. Rev. Lett.}\ }\textbf {\bibinfo
  {volume} {53}},\ \bibinfo {pages} {624} (\bibinfo {year} {1984})}\BibitemShut
  {NoStop}%
\bibitem [{\citenamefont {Stecker}\ \emph {et~al.}(1985)\citenamefont
  {Stecker}, \citenamefont {Rudaz},\ and\ \citenamefont
  {Walsh}}]{Stecker:1985jc}%
  \BibitemOpen
  \bibfield  {author} {\bibinfo {author} {\bibfnamefont {F.~W.}\ \bibnamefont
  {Stecker}}, \bibinfo {author} {\bibfnamefont {S.}~\bibnamefont {Rudaz}}, \
  and\ \bibinfo {author} {\bibfnamefont {T.~F.}\ \bibnamefont {Walsh}},\ }\href
  {\doibase 10.1103/PhysRevLett.55.2622} {\bibfield  {journal} {\bibinfo
  {journal} {Phys. Rev. Lett.}\ }\textbf {\bibinfo {volume} {55}},\ \bibinfo
  {pages} {2622} (\bibinfo {year} {1985})}\BibitemShut {NoStop}%
\bibitem [{\citenamefont {Aguilar}\ \emph
  {et~al.}(2016{\natexlab{a}})\citenamefont {Aguilar} \emph
  {et~al.}}]{Aguilar:2016kjl}%
  \BibitemOpen
  \bibfield  {author} {\bibinfo {author} {\bibfnamefont {M.}~\bibnamefont
  {Aguilar}} \emph {et~al.} (\bibinfo {collaboration} {AMS collaboration}),\
  }\href {\doibase 10.1103/PhysRevLett.117.091103} {\bibfield  {journal}
  {\bibinfo  {journal} {Phys. Rev. Lett.}\ }\textbf {\bibinfo {volume} {117}},\
  \bibinfo {pages} {091103} (\bibinfo {year} {2016}{\natexlab{a}})}\BibitemShut
  {NoStop}%
\bibitem [{\citenamefont {Cuoco}\ \emph
  {et~al.}(2017{\natexlab{a}})\citenamefont {Cuoco}, \citenamefont
  {Kr{\"a}mer},\ and\ \citenamefont {Korsmeier}}]{Cuoco:2016eej}%
  \BibitemOpen
  \bibfield  {author} {\bibinfo {author} {\bibfnamefont {A.}~\bibnamefont
  {Cuoco}}, \bibinfo {author} {\bibfnamefont {M.}~\bibnamefont {Kr{\"a}mer}}, \
  and\ \bibinfo {author} {\bibfnamefont {M.}~\bibnamefont {Korsmeier}},\ }\href
  {\doibase 10.1103/PhysRevLett.118.191102} {\bibfield  {journal} {\bibinfo
  {journal} {Phys. Rev. Lett.}\ }\textbf {\bibinfo {volume} {118}},\ \bibinfo
  {pages} {191102} (\bibinfo {year} {2017}{\natexlab{a}})},\ \Eprint
  {http://arxiv.org/abs/1610.03071} {arXiv:1610.03071 [astro-ph.HE]}
  \BibitemShut {NoStop}%
\bibitem [{\citenamefont {Cui}\ \emph {et~al.}(2017)\citenamefont {Cui},
  \citenamefont {Yuan}, \citenamefont {Tsai},\ and\ \citenamefont
  {Fan}}]{Cui:2016ppb}%
  \BibitemOpen
  \bibfield  {author} {\bibinfo {author} {\bibfnamefont {M.-Y.}\ \bibnamefont
  {Cui}}, \bibinfo {author} {\bibfnamefont {Q.}~\bibnamefont {Yuan}}, \bibinfo
  {author} {\bibfnamefont {Y.-L.~S.}\ \bibnamefont {Tsai}}, \ and\ \bibinfo
  {author} {\bibfnamefont {Y.-Z.}\ \bibnamefont {Fan}},\ }\href {\doibase
  10.1103/PhysRevLett.118.191101} {\bibfield  {journal} {\bibinfo  {journal}
  {Phys. Rev. Lett.}\ }\textbf {\bibinfo {volume} {118}},\ \bibinfo {pages}
  {191101} (\bibinfo {year} {2017})},\ \Eprint
  {http://arxiv.org/abs/1610.03840} {arXiv:1610.03840 [astro-ph.HE]}
  \BibitemShut {NoStop}%
\bibitem [{\citenamefont {Cuoco}\ \emph
  {et~al.}(2017{\natexlab{b}})\citenamefont {Cuoco}, \citenamefont {Heisig},
  \citenamefont {Korsmeier},\ and\ \citenamefont {Kr{\"a}mer}}]{Cuoco:2017rxb}%
  \BibitemOpen
  \bibfield  {author} {\bibinfo {author} {\bibfnamefont {A.}~\bibnamefont
  {Cuoco}}, \bibinfo {author} {\bibfnamefont {J.}~\bibnamefont {Heisig}},
  \bibinfo {author} {\bibfnamefont {M.}~\bibnamefont {Korsmeier}}, \ and\
  \bibinfo {author} {\bibfnamefont {M.}~\bibnamefont {Kr{\"a}mer}},\ }\href
  {\doibase 10.1088/1475-7516/2017/10/053} {\bibfield  {journal} {\bibinfo
  {journal} {JCAP}\ }\textbf {\bibinfo {volume} {10}},\ \bibinfo {pages} {053}
  (\bibinfo {year} {2017}{\natexlab{b}})},\ \Eprint
  {http://arxiv.org/abs/1704.08258} {arXiv:1704.08258 [astro-ph.HE]}
  \BibitemShut {NoStop}%
\bibitem [{\citenamefont {Reinert}\ and\ \citenamefont
  {Winkler}(2018)}]{Reinert:2017aga}%
  \BibitemOpen
  \bibfield  {author} {\bibinfo {author} {\bibfnamefont {A.}~\bibnamefont
  {Reinert}}\ and\ \bibinfo {author} {\bibfnamefont {M.~W.}\ \bibnamefont
  {Winkler}},\ }\href {\doibase 10.1088/1475-7516/2018/01/055} {\bibfield
  {journal} {\bibinfo  {journal} {JCAP}\ }\textbf {\bibinfo {volume} {01}},\
  \bibinfo {pages} {055} (\bibinfo {year} {2018})},\ \Eprint
  {http://arxiv.org/abs/1712.00002} {arXiv:1712.00002 [astro-ph.HE]}
  \BibitemShut {NoStop}%
\bibitem [{\citenamefont {Cui}\ \emph {et~al.}(2018)\citenamefont {Cui},
  \citenamefont {Pan}, \citenamefont {Yuan}, \citenamefont {Fan},\ and\
  \citenamefont {Zong}}]{Cui:2018klo}%
  \BibitemOpen
  \bibfield  {author} {\bibinfo {author} {\bibfnamefont {M.-Y.}\ \bibnamefont
  {Cui}}, \bibinfo {author} {\bibfnamefont {X.}~\bibnamefont {Pan}}, \bibinfo
  {author} {\bibfnamefont {Q.}~\bibnamefont {Yuan}}, \bibinfo {author}
  {\bibfnamefont {Y.-Z.}\ \bibnamefont {Fan}}, \ and\ \bibinfo {author}
  {\bibfnamefont {H.-S.}\ \bibnamefont {Zong}},\ }\href {\doibase
  10.1088/1475-7516/2018/06/024} {\bibfield  {journal} {\bibinfo  {journal}
  {JCAP}\ }\textbf {\bibinfo {volume} {06}},\ \bibinfo {pages} {024} (\bibinfo
  {year} {2018})},\ \Eprint {http://arxiv.org/abs/1803.02163} {arXiv:1803.02163
  [astro-ph.HE]} \BibitemShut {NoStop}%
\bibitem [{\citenamefont {Cuoco}\ \emph {et~al.}(2019)\citenamefont {Cuoco},
  \citenamefont {Heisig}, \citenamefont {Klamt}, \citenamefont {Korsmeier},\
  and\ \citenamefont {Kr{\"a}mer}}]{Cuoco:2019kuu}%
  \BibitemOpen
  \bibfield  {author} {\bibinfo {author} {\bibfnamefont {A.}~\bibnamefont
  {Cuoco}}, \bibinfo {author} {\bibfnamefont {J.}~\bibnamefont {Heisig}},
  \bibinfo {author} {\bibfnamefont {L.}~\bibnamefont {Klamt}}, \bibinfo
  {author} {\bibfnamefont {M.}~\bibnamefont {Korsmeier}}, \ and\ \bibinfo
  {author} {\bibfnamefont {M.}~\bibnamefont {Kr{\"a}mer}},\ }\href {\doibase
  10.1103/PhysRevD.99.103014} {\bibfield  {journal} {\bibinfo  {journal} {Phys.
  Rev. D}\ }\textbf {\bibinfo {volume} {99}},\ \bibinfo {pages} {103014}
  (\bibinfo {year} {2019})},\ \Eprint {http://arxiv.org/abs/1903.01472}
  {arXiv:1903.01472 [astro-ph.HE]} \BibitemShut {NoStop}%
\bibitem [{\citenamefont {Cholis}\ \emph {et~al.}(2019)\citenamefont {Cholis},
  \citenamefont {Linden},\ and\ \citenamefont {Hooper}}]{Cholis:2019ejx}%
  \BibitemOpen
  \bibfield  {author} {\bibinfo {author} {\bibfnamefont {I.}~\bibnamefont
  {Cholis}}, \bibinfo {author} {\bibfnamefont {T.}~\bibnamefont {Linden}}, \
  and\ \bibinfo {author} {\bibfnamefont {D.}~\bibnamefont {Hooper}},\ }\href
  {\doibase 10.1103/PhysRevD.99.103026} {\bibfield  {journal} {\bibinfo
  {journal} {Phys. Rev.}\ }\textbf {\bibinfo {volume} {D99}},\ \bibinfo {pages}
  {103026} (\bibinfo {year} {2019})},\ \Eprint
  {http://arxiv.org/abs/1903.02549} {arXiv:1903.02549 [astro-ph.HE]}
  \BibitemShut {NoStop}%
\bibitem [{\citenamefont {Lin}\ \emph {et~al.}(2019)\citenamefont {Lin},
  \citenamefont {Bi},\ and\ \citenamefont {Yin}}]{Lin:2019ljc}%
  \BibitemOpen
  \bibfield  {author} {\bibinfo {author} {\bibfnamefont {S.-J.}\ \bibnamefont
  {Lin}}, \bibinfo {author} {\bibfnamefont {X.-J.}\ \bibnamefont {Bi}}, \ and\
  \bibinfo {author} {\bibfnamefont {P.-F.}\ \bibnamefont {Yin}},\ }\href
  {\doibase 10.1103/PhysRevD.100.103014} {\bibfield  {journal} {\bibinfo
  {journal} {Phys. Rev.}\ }\textbf {\bibinfo {volume} {D100}},\ \bibinfo
  {pages} {103014} (\bibinfo {year} {2019})},\ \Eprint
  {http://arxiv.org/abs/1903.09545} {arXiv:1903.09545 [astro-ph.HE]}
  \BibitemShut {NoStop}%
\bibitem [{\citenamefont {Goodenough}\ and\ \citenamefont
  {Hooper}(2009)}]{Goodenough:2009gk}%
  \BibitemOpen
  \bibfield  {author} {\bibinfo {author} {\bibfnamefont {L.}~\bibnamefont
  {Goodenough}}\ and\ \bibinfo {author} {\bibfnamefont {D.}~\bibnamefont
  {Hooper}},\ }\href@noop {} {\  (\bibinfo {year} {2009})},\ \Eprint
  {http://arxiv.org/abs/0910.2998} {arXiv:0910.2998 [hep-ph]} \BibitemShut
  {NoStop}%
\bibitem [{\citenamefont {di~Mauro}\ \emph {et~al.}(2014)\citenamefont
  {di~Mauro}, \citenamefont {Donato}, \citenamefont {Goudelis},\ and\
  \citenamefont {Serpico}}]{diMauro:2014zea}%
  \BibitemOpen
  \bibfield  {author} {\bibinfo {author} {\bibfnamefont {M.}~\bibnamefont
  {di~Mauro}}, \bibinfo {author} {\bibfnamefont {F.}~\bibnamefont {Donato}},
  \bibinfo {author} {\bibfnamefont {A.}~\bibnamefont {Goudelis}}, \ and\
  \bibinfo {author} {\bibfnamefont {P.~D.}\ \bibnamefont {Serpico}},\ }\href
  {\doibase 10.1103/PhysRevD.90.085017} {\bibfield  {journal} {\bibinfo
  {journal} {Phys. Rev. D}\ }\textbf {\bibinfo {volume} {90}},\ \bibinfo
  {pages} {085017} (\bibinfo {year} {2014})},\ \bibinfo {note} {[Erratum:
  Phys.Rev.D 98, 049901 (2018)]},\ \Eprint {http://arxiv.org/abs/1408.0288}
  {arXiv:1408.0288 [hep-ph]} \BibitemShut {NoStop}%
\bibitem [{\citenamefont {Kappl}\ and\ \citenamefont
  {Winkler}(2014)}]{Kappl:2014hha}%
  \BibitemOpen
  \bibfield  {author} {\bibinfo {author} {\bibfnamefont {R.}~\bibnamefont
  {Kappl}}\ and\ \bibinfo {author} {\bibfnamefont {M.~W.}\ \bibnamefont
  {Winkler}},\ }\href {\doibase 10.1088/1475-7516/2014/09/051} {\bibfield
  {journal} {\bibinfo  {journal} {JCAP}\ }\textbf {\bibinfo {volume} {1409}},\
  \bibinfo {pages} {051} (\bibinfo {year} {2014})},\ \Eprint
  {http://arxiv.org/abs/1408.0299} {arXiv:1408.0299 [hep-ph]} \BibitemShut
  {NoStop}%
\bibitem [{\citenamefont {Kachelriess}\ \emph {et~al.}(2015)\citenamefont
  {Kachelriess}, \citenamefont {Moskalenko},\ and\ \citenamefont
  {Ostapchenko}}]{Kachelriess:2015wpa}%
  \BibitemOpen
  \bibfield  {author} {\bibinfo {author} {\bibfnamefont {M.}~\bibnamefont
  {Kachelriess}}, \bibinfo {author} {\bibfnamefont {I.~V.}\ \bibnamefont
  {Moskalenko}}, \ and\ \bibinfo {author} {\bibfnamefont {S.~S.}\ \bibnamefont
  {Ostapchenko}},\ }\href {\doibase 10.1088/0004-637X/803/2/54} {\bibfield
  {journal} {\bibinfo  {journal} {Astrophys. J.}\ }\textbf {\bibinfo {volume}
  {803}},\ \bibinfo {pages} {54} (\bibinfo {year} {2015})},\ \Eprint
  {http://arxiv.org/abs/1502.04158} {arXiv:1502.04158 [astro-ph.HE]}
  \BibitemShut {NoStop}%
\bibitem [{\citenamefont {Winkler}(2017)}]{Winkler:2017xor}%
  \BibitemOpen
  \bibfield  {author} {\bibinfo {author} {\bibfnamefont {M.~W.}\ \bibnamefont
  {Winkler}},\ }\href {\doibase 10.1088/1475-7516/2017/02/048} {\bibfield
  {journal} {\bibinfo  {journal} {JCAP}\ }\textbf {\bibinfo {volume} {1702}},\
  \bibinfo {pages} {048} (\bibinfo {year} {2017})},\ \Eprint
  {http://arxiv.org/abs/1701.04866} {arXiv:1701.04866 [hep-ph]} \BibitemShut
  {NoStop}%
\bibitem [{\citenamefont {Korsmeier}\ \emph {et~al.}(2018)\citenamefont
  {Korsmeier}, \citenamefont {Donato},\ and\ \citenamefont
  {Di~Mauro}}]{Korsmeier:2018gcy}%
  \BibitemOpen
  \bibfield  {author} {\bibinfo {author} {\bibfnamefont {M.}~\bibnamefont
  {Korsmeier}}, \bibinfo {author} {\bibfnamefont {F.}~\bibnamefont {Donato}}, \
  and\ \bibinfo {author} {\bibfnamefont {M.}~\bibnamefont {Di~Mauro}},\ }\href
  {\doibase 10.1103/PhysRevD.97.103019} {\bibfield  {journal} {\bibinfo
  {journal} {Phys. Rev.}\ }\textbf {\bibinfo {volume} {D97}},\ \bibinfo {pages}
  {103019} (\bibinfo {year} {2018})},\ \Eprint
  {http://arxiv.org/abs/1802.03030} {arXiv:1802.03030 [astro-ph.HE]}
  \BibitemShut {NoStop}%
\bibitem [{\citenamefont {Derome}\ \emph {et~al.}(2019)\citenamefont {Derome},
  \citenamefont {Maurin}, \citenamefont {Salati}, \citenamefont {Boudaud},
  \citenamefont {G{\'e}nolini},\ and\ \citenamefont
  {Kunz{\'e}}}]{Derome:2019jfs}%
  \BibitemOpen
  \bibfield  {author} {\bibinfo {author} {\bibfnamefont {L.}~\bibnamefont
  {Derome}}, \bibinfo {author} {\bibfnamefont {D.}~\bibnamefont {Maurin}},
  \bibinfo {author} {\bibfnamefont {P.}~\bibnamefont {Salati}}, \bibinfo
  {author} {\bibfnamefont {M.}~\bibnamefont {Boudaud}}, \bibinfo {author}
  {\bibfnamefont {Y.}~\bibnamefont {G{\'e}nolini}}, \ and\ \bibinfo {author}
  {\bibfnamefont {P.}~\bibnamefont {Kunz{\'e}}},\ }\href {\doibase
  10.1051/0004-6361/201935717} {\bibfield  {journal} {\bibinfo  {journal}
  {Astron. Astrophys.}\ }\textbf {\bibinfo {volume} {627}},\ \bibinfo {pages}
  {A158} (\bibinfo {year} {2019})},\ \Eprint {http://arxiv.org/abs/1904.08210}
  {arXiv:1904.08210 [astro-ph.HE]} \BibitemShut {NoStop}%
\bibitem [{\citenamefont {Boudaud}\ \emph {et~al.}(2020)\citenamefont
  {Boudaud}, \citenamefont {G{\'e}nolini}, \citenamefont {Derome}, \citenamefont
  {Lavalle}, \citenamefont {Maurin}, \citenamefont {Salati},\ and\
  \citenamefont {Serpico}}]{Boudaud:2019efq}%
  \BibitemOpen
  \bibfield  {author} {\bibinfo {author} {\bibfnamefont {M.}~\bibnamefont
  {Boudaud}}, \bibinfo {author} {\bibfnamefont {Y.}~\bibnamefont {G{\'e}nolini}},
  \bibinfo {author} {\bibfnamefont {L.}~\bibnamefont {Derome}}, \bibinfo
  {author} {\bibfnamefont {J.}~\bibnamefont {Lavalle}}, \bibinfo {author}
  {\bibfnamefont {D.}~\bibnamefont {Maurin}}, \bibinfo {author} {\bibfnamefont
  {P.}~\bibnamefont {Salati}}, \ and\ \bibinfo {author} {\bibfnamefont {P.~D.}\
  \bibnamefont {Serpico}},\ }\href {\doibase 10.1103/PhysRevResearch.2.023022}
  {\bibfield  {journal} {\bibinfo  {journal} {Phys. Rev. Res.}\ }\textbf
  {\bibinfo {volume} {2}},\ \bibinfo {pages} {023022} (\bibinfo {year}
  {2020})},\ \Eprint {http://arxiv.org/abs/1906.07119} {arXiv:1906.07119
  [astro-ph.HE]} \BibitemShut {NoStop}%
\bibitem [{\citenamefont {Glauber}(1959)}]{Glauber1959}%
  \BibitemOpen
  \bibfield  {author} {\bibinfo {author} {\bibfnamefont {R.~J.}\ \bibnamefont
  {Glauber}},\ }in\ \href@noop {} {\emph {\bibinfo {booktitle} {Lectures in
  Theoretical Physics, Vol. 1}}},\ \bibinfo {editor} {edited by\ \bibinfo
  {editor} {\bibfnamefont {W.}~\bibnamefont {Brittin}}\ and\ \bibinfo {editor}
  {\bibfnamefont {L.~G.}\ \bibnamefont {Dunham}}}\ (\bibinfo  {publisher}
  {Interscience Publishers},\ \bibinfo {address} {New York},\ \bibinfo {year}
  {1959})\ pp.\ \bibinfo {pages} {315--414}\BibitemShut {NoStop}%
\bibitem [{\citenamefont {Sitenko}(1959)}]{Sitenko:1959zh}%
  \BibitemOpen
  \bibfield  {author} {\bibinfo {author} {\bibfnamefont {A.~G.}\ \bibnamefont
  {Sitenko}},\ }\href@noop {} {\bibfield  {journal} {\bibinfo  {journal} {Ukr.
  Fiz. Zh.}\ }\textbf {\bibinfo {volume} {4}},\ \bibinfo {pages} {152}
  (\bibinfo {year} {1959})},\ \bibinfo {note} {[ISSN 2071-0194. Ukr. J. Phys.
  2008. V. 53, Special Issue]}\BibitemShut {NoStop}%
\bibitem [{\citenamefont {Glauber}\ and\ \citenamefont
  {Matthiae}(1970)}]{Glauber:1970jm}%
  \BibitemOpen
  \bibfield  {author} {\bibinfo {author} {\bibfnamefont {R.~J.}\ \bibnamefont
  {Glauber}}\ and\ \bibinfo {author} {\bibfnamefont {G.}~\bibnamefont
  {Matthiae}},\ }\href {\doibase 10.1016/0550-3213(70)90511-0} {\bibfield
  {journal} {\bibinfo  {journal} {Nucl. Phys.}\ }\textbf {\bibinfo {volume}
  {B21}},\ \bibinfo {pages} {135} (\bibinfo {year} {1970})}\BibitemShut
  {NoStop}%
\bibitem [{\citenamefont {Gribov}(1969)}]{Gribov:1968jf}%
  \BibitemOpen
  \bibfield  {author} {\bibinfo {author} {\bibfnamefont {V.~N.}\ \bibnamefont
  {Gribov}},\ }\href@noop {} {\bibfield  {journal} {\bibinfo  {journal} {Sov.
  Phys. JETP}\ }\textbf {\bibinfo {volume} {29}},\ \bibinfo {pages} {483}
  (\bibinfo {year} {1969})},\ \bibinfo {note} {[Zh. Eksp. Teor.
  Fiz.56,892(1969)]}\BibitemShut {NoStop}%
\bibitem [{\citenamefont {Pumplin}\ and\ \citenamefont
  {Ross}(1968)}]{Pumplin:1968bi}%
  \BibitemOpen
  \bibfield  {author} {\bibinfo {author} {\bibfnamefont {J.}~\bibnamefont
  {Pumplin}}\ and\ \bibinfo {author} {\bibfnamefont {M.~H.}\ \bibnamefont
  {Ross}},\ }\href {\doibase 10.1103/PhysRevLett.21.1778} {\bibfield  {journal}
  {\bibinfo  {journal} {Phys. Rev. Lett.}\ }\textbf {\bibinfo {volume} {21}},\
  \bibinfo {pages} {1778} (\bibinfo {year} {1968})}\BibitemShut {NoStop}%
\bibitem [{\citenamefont {Elton}(1961)}]{elton1961nuclear}%
  \BibitemOpen
  \bibfield  {author} {\bibinfo {author} {\bibfnamefont {L.}~\bibnamefont
  {Elton}},\ }\href@noop {} {\emph {\bibinfo {title} {Nuclear Sizes}}},\ Oxford
  Library of the Physical Sciences\ (\bibinfo  {publisher} {Oxford University
  Press},\ \bibinfo {year} {1961})\BibitemShut {NoStop}%
\bibitem [{\citenamefont {De~Vries}\ \emph {et~al.}(1987)\citenamefont
  {De~Vries}, \citenamefont {De~Jager},\ and\ \citenamefont
  {De~Vries}}]{DeJager:1987qc}%
  \BibitemOpen
  \bibfield  {author} {\bibinfo {author} {\bibfnamefont {H.}~\bibnamefont
  {De~Vries}}, \bibinfo {author} {\bibfnamefont {C.~W.}\ \bibnamefont
  {De~Jager}}, \ and\ \bibinfo {author} {\bibfnamefont {C.}~\bibnamefont
  {De~Vries}},\ }\href {\doibase 10.1016/0092-640X(87)90013-1} {\bibfield
  {journal} {\bibinfo  {journal} {Atom. Data Nucl. Data Tabl.}\ }\textbf
  {\bibinfo {volume} {36}},\ \bibinfo {pages} {495} (\bibinfo {year}
  {1987})}\BibitemShut {NoStop}%
\bibitem [{\citenamefont {Pi}(1992)}]{Pi:1992ug}%
  \BibitemOpen
  \bibfield  {author} {\bibinfo {author} {\bibfnamefont {H.}~\bibnamefont
  {Pi}},\ }\href {\doibase 10.1016/0010-4655(92)90082-A} {\bibfield  {journal}
  {\bibinfo  {journal} {Comput. Phys. Commun.}\ }\textbf {\bibinfo {volume}
  {71}},\ \bibinfo {pages} {173} (\bibinfo {year} {1992})}\BibitemShut
  {NoStop}%
\bibitem [{\citenamefont {Rybczynski}\ \emph {et~al.}(2014)\citenamefont
  {Rybczynski}, \citenamefont {Stefanek}, \citenamefont {Broniowski},\ and\
  \citenamefont {Bozek}}]{Rybczynski:2013yba}%
  \BibitemOpen
  \bibfield  {author} {\bibinfo {author} {\bibfnamefont {M.}~\bibnamefont
  {Rybczynski}}, \bibinfo {author} {\bibfnamefont {G.}~\bibnamefont
  {Stefanek}}, \bibinfo {author} {\bibfnamefont {W.}~\bibnamefont
  {Broniowski}}, \ and\ \bibinfo {author} {\bibfnamefont {P.}~\bibnamefont
  {Bozek}},\ }\href {\doibase 10.1016/j.cpc.2014.02.016} {\bibfield  {journal}
  {\bibinfo  {journal} {Comput. Phys. Commun.}\ }\textbf {\bibinfo {volume}
  {185}},\ \bibinfo {pages} {1759} (\bibinfo {year} {2014})},\ \Eprint
  {http://arxiv.org/abs/1310.5475} {arXiv:1310.5475 [nucl-th]} \BibitemShut
  {NoStop}%
\bibitem [{\citenamefont {Angeli}\ and\ \citenamefont
  {Marinova}(2013)}]{Angeli:2013epw}%
  \BibitemOpen
  \bibfield  {author} {\bibinfo {author} {\bibfnamefont {I.}~\bibnamefont
  {Angeli}}\ and\ \bibinfo {author} {\bibfnamefont {K.~P.}\ \bibnamefont
  {Marinova}},\ }\href {\doibase 10.1016/j.adt.2011.12.006} {\bibfield
  {journal} {\bibinfo  {journal} {Atom. Data Nucl. Data Tabl.}\ }\textbf
  {\bibinfo {volume} {99}},\ \bibinfo {pages} {69} (\bibinfo {year}
  {2013})}\BibitemShut {NoStop}%
\bibitem [{\citenamefont {Kopeliovich}\ \emph {et~al.}(1989)\citenamefont
  {Kopeliovich}, \citenamefont {Nikolaev},\ and\ \citenamefont
  {Potashnikova}}]{Kopeliovich:1989iy}%
  \BibitemOpen
  \bibfield  {author} {\bibinfo {author} {\bibfnamefont {B.~Z.}\ \bibnamefont
  {Kopeliovich}}, \bibinfo {author} {\bibfnamefont {N.~N.}\ \bibnamefont
  {Nikolaev}}, \ and\ \bibinfo {author} {\bibfnamefont {I.~K.}\ \bibnamefont
  {Potashnikova}},\ }\href {\doibase 10.1103/PhysRevD.39.769} {\bibfield
  {journal} {\bibinfo  {journal} {Phys. Rev.}\ }\textbf {\bibinfo {volume}
  {D39}},\ \bibinfo {pages} {769} (\bibinfo {year} {1989})}\BibitemShut
  {NoStop}%
\bibitem [{\citenamefont {Faldt}\ and\ \citenamefont
  {Hulthage}(1978)}]{Faldt:1977qs}%
  \BibitemOpen
  \bibfield  {author} {\bibinfo {author} {\bibfnamefont {G.}~\bibnamefont
  {Faldt}}\ and\ \bibinfo {author} {\bibfnamefont {I.}~\bibnamefont
  {Hulthage}},\ }\href {\doibase 10.1088/0305-4616/4/3/012} {\bibfield
  {journal} {\bibinfo  {journal} {J. Phys.}\ }\textbf {\bibinfo {volume}
  {G4}},\ \bibinfo {pages} {363} (\bibinfo {year} {1978})}\BibitemShut
  {NoStop}%
\bibitem [{\citenamefont {Glauber}\ and\ \citenamefont
  {Osland}(1979)}]{Glauber:1978fw}%
  \BibitemOpen
  \bibfield  {author} {\bibinfo {author} {\bibfnamefont {R.~J.}\ \bibnamefont
  {Glauber}}\ and\ \bibinfo {author} {\bibfnamefont {P.}~\bibnamefont
  {Osland}},\ }\href {\doibase 10.1016/0370-2693(79)91200-0} {\bibfield
  {journal} {\bibinfo  {journal} {Phys. Lett.}\ }\textbf {\bibinfo {volume}
  {80B}},\ \bibinfo {pages} {401} (\bibinfo {year} {1979})}\BibitemShut
  {NoStop}%
\bibitem [{\citenamefont {Larionov}\ and\ \citenamefont
  {Lenske}(2017)}]{Larionov:2016xeb}%
  \BibitemOpen
  \bibfield  {author} {\bibinfo {author} {\bibfnamefont {A.~B.}\ \bibnamefont
  {Larionov}}\ and\ \bibinfo {author} {\bibfnamefont {H.}~\bibnamefont
  {Lenske}},\ }\href {\doibase 10.1016/j.nuclphysa.2016.10.006} {\bibfield
  {journal} {\bibinfo  {journal} {Nucl. Phys.}\ }\textbf {\bibinfo {volume}
  {A957}},\ \bibinfo {pages} {450} (\bibinfo {year} {2017})},\ \Eprint
  {http://arxiv.org/abs/1609.03343} {arXiv:1609.03343 [nucl-th]} \BibitemShut
  {NoStop}%
\bibitem [{\citenamefont {Kopeliovich}\ \emph {et~al.}(2006)\citenamefont
  {Kopeliovich}, \citenamefont {Potashnikova},\ and\ \citenamefont
  {Schmidt}}]{Kopeliovich:2005us}%
  \BibitemOpen
  \bibfield  {author} {\bibinfo {author} {\bibfnamefont {B.~Z.}\ \bibnamefont
  {Kopeliovich}}, \bibinfo {author} {\bibfnamefont {I.~K.}\ \bibnamefont
  {Potashnikova}}, \ and\ \bibinfo {author} {\bibfnamefont {I.}~\bibnamefont
  {Schmidt}},\ }\href {\doibase 10.1103/PhysRevC.73.034901} {\bibfield
  {journal} {\bibinfo  {journal} {Phys. Rev.}\ }\textbf {\bibinfo {volume}
  {C73}},\ \bibinfo {pages} {034901} (\bibinfo {year} {2006})},\ \Eprint
  {http://arxiv.org/abs/hep-ph/0508277} {arXiv:hep-ph/0508277}
  \BibitemShut {NoStop}%
\bibitem [{\citenamefont {Kaidalov}\ and\ \citenamefont
  {Kondratyuk}(1973)}]{Kaidalov:1973fz}%
  \BibitemOpen
  \bibfield  {author} {\bibinfo {author} {\bibfnamefont {A.~B.}\ \bibnamefont
  {Kaidalov}}\ and\ \bibinfo {author} {\bibfnamefont {L.~A.}\ \bibnamefont
  {Kondratyuk}},\ }\href {\doibase 10.1016/0550-3213(73)90222-8} {\bibfield
  {journal} {\bibinfo  {journal} {Nucl. Phys.}\ }\textbf {\bibinfo {volume}
  {B56}},\ \bibinfo {pages} {90} (\bibinfo {year} {1973})}\BibitemShut
  {NoStop}%
\bibitem [{\citenamefont {Von~Bochmann}\ \emph {et~al.}(1970)\citenamefont
  {Von~Bochmann}, \citenamefont {Kofoed-Hansen},\ and\ \citenamefont
  {Margolis}}]{VonBochmann:1970xx}%
  \BibitemOpen
  \bibfield  {author} {\bibinfo {author} {\bibfnamefont {G.}~\bibnamefont
  {Von~Bochmann}}, \bibinfo {author} {\bibfnamefont {O.}~\bibnamefont
  {Kofoed-Hansen}}, \ and\ \bibinfo {author} {\bibfnamefont {B.}~\bibnamefont
  {Margolis}},\ }\href {\doibase 10.1016/0370-2693(70)90578-2} {\bibfield
  {journal} {\bibinfo  {journal} {Phys. Lett.}\ }\textbf {\bibinfo {volume}
  {33B}},\ \bibinfo {pages} {222} (\bibinfo {year} {1970})}\BibitemShut
  {NoStop}%
\bibitem [{\citenamefont {Karmanov}\ and\ \citenamefont
  {Kondratyuk}(1973)}]{Karmanov:1973va}%
  \BibitemOpen
  \bibfield  {author} {\bibinfo {author} {\bibfnamefont {V.~A.}\ \bibnamefont
  {Karmanov}}\ and\ \bibinfo {author} {\bibfnamefont {L.~A.}\ \bibnamefont
  {Kondratyuk}},\ }\href@noop {} {\bibfield  {journal} {\bibinfo  {journal}
  {Pisma Zh. Eksp. Teor. Fiz.}\ }\textbf {\bibinfo {volume} {18}},\ \bibinfo
  {pages} {451} (\bibinfo {year} {1973})}\BibitemShut {NoStop}%
\bibitem [{\citenamefont {Kopeliovich}(2003)}]{Kopeliovich:2003tz}%
  \BibitemOpen
  \bibfield  {author} {\bibinfo {author} {\bibfnamefont {B.~Z.}\ \bibnamefont
  {Kopeliovich}},\ }\href {\doibase 10.1103/PhysRevC.68.044906} {\bibfield
  {journal} {\bibinfo  {journal} {Phys. Rev.}\ }\textbf {\bibinfo {volume}
  {C68}},\ \bibinfo {pages} {044906} (\bibinfo {year} {2003})},\ \Eprint
  {http://arxiv.org/abs/nucl-th/0306044} {arXiv:nucl-th/0306044}
  \BibitemShut {NoStop}%
\bibitem [{\citenamefont {Frankfurt}\ \emph {et~al.}(2012)\citenamefont
  {Frankfurt}, \citenamefont {Guzey},\ and\ \citenamefont
  {Strikman}}]{Frankfurt:2011cs}%
  \BibitemOpen
  \bibfield  {author} {\bibinfo {author} {\bibfnamefont {L.}~\bibnamefont
  {Frankfurt}}, \bibinfo {author} {\bibfnamefont {V.}~\bibnamefont {Guzey}}, \
  and\ \bibinfo {author} {\bibfnamefont {M.}~\bibnamefont {Strikman}},\ }\href
  {\doibase 10.1016/j.physrep.2011.12.002} {\bibfield  {journal} {\bibinfo
  {journal} {Phys. Rept.}\ }\textbf {\bibinfo {volume} {512}},\ \bibinfo
  {pages} {255} (\bibinfo {year} {2012})},\ \Eprint
  {http://arxiv.org/abs/1106.2091} {arXiv:1106.2091 [hep-ph]} \BibitemShut
  {NoStop}%
\bibitem [{\citenamefont {Ramana~Murthy}\ \emph {et~al.}(1975)\citenamefont
  {Ramana~Murthy}, \citenamefont {Ayre}, \citenamefont {Gustafson},
  \citenamefont {Jones},\ and\ \citenamefont {Longo}}]{RamanaMurthy:1975vfu}%
  \BibitemOpen
  \bibfield  {author} {\bibinfo {author} {\bibfnamefont {P.~V.}\ \bibnamefont
  {Ramana~Murthy}}, \bibinfo {author} {\bibfnamefont {C.~A.}\ \bibnamefont
  {Ayre}}, \bibinfo {author} {\bibfnamefont {H.~R.}\ \bibnamefont {Gustafson}},
  \bibinfo {author} {\bibfnamefont {L.~W.}\ \bibnamefont {Jones}}, \ and\
  \bibinfo {author} {\bibfnamefont {M.~J.}\ \bibnamefont {Longo}},\ }\href
  {\doibase 10.1016/0550-3213(75)90182-0} {\bibfield  {journal} {\bibinfo
  {journal} {Nucl. Phys.}\ }\textbf {\bibinfo {volume} {B92}},\ \bibinfo
  {pages} {269} (\bibinfo {year} {1975})}\BibitemShut {NoStop}%
\bibitem [{\citenamefont {Bartenev}\ \emph {et~al.}(1974)\citenamefont
  {Bartenev} \emph {et~al.}}]{Bartenev:1974gz}%
  \BibitemOpen
  \bibfield  {author} {\bibinfo {author} {\bibfnamefont {V.}~\bibnamefont
  {Bartenev}} \emph {et~al.},\ }\href {\doibase 10.1016/0370-2693(74)90296-2}
  {\bibfield  {journal} {\bibinfo  {journal} {Phys. Lett.}\ }\textbf {\bibinfo
  {volume} {51B}},\ \bibinfo {pages} {299} (\bibinfo {year}
  {1974})}\BibitemShut {NoStop}%
\bibitem [{\citenamefont {Jenkovszky}\ \emph {et~al.}(2014)\citenamefont
  {Jenkovszky}, \citenamefont {Kuprash}, \citenamefont {Orava},\ and\
  \citenamefont {Salii}}]{Jenkovszky:2012hf}%
  \BibitemOpen
  \bibfield  {author} {\bibinfo {author} {\bibfnamefont {L.}~\bibnamefont
  {Jenkovszky}}, \bibinfo {author} {\bibfnamefont {O.}~\bibnamefont {Kuprash}},
  \bibinfo {author} {\bibfnamefont {R.}~\bibnamefont {Orava}}, \ and\ \bibinfo
  {author} {\bibfnamefont {A.}~\bibnamefont {Salii}},\ }\href {\doibase
  10.1134/S1063778814120072} {\bibfield  {journal} {\bibinfo  {journal} {Phys.
  Atom. Nucl.}\ }\textbf {\bibinfo {volume} {77}},\ \bibinfo {pages} {1463}
  (\bibinfo {year} {2014})},\ \bibinfo {note} {[Odessa Astron.
  Pub.25,102(2012)]},\ \Eprint {http://arxiv.org/abs/1211.5841}
  {arXiv:1211.5841 [hep-ph]} \BibitemShut {NoStop}%
\bibitem [{\citenamefont {Ishida}\ and\ \citenamefont
  {Igi}(2009)}]{Ishida:2009fp}%
  \BibitemOpen
  \bibfield  {author} {\bibinfo {author} {\bibfnamefont {M.}~\bibnamefont
  {Ishida}}\ and\ \bibinfo {author} {\bibfnamefont {K.}~\bibnamefont {Igi}},\
  }\href {\doibase 10.1103/PhysRevD.79.096003} {\bibfield  {journal} {\bibinfo
  {journal} {Phys. Rev.}\ }\textbf {\bibinfo {volume} {D79}},\ \bibinfo {pages}
  {096003} (\bibinfo {year} {2009})},\ \Eprint {http://arxiv.org/abs/0903.1889}
  {arXiv:0903.1889 [hep-ph]} \BibitemShut {NoStop}%
\bibitem [{\citenamefont {Tanabashi}\ \emph {et~al.}(2018)\citenamefont
  {Tanabashi} \emph {et~al.}}]{Tanabashi:2018oca}%
  \BibitemOpen
  \bibfield  {author} {\bibinfo {author} {\bibfnamefont {M.}~\bibnamefont
  {Tanabashi}} \emph {et~al.} (\bibinfo {collaboration} {Particle Data
  Group}),\ }\href {\doibase 10.1103/PhysRevD.98.030001} {\bibfield  {journal}
  {\bibinfo  {journal} {Phys. Rev.}\ }\textbf {\bibinfo {volume} {D98}},\
  \bibinfo {pages} {030001} (\bibinfo {year} {2018})}\BibitemShut {NoStop}%
\bibitem [{\citenamefont {Okorokov}(2015)}]{Okorokov:2015bha}%
  \BibitemOpen
  \bibfield  {author} {\bibinfo {author} {\bibfnamefont {V.~A.}\ \bibnamefont
  {Okorokov}},\ }\href {\doibase 10.1155/2015/914170} {\bibfield  {journal}
  {\bibinfo  {journal} {Adv. High Energy Phys.}\ }\textbf {\bibinfo {volume}
  {2015}},\ \bibinfo {pages} {914170} (\bibinfo {year} {2015})},\ \Eprint
  {http://arxiv.org/abs/1501.01142} {arXiv:1501.01142 [hep-ph]} \BibitemShut
  {NoStop}%
\bibitem [{\citenamefont {Allaby}\ \emph {et~al.}(1970)\citenamefont {Allaby}
  \emph {et~al.}}]{Allaby:1970pv}%
  \BibitemOpen
  \bibfield  {author} {\bibinfo {author} {\bibfnamefont {J.~V.}\ \bibnamefont
  {Allaby}} \emph {et~al.},\ }\href@noop {} {\bibfield  {journal} {\bibinfo
  {journal} {Yad. Fiz.}\ }\textbf {\bibinfo {volume} {12}},\ \bibinfo {pages}
  {538} (\bibinfo {year} {1970})}\BibitemShut {NoStop}%
\bibitem [{\citenamefont {Abrams}\ \emph {et~al.}(1971)\citenamefont {Abrams},
  \citenamefont {Cool}, \citenamefont {Giacomelli}, \citenamefont {Kycia},
  \citenamefont {Leontic}, \citenamefont {Li}, \citenamefont {Lundby},
  \citenamefont {Michael},\ and\ \citenamefont {Teiger}}]{Abrams:1972ab}%
  \BibitemOpen
  \bibfield  {author} {\bibinfo {author} {\bibfnamefont {R.~J.}\ \bibnamefont
  {Abrams}}, \bibinfo {author} {\bibfnamefont {R.~L.}\ \bibnamefont {Cool}},
  \bibinfo {author} {\bibfnamefont {G.}~\bibnamefont {Giacomelli}}, \bibinfo
  {author} {\bibfnamefont {T.~F.}\ \bibnamefont {Kycia}}, \bibinfo {author}
  {\bibfnamefont {B.~A.}\ \bibnamefont {Leontic}}, \bibinfo {author}
  {\bibfnamefont {K.~K.}\ \bibnamefont {Li}}, \bibinfo {author} {\bibfnamefont
  {A.}~\bibnamefont {Lundby}}, \bibinfo {author} {\bibfnamefont {D.~N.}\
  \bibnamefont {Michael}}, \ and\ \bibinfo {author} {\bibfnamefont
  {J.}~\bibnamefont {Teiger}},\ }\href {\doibase 10.1103/PhysRevD.4.3235}
  {\bibfield  {journal} {\bibinfo  {journal} {Phys. Rev.}\ }\textbf {\bibinfo
  {volume} {D4}},\ \bibinfo {pages} {3235} (\bibinfo {year}
  {1971})}\BibitemShut {NoStop}%
\bibitem [{\citenamefont {Denisov}\ \emph {et~al.}(1973)\citenamefont
  {Denisov}, \citenamefont {Donskov}, \citenamefont {Gorin}, \citenamefont
  {Krasnokutsky}, \citenamefont {Petrukhin}, \citenamefont {Prokoshkin},\ and\
  \citenamefont {Stoyanova}}]{Denisov:1973zv}%
  \BibitemOpen
  \bibfield  {author} {\bibinfo {author} {\bibfnamefont {S.~P.}\ \bibnamefont
  {Denisov}}, \bibinfo {author} {\bibfnamefont {S.~V.}\ \bibnamefont
  {Donskov}}, \bibinfo {author} {\bibfnamefont {{\relax Yu}.~P.}\ \bibnamefont
  {Gorin}}, \bibinfo {author} {\bibfnamefont {R.~N.}\ \bibnamefont
  {Krasnokutsky}}, \bibinfo {author} {\bibfnamefont {A.~I.}\ \bibnamefont
  {Petrukhin}}, \bibinfo {author} {\bibfnamefont {{\relax Yu}.~D.}\
  \bibnamefont {Prokoshkin}}, \ and\ \bibinfo {author} {\bibfnamefont {D.~A.}\
  \bibnamefont {Stoyanova}},\ }\href {\doibase 10.1016/0550-3213(73)90351-9}
  {\bibfield  {journal} {\bibinfo  {journal} {Nucl. Phys.}\ }\textbf {\bibinfo
  {volume} {B61}},\ \bibinfo {pages} {62} (\bibinfo {year} {1973})}\BibitemShut
  {NoStop}%
\bibitem [{\citenamefont {Carroll}\ \emph {et~al.}(1979)\citenamefont {Carroll}
  \emph {et~al.}}]{Carroll:1978hc}%
  \BibitemOpen
  \bibfield  {author} {\bibinfo {author} {\bibfnamefont {A.~S.}\ \bibnamefont
  {Carroll}} \emph {et~al.},\ }\href {\doibase 10.1016/0370-2693(79)90226-0}
  {\bibfield  {journal} {\bibinfo  {journal} {Phys. Lett.}\ }\textbf {\bibinfo
  {volume} {80B}},\ \bibinfo {pages} {319} (\bibinfo {year}
  {1979})}\BibitemShut {NoStop}%
\bibitem [{\citenamefont {Nakamura}\ \emph {et~al.}(1984)\citenamefont
  {Nakamura}, \citenamefont {Chiba}, \citenamefont {Fujii}, \citenamefont
  {Iwasaki}, \citenamefont {Kageyama}, \citenamefont {Kuribayashi},
  \citenamefont {Sumiyoshi}, \citenamefont {Takeda}, \citenamefont {Ikeda},\
  and\ \citenamefont {Takada}}]{Nakamura:1984xw}%
  \BibitemOpen
  \bibfield  {author} {\bibinfo {author} {\bibfnamefont {K.}~\bibnamefont
  {Nakamura}}, \bibinfo {author} {\bibfnamefont {J.}~\bibnamefont {Chiba}},
  \bibinfo {author} {\bibfnamefont {T.}~\bibnamefont {Fujii}}, \bibinfo
  {author} {\bibfnamefont {H.}~\bibnamefont {Iwasaki}}, \bibinfo {author}
  {\bibfnamefont {T.}~\bibnamefont {Kageyama}}, \bibinfo {author}
  {\bibfnamefont {S.}~\bibnamefont {Kuribayashi}}, \bibinfo {author}
  {\bibfnamefont {T.}~\bibnamefont {Sumiyoshi}}, \bibinfo {author}
  {\bibfnamefont {T.}~\bibnamefont {Takeda}}, \bibinfo {author} {\bibfnamefont
  {H.}~\bibnamefont {Ikeda}}, \ and\ \bibinfo {author} {\bibfnamefont
  {Y.}~\bibnamefont {Takada}},\ }\href {\doibase 10.1103/PhysRevLett.52.731}
  {\bibfield  {journal} {\bibinfo  {journal} {Phys. Rev. Lett.}\ }\textbf
  {\bibinfo {volume} {52}},\ \bibinfo {pages} {731} (\bibinfo {year}
  {1984})}\BibitemShut {NoStop}%
\bibitem [{\citenamefont {Chen}\ \emph {et~al.}(1955)\citenamefont {Chen},
  \citenamefont {Leavitt},\ and\ \citenamefont {Shapiro}}]{Chen:1955nkq}%
  \BibitemOpen
  \bibfield  {author} {\bibinfo {author} {\bibfnamefont {F.~F.}\ \bibnamefont
  {Chen}}, \bibinfo {author} {\bibfnamefont {C.~P.}\ \bibnamefont {Leavitt}}, \
  and\ \bibinfo {author} {\bibfnamefont {A.~M.}\ \bibnamefont {Shapiro}},\
  }\href {\doibase 10.1103/PhysRev.99.857} {\bibfield  {journal} {\bibinfo
  {journal} {Phys. Rev.}\ }\textbf {\bibinfo {volume} {99}},\ \bibinfo {pages}
  {857} (\bibinfo {year} {1955})}\BibitemShut {NoStop}%
\bibitem [{\citenamefont {Moskalev}\ and\ \citenamefont
  {Gavrilovskii}(1956)}]{Moskalev:1956}%
  \BibitemOpen
  \bibfield  {author} {\bibinfo {author} {\bibfnamefont {V.~N.}\ \bibnamefont
  {Moskalev}}\ and\ \bibinfo {author} {\bibfnamefont {B.~V.}\ \bibnamefont
  {Gavrilovskii}},\ }\href@noop {} {\bibfield  {journal} {\bibinfo  {journal}
  {Dokl. Akad. Nauk SSSR}\ }\textbf {\bibinfo {volume} {110}},\ \bibinfo
  {pages} {972} (\bibinfo {year} {1956})}\BibitemShut {NoStop}%
\bibitem [{\citenamefont {Booth}\ \emph {et~al.}(1957)\citenamefont {Booth},
  \citenamefont {Ledley}, \citenamefont {Walker},\ and\ \citenamefont
  {White}}]{Booth:1957}%
  \BibitemOpen
  \bibfield  {author} {\bibinfo {author} {\bibfnamefont {N.~E.}\ \bibnamefont
  {Booth}}, \bibinfo {author} {\bibfnamefont {B.}~\bibnamefont {Ledley}},
  \bibinfo {author} {\bibfnamefont {D.}~\bibnamefont {Walker}}, \ and\ \bibinfo
  {author} {\bibfnamefont {D.~H.}\ \bibnamefont {White}},\ }\href@noop {}
  {\bibfield  {journal} {\bibinfo  {journal} {Proc. Phys. Soc.}\ }\textbf
  {\bibinfo {volume} {A70}},\ \bibinfo {pages} {209} (\bibinfo {year}
  {1957})}\BibitemShut {NoStop}%
\bibitem [{\citenamefont {Bowen}\ \emph {et~al.}(1958)\citenamefont {Bowen},
  \citenamefont {Di~Corato}, \citenamefont {H.},\ and\ \citenamefont
  {Tagliaferri}}]{Bowen:1958}%
  \BibitemOpen
  \bibfield  {author} {\bibinfo {author} {\bibfnamefont {T.}~\bibnamefont
  {Bowen}}, \bibinfo {author} {\bibfnamefont {M.}~\bibnamefont {Di~Corato}},
  \bibinfo {author} {\bibfnamefont {M.~W.}\ \bibnamefont {H.}}, \ and\ \bibinfo
  {author} {\bibfnamefont {G.}~\bibnamefont {Tagliaferri}},\ }\href@noop {}
  {\bibfield  {journal} {\bibinfo  {journal} {Nuovo Cim.}\ }\textbf {\bibinfo
  {volume} {9}},\ \bibinfo {pages} {908} (\bibinfo {year} {1958})}\BibitemShut
  {NoStop}%
\bibitem [{\citenamefont {Batty}\ \emph {et~al.}(1958)\citenamefont {Batty},
  \citenamefont {Lock},\ and\ \citenamefont {March}}]{Batty:1958}%
  \BibitemOpen
  \bibfield  {author} {\bibinfo {author} {\bibfnamefont {C.~J.}\ \bibnamefont
  {Batty}}, \bibinfo {author} {\bibfnamefont {W.~O.}\ \bibnamefont {Lock}}, \
  and\ \bibinfo {author} {\bibfnamefont {P.~V.}\ \bibnamefont {March}},\
  }\href@noop {} {\bibfield  {journal} {\bibinfo  {journal} {Proc. Phys. Soc.}\
  }\textbf {\bibinfo {volume} {73}},\ \bibinfo {pages} {100} (\bibinfo {year}
  {1958})}\BibitemShut {NoStop}%
\bibitem [{\citenamefont {Law}\ \emph {et~al.}(1958)\citenamefont {Law},
  \citenamefont {Hutchinson},\ and\ \citenamefont {White}}]{Low:1958}%
  \BibitemOpen
  \bibfield  {author} {\bibinfo {author} {\bibfnamefont {M.~E.}\ \bibnamefont
  {Law}}, \bibinfo {author} {\bibfnamefont {G.}~\bibnamefont {Hutchinson}}, \
  and\ \bibinfo {author} {\bibfnamefont {D.}~\bibnamefont {White}},\ }\href
  {\doibase 10.1016/0029-5582(58)90344-4} {\bibfield  {journal} {\bibinfo
  {journal} {Nucl. Phys.}\ }\textbf {\bibinfo {volume} {9}},\ \bibinfo {pages}
  {600} (\bibinfo {year} {1958})}\BibitemShut {NoStop}%
\bibitem [{\citenamefont {Longo}\ and\ \citenamefont
  {Moyer}(1962)}]{Longo:1962zz}%
  \BibitemOpen
  \bibfield  {author} {\bibinfo {author} {\bibfnamefont {M.~J.}\ \bibnamefont
  {Longo}}\ and\ \bibinfo {author} {\bibfnamefont {B.~J.}\ \bibnamefont
  {Moyer}},\ }\href {\doibase 10.1103/PhysRev.125.701} {\bibfield  {journal}
  {\bibinfo  {journal} {Phys. Rev.}\ }\textbf {\bibinfo {volume} {125}},\
  \bibinfo {pages} {701} (\bibinfo {year} {1962})}\BibitemShut {NoStop}%
\bibitem [{\citenamefont {Grigorov}\ \emph {et~al.}(1964)\citenamefont
  {Grigorov}, \citenamefont {Erofeeva}, \citenamefont {Mishchenko},
  \citenamefont {Murzin}, \citenamefont {Rapoport}, \citenamefont {Sarycheva},\
  and\ \citenamefont {A.}}]{Grigorov:1964}%
  \BibitemOpen
  \bibfield  {author} {\bibinfo {author} {\bibfnamefont {N.~L.}\ \bibnamefont
  {Grigorov}}, \bibinfo {author} {\bibfnamefont {I.~N.}\ \bibnamefont
  {Erofeeva}}, \bibinfo {author} {\bibfnamefont {L.~G.}\ \bibnamefont
  {Mishchenko}}, \bibinfo {author} {\bibfnamefont {V.~S.}\ \bibnamefont
  {Murzin}}, \bibinfo {author} {\bibfnamefont {I.~D.}\ \bibnamefont
  {Rapoport}}, \bibinfo {author} {\bibfnamefont {L.~I.}\ \bibnamefont
  {Sarycheva}}, \ and\ \bibinfo {author} {\bibfnamefont {S.~V.}\ \bibnamefont
  {A.}},\ }\href@noop {} {\bibfield  {journal} {\bibinfo  {journal} {Izv. Akad.
  Nauk SSSR Ser. Fiz.}\ }\textbf {\bibinfo {volume} {28}},\ \bibinfo {pages}
  {1798} (\bibinfo {year} {1964})}\BibitemShut {NoStop}%
\bibitem [{\citenamefont {Basilova}\ \emph {et~al.}(1966)\citenamefont
  {Basilova}, \citenamefont {Grigorov}, \citenamefont {Kakhidze}, \citenamefont
  {Kovrizhnykh}, \citenamefont {Savenko}, \citenamefont {Nesterov},
  \citenamefont {Rapoport}, \citenamefont {Skuridin},\ and\ \citenamefont
  {Titenkov}}]{Basilova:1966}%
  \BibitemOpen
  \bibfield  {author} {\bibinfo {author} {\bibfnamefont {R.~N.}\ \bibnamefont
  {Basilova}}, \bibinfo {author} {\bibfnamefont {N.~L.}\ \bibnamefont
  {Grigorov}}, \bibinfo {author} {\bibfnamefont {G.~P.}\ \bibnamefont
  {Kakhidze}}, \bibinfo {author} {\bibfnamefont {O.~M.}\ \bibnamefont
  {Kovrizhnykh}}, \bibinfo {author} {\bibfnamefont {I.~A.}\ \bibnamefont
  {Savenko}}, \bibinfo {author} {\bibfnamefont {V.~E.}\ \bibnamefont
  {Nesterov}}, \bibinfo {author} {\bibfnamefont {I.~D.}\ \bibnamefont
  {Rapoport}}, \bibinfo {author} {\bibfnamefont {G.~A.}\ \bibnamefont
  {Skuridin}}, \ and\ \bibinfo {author} {\bibfnamefont {A.~F.}\ \bibnamefont
  {Titenkov}},\ }\href@noop {} {\bibfield  {journal} {\bibinfo  {journal} {Izv.
  Akad. Nauk SSSR Ser. Fiz.}\ }\textbf {\bibinfo {volume} {30}},\ \bibinfo
  {pages} {1610} (\bibinfo {year} {1966})}\BibitemShut {NoStop}%
\bibitem [{\citenamefont {Bellettini}\ \emph {et~al.}(1966)\citenamefont
  {Bellettini}, \citenamefont {Cocconi}, \citenamefont {Diddens}, \citenamefont
  {Lillethun}, \citenamefont {Matthiae}, \citenamefont {Scanlon},\ and\
  \citenamefont {Wetherell}}]{Bellettini:1966zz}%
  \BibitemOpen
  \bibfield  {author} {\bibinfo {author} {\bibfnamefont {G.}~\bibnamefont
  {Bellettini}}, \bibinfo {author} {\bibfnamefont {G.}~\bibnamefont {Cocconi}},
  \bibinfo {author} {\bibfnamefont {A.~N.}\ \bibnamefont {Diddens}}, \bibinfo
  {author} {\bibfnamefont {E.}~\bibnamefont {Lillethun}}, \bibinfo {author}
  {\bibfnamefont {G.}~\bibnamefont {Matthiae}}, \bibinfo {author}
  {\bibfnamefont {J.~P.}\ \bibnamefont {Scanlon}}, \ and\ \bibinfo {author}
  {\bibfnamefont {A.~M.}\ \bibnamefont {Wetherell}},\ }\href {\doibase
  10.1016/0029-5582(66)90267-7} {\bibfield  {journal} {\bibinfo  {journal}
  {Nucl. Phys.}\ }\textbf {\bibinfo {volume} {79}},\ \bibinfo {pages} {609}
  (\bibinfo {year} {1966})}\BibitemShut {NoStop}%
\bibitem [{\citenamefont {Grigorov}\ \emph {et~al.}(1970)\citenamefont
  {Grigorov}, \citenamefont {Nesterov}, \citenamefont {Rapoport}, \citenamefont
  {Savenko},\ and\ \citenamefont {Skuridin}}]{Grigorov:1970yt}%
  \BibitemOpen
  \bibfield  {author} {\bibinfo {author} {\bibfnamefont {N.~L.}\ \bibnamefont
  {Grigorov}}, \bibinfo {author} {\bibfnamefont {V.~E.}\ \bibnamefont
  {Nesterov}}, \bibinfo {author} {\bibfnamefont {I.~D.}\ \bibnamefont
  {Rapoport}}, \bibinfo {author} {\bibfnamefont {I.~A.}\ \bibnamefont
  {Savenko}}, \ and\ \bibinfo {author} {\bibfnamefont {G.~A.}\ \bibnamefont
  {Skuridin}},\ }\href@noop {} {\bibfield  {journal} {\bibinfo  {journal} {Yad.
  Fiz.}\ }\textbf {\bibinfo {volume} {11}},\ \bibinfo {pages} {814} (\bibinfo
  {year} {1970})}\BibitemShut {NoStop}%
\bibitem [{\citenamefont {Alakoz}\ \emph {et~al.}(1971)\citenamefont {Alakoz},
  \citenamefont {Vasilev}, \citenamefont {Vasoleva},\ and\ \citenamefont
  {Shmeleva}}]{Alakoz:1971}%
  \BibitemOpen
  \bibfield  {author} {\bibinfo {author} {\bibfnamefont {A.~V.}\ \bibnamefont
  {Alakoz}}, \bibinfo {author} {\bibfnamefont {P.~S.}\ \bibnamefont {Vasilev}},
  \bibinfo {author} {\bibfnamefont {L.~F.}\ \bibnamefont {Vasoleva}}, \ and\
  \bibinfo {author} {\bibfnamefont {A.~P.}\ \bibnamefont {Shmeleva}},\
  }\href@noop {} {\bibfield  {journal} {\bibinfo  {journal} {Izv. Akad. Nauk
  SSSR Ser. Fiz}\ }\textbf {\bibinfo {volume} {35}},\ \bibinfo {pages} {2069}
  (\bibinfo {year} {1971})}\BibitemShut {NoStop}%
\bibitem [{\citenamefont {Renberg}\ \emph {et~al.}(1972)\citenamefont
  {Renberg}, \citenamefont {Measday}, \citenamefont {Pepin}, \citenamefont
  {Schwaller}, \citenamefont {Favier},\ and\ \citenamefont
  {Richard-Serre}}]{Renberg:1972jf}%
  \BibitemOpen
  \bibfield  {author} {\bibinfo {author} {\bibfnamefont {P.~U.}\ \bibnamefont
  {Renberg}}, \bibinfo {author} {\bibfnamefont {D.~F.}\ \bibnamefont
  {Measday}}, \bibinfo {author} {\bibfnamefont {M.}~\bibnamefont {Pepin}},
  \bibinfo {author} {\bibfnamefont {P.}~\bibnamefont {Schwaller}}, \bibinfo
  {author} {\bibfnamefont {B.}~\bibnamefont {Favier}}, \ and\ \bibinfo {author}
  {\bibfnamefont {C.}~\bibnamefont {Richard-Serre}},\ }\href {\doibase
  10.1016/0375-9474(72)90932-3} {\bibfield  {journal} {\bibinfo  {journal}
  {Nucl. Phys.}\ }\textbf {\bibinfo {volume} {A183}},\ \bibinfo {pages} {81}
  (\bibinfo {year} {1972})}\BibitemShut {NoStop}%
\bibitem [{\citenamefont {McGill}\ \emph {et~al.}(1974)\citenamefont {McGill},
  \citenamefont {Carlson}, \citenamefont {Short}, \citenamefont {Cameron},
  \citenamefont {Richardson}, \citenamefont {Slaus}, \citenamefont {Van~Oers},
  \citenamefont {Verba}, \citenamefont {Margaziotis},\ and\ \citenamefont
  {Doherty}}]{McGill:1974zz}%
  \BibitemOpen
  \bibfield  {author} {\bibinfo {author} {\bibfnamefont {W.~F.}\ \bibnamefont
  {McGill}}, \bibinfo {author} {\bibfnamefont {R.~F.}\ \bibnamefont {Carlson}},
  \bibinfo {author} {\bibfnamefont {T.~H.}\ \bibnamefont {Short}}, \bibinfo
  {author} {\bibfnamefont {J.~M.}\ \bibnamefont {Cameron}}, \bibinfo {author}
  {\bibfnamefont {J.~R.}\ \bibnamefont {Richardson}}, \bibinfo {author}
  {\bibfnamefont {I.}~\bibnamefont {Slaus}}, \bibinfo {author} {\bibfnamefont
  {W.~T.~H.}\ \bibnamefont {Van~Oers}}, \bibinfo {author} {\bibfnamefont
  {J.~W.}\ \bibnamefont {Verba}}, \bibinfo {author} {\bibfnamefont {D.~J.}\
  \bibnamefont {Margaziotis}}, \ and\ \bibinfo {author} {\bibfnamefont
  {P.}~\bibnamefont {Doherty}},\ }\href {\doibase 10.1103/PhysRevC.10.2237}
  {\bibfield  {journal} {\bibinfo  {journal} {Phys. Rev.}\ }\textbf {\bibinfo
  {volume} {C10}},\ \bibinfo {pages} {2237} (\bibinfo {year}
  {1974})}\BibitemShut {NoStop}%
\bibitem [{\citenamefont {Lindstrom}\ \emph {et~al.}(1975)\citenamefont
  {Lindstrom}, \citenamefont {Greiner}, \citenamefont {Heckman}, \citenamefont
  {Cork},\ and\ \citenamefont {Bieser}}]{Lindstrom:1975xr}%
  \BibitemOpen
  \bibfield  {author} {\bibinfo {author} {\bibfnamefont {P.~J.}\ \bibnamefont
  {Lindstrom}}, \bibinfo {author} {\bibfnamefont {D.~E.}\ \bibnamefont
  {Greiner}}, \bibinfo {author} {\bibfnamefont {H.~H.}\ \bibnamefont
  {Heckman}}, \bibinfo {author} {\bibfnamefont {B.}~\bibnamefont {Cork}}, \
  and\ \bibinfo {author} {\bibfnamefont {F.~S.}\ \bibnamefont {Bieser}},\ }in\
  \href@noop {} {\emph {\bibinfo {booktitle} {{14th International Cosmic Ray
  Conference (ICRC 1975) Munich, Germany, August 15-29, 1975}}}}\ (\bibinfo
  {year} {1975})\ pp.\ \bibinfo {pages} {2315--2318}\BibitemShut {NoStop}%
\bibitem [{\citenamefont {Baros}\ \emph {et~al.}(1978)\citenamefont {Baros},
  \citenamefont {Wagner}, \citenamefont {Anderson} \emph
  {et~al.}}]{Baros:1978}%
  \BibitemOpen
  \bibfield  {author} {\bibinfo {author} {\bibfnamefont {J.}~\bibnamefont
  {Baros}}, \bibinfo {author} {\bibfnamefont {A.}~\bibnamefont {Wagner}},
  \bibinfo {author} {\bibfnamefont {L.}~\bibnamefont {Anderson}},  \emph
  {et~al.},\ }\href@noop {} {\bibfield  {journal} {\bibinfo  {journal} {Phys.
  Rev.}\ }\textbf {\bibinfo {volume} {C18}},\ \bibinfo {pages} {2273} (\bibinfo
  {year} {1978})}\BibitemShut {NoStop}%
\bibitem [{\citenamefont {Heckman}\ \emph {et~al.}(1978)\citenamefont
  {Heckman}, \citenamefont {Greiner}, \citenamefont {Lindstrom},\ and\
  \citenamefont {Shwe}}]{Heckman:1978ib}%
  \BibitemOpen
  \bibfield  {author} {\bibinfo {author} {\bibfnamefont {H.~H.}\ \bibnamefont
  {Heckman}}, \bibinfo {author} {\bibfnamefont {D.~E.}\ \bibnamefont
  {Greiner}}, \bibinfo {author} {\bibfnamefont {P.~J.}\ \bibnamefont
  {Lindstrom}}, \ and\ \bibinfo {author} {\bibfnamefont {H.}~\bibnamefont
  {Shwe}},\ }\href {\doibase 10.1103/PhysRevC.17.1735} {\bibfield  {journal}
  {\bibinfo  {journal} {Phys. Rev.}\ }\textbf {\bibinfo {volume} {C17}},\
  \bibinfo {pages} {1735} (\bibinfo {year} {1978})}\BibitemShut {NoStop}%
\bibitem [{\citenamefont {Akhababyan}\ \emph {et~al.}(1979)\citenamefont
  {Akhababyan}, \citenamefont {Baatar}, \citenamefont {Gasparyan},
  \citenamefont {Grigalashvili}, \citenamefont {Gulkanyan}, \citenamefont
  {Ivanovskaya},\ and\ \citenamefont {Cheplakov}}]{Akhababyan:1979}%
  \BibitemOpen
  \bibfield  {author} {\bibinfo {author} {\bibfnamefont {N.}~\bibnamefont
  {Akhababyan}}, \bibinfo {author} {\bibfnamefont {T.}~\bibnamefont {Baatar}},
  \bibinfo {author} {\bibfnamefont {A.~P.}\ \bibnamefont {Gasparyan}}, \bibinfo
  {author} {\bibfnamefont {N.~S.}\ \bibnamefont {Grigalashvili}}, \bibinfo
  {author} {\bibfnamefont {T.~R.}\ \bibnamefont {Gulkanyan}}, \bibinfo {author}
  {\bibfnamefont {I.~A.}\ \bibnamefont {Ivanovskaya}}, \ and\ \bibinfo {author}
  {\bibfnamefont {A.~A.}\ \bibnamefont {Cheplakov}},\ }\href@noop {} {\bibfield
   {journal} {\bibinfo  {journal} {JINR Communication}\ }\textbf {\bibinfo
  {volume} {1-12114}} (\bibinfo {year} {1979})}\BibitemShut {NoStop}%
\bibitem [{\citenamefont {Bobchenko}\ \emph {et~al.}(1979)\citenamefont
  {Bobchenko} \emph {et~al.}}]{Bobchenko:1979hp}%
  \BibitemOpen
  \bibfield  {author} {\bibinfo {author} {\bibfnamefont {B.~M.}\ \bibnamefont
  {Bobchenko}} \emph {et~al.},\ }\href@noop {} {\bibfield  {journal} {\bibinfo
  {journal} {Sov. J. Nucl. Phys.}\ }\textbf {\bibinfo {volume} {30}},\ \bibinfo
  {pages} {805} (\bibinfo {year} {1979})},\ \bibinfo {note} {[Yad.
  Fiz.30,1553(1979)]}\BibitemShut {NoStop}%
\bibitem [{\citenamefont {Afanasev}\ \emph {et~al.}(1984)\citenamefont
  {Afanasev}, \citenamefont {Borisov}, \citenamefont {Borodina} \emph
  {et~al.}}]{Afanasev:1984}%
  \BibitemOpen
  \bibfield  {author} {\bibinfo {author} {\bibfnamefont {V.~N.}\ \bibnamefont
  {Afanasev}}, \bibinfo {author} {\bibfnamefont {V.~S.}\ \bibnamefont
  {Borisov}}, \bibinfo {author} {\bibfnamefont {I.~N.}\ \bibnamefont
  {Borodina}},  \emph {et~al.},\ }\href@noop {} {\bibfield  {journal} {\bibinfo
   {journal} {Yad. Fiz.}\ }\textbf {\bibinfo {volume} {40}},\ \bibinfo {pages}
  {34} (\bibinfo {year} {1984})}\BibitemShut {NoStop}%
\bibitem [{\citenamefont {Grchurin}\ \emph {et~al.}(1985)\citenamefont
  {Grchurin}, \citenamefont {Druzhinin}, \citenamefont {A.} \emph
  {et~al.}}]{Grchurin:1985}%
  \BibitemOpen
  \bibfield  {author} {\bibinfo {author} {\bibfnamefont {V.~V.}\ \bibnamefont
  {Grchurin}}, \bibinfo {author} {\bibfnamefont {B.~L.}\ \bibnamefont
  {Druzhinin}}, \bibinfo {author} {\bibfnamefont {B.~A.}\ \bibnamefont {Ezhov}},
  \emph {et~al.},\ }\href@noop {} {\bibfield  {journal} {\bibinfo  {journal}
  {Preprint ITEP}\ }\textbf {\bibinfo {volume} {No. 59}} (\bibinfo {year}
  {1985})}\BibitemShut {NoStop}%
\bibitem [{\citenamefont {Abgrall}\ \emph {et~al.}(2011)\citenamefont {Abgrall}
  \emph {et~al.}}]{Abgrall:2011ae}%
  \BibitemOpen
  \bibfield  {author} {\bibinfo {author} {\bibfnamefont {N.}~\bibnamefont
  {Abgrall}} \emph {et~al.} (\bibinfo {collaboration} {NA61/SHINE}),\ }\href
  {\doibase 10.1103/PhysRevC.84.034604} {\bibfield  {journal} {\bibinfo
  {journal} {Phys. Rev.}\ }\textbf {\bibinfo {volume} {C84}},\ \bibinfo {pages}
  {034604} (\bibinfo {year} {2011})},\ \Eprint {http://arxiv.org/abs/1102.0983}
  {arXiv:1102.0983 [hep-ex]} \BibitemShut {NoStop}%
\bibitem [{\citenamefont {Feroz}\ \emph {et~al.}(2009)\citenamefont {Feroz},
  \citenamefont {Hobson},\ and\ \citenamefont {Bridges}}]{Feroz:2008xx}%
  \BibitemOpen
  \bibfield  {author} {\bibinfo {author} {\bibfnamefont {F.}~\bibnamefont
  {Feroz}}, \bibinfo {author} {\bibfnamefont {M.~P.}\ \bibnamefont {Hobson}}, \
  and\ \bibinfo {author} {\bibfnamefont {M.}~\bibnamefont {Bridges}},\ }\href
  {\doibase 10.1111/j.1365-2966.2009.14548.x} {\bibfield  {journal} {\bibinfo
  {journal} {Mon. Not. Roy. Astron. Soc.}\ }\textbf {\bibinfo {volume} {398}},\
  \bibinfo {pages} {1601} (\bibinfo {year} {2009})},\ \Eprint
  {http://arxiv.org/abs/0809.3437} {arXiv:0809.3437 [astro-ph]} \BibitemShut
  {NoStop}%
\bibitem [{\citenamefont {Zuccon}(2019)}]{Zuccon:2019}%
  \BibitemOpen
  \bibfield  {author} {\bibinfo {author} {\bibfnamefont {P.}~\bibnamefont
  {Zuccon}},\ }\href@noop {} {\bibfield  {journal} {\bibinfo  {journal} {Talk
  at {`Antideuteron 2019'}}} (\bibinfo {year} {2019})}\BibitemShut {NoStop}%
\bibitem [{\citenamefont {Alvioli}\ \emph {et~al.}(2010)\citenamefont
  {Alvioli}, \citenamefont {Ciofi~degli Atti}, \citenamefont {Kopeliovich},
  \citenamefont {Potashnikova},\ and\ \citenamefont
  {Schmidt}}]{Alvioli:2009iw}%
  \BibitemOpen
  \bibfield  {author} {\bibinfo {author} {\bibfnamefont {M.}~\bibnamefont
  {Alvioli}}, \bibinfo {author} {\bibfnamefont {C.}~\bibnamefont {Ciofi~degli
  Atti}}, \bibinfo {author} {\bibfnamefont {B.~Z.}\ \bibnamefont
  {Kopeliovich}}, \bibinfo {author} {\bibfnamefont {I.~K.}\ \bibnamefont
  {Potashnikova}}, \ and\ \bibinfo {author} {\bibfnamefont {I.}~\bibnamefont
  {Schmidt}},\ }\href {\doibase 10.1103/PhysRevC.81.025204} {\bibfield
  {journal} {\bibinfo  {journal} {Phys. Rev.}\ }\textbf {\bibinfo {volume}
  {C81}},\ \bibinfo {pages} {025204} (\bibinfo {year} {2010})},\ \Eprint
  {http://arxiv.org/abs/0911.1382} {arXiv:0911.1382 [nucl-th]} \BibitemShut
  {NoStop}%
\bibitem [{\citenamefont {Ganguli}\ \emph {et~al.}(1974)\citenamefont
  {Ganguli}, \citenamefont {Raghavan},\ and\ \citenamefont
  {Subramanian}}]{Ganguli:1973rja}%
  \BibitemOpen
  \bibfield  {author} {\bibinfo {author} {\bibfnamefont {S.~N.}\ \bibnamefont
  {Ganguli}}, \bibinfo {author} {\bibfnamefont {R.}~\bibnamefont {Raghavan}}, \
  and\ \bibinfo {author} {\bibfnamefont {A.}~\bibnamefont {Subramanian}},\
  }\href {\doibase 10.1007/BF02847248} {\bibfield  {journal} {\bibinfo
  {journal} {Pramana}\ }\textbf {\bibinfo {volume} {2}},\ \bibinfo {pages}
  {341} (\bibinfo {year} {1974})}\BibitemShut {NoStop}%
\bibitem [{\citenamefont {Chen}(2017)}]{Chen:2017}%
  \BibitemOpen
  \bibfield  {author} {\bibinfo {author} {\bibfnamefont {A.~I.}\ \bibnamefont
  {Chen}},\ }\href@noop {} {\bibfield  {journal} {\bibinfo  {journal} {Ph.D.
  thesis, Massachusetts Institute of Technology} } (\bibinfo {year}
  {2017})}\BibitemShut {NoStop}%
\bibitem [{\citenamefont {Konak}(2019)}]{Konak:2019}%
  \BibitemOpen
  \bibfield  {author} {\bibinfo {author} {\bibfnamefont {K.}~\bibnamefont
  {Konak}},\ }\href@noop {} {\bibfield  {journal} {\bibinfo  {journal} {Ph.D.
  thesis, Middle East Technical University} } (\bibinfo {year}
  {2019})}\BibitemShut {NoStop}%
\bibitem [{\citenamefont {Aguilar}\ \emph
  {et~al.}(2015{\natexlab{a}})\citenamefont {Aguilar} \emph
  {et~al.}}]{Aguilar:2015ooa}%
  \BibitemOpen
  \bibfield  {author} {\bibinfo {author} {\bibfnamefont {M.}~\bibnamefont
  {Aguilar}} \emph {et~al.} (\bibinfo {collaboration} {AMS collaboration}),\
  }\href {\doibase 10.1103/PhysRevLett.114.171103} {\bibfield  {journal}
  {\bibinfo  {journal} {Phys. Rev. Lett.}\ }\textbf {\bibinfo {volume} {114}},\
  \bibinfo {pages} {171103} (\bibinfo {year} {2015}{\natexlab{a}})}\BibitemShut
  {NoStop}%
\bibitem [{\citenamefont {Alaoui}(2016)}]{Alaoui:2016}%
  \BibitemOpen
  \bibfield  {author} {\bibinfo {author} {\bibfnamefont {M.~H.}\ \bibnamefont
  {Alaoui}},\ }\href@noop {} {\bibfield  {journal} {\bibinfo  {journal} {Ph.D.
  thesis, University of Geneva}\ } (\bibinfo {year} {2016})}\BibitemShut
  {NoStop}%
\bibitem [{\citenamefont {Genolini}\ \emph {et~al.}(2018)\citenamefont
  {Genolini}, \citenamefont {Maurin}, \citenamefont {Moskalenko},\ and\
  \citenamefont {Unger}}]{Genolini:2018ekk}%
  \BibitemOpen
  \bibfield  {author} {\bibinfo {author} {\bibfnamefont {Y.}~\bibnamefont
  {Genolini}}, \bibinfo {author} {\bibfnamefont {D.}~\bibnamefont {Maurin}},
  \bibinfo {author} {\bibfnamefont {I.~V.}\ \bibnamefont {Moskalenko}}, \ and\
  \bibinfo {author} {\bibfnamefont {M.}~\bibnamefont {Unger}},\ }\href
  {\doibase 10.1103/PhysRevC.98.034611} {\bibfield  {journal} {\bibinfo
  {journal} {Phys. Rev. C}\ }\textbf {\bibinfo {volume} {98}},\ \bibinfo
  {pages} {034611} (\bibinfo {year} {2018})},\ \Eprint
  {http://arxiv.org/abs/1803.04686} {arXiv:1803.04686 [astro-ph.HE]}
  \BibitemShut {NoStop}%
\bibitem [{\citenamefont {Evoli}\ \emph {et~al.}(2019)\citenamefont {Evoli},
  \citenamefont {Aloisio},\ and\ \citenamefont {Blasi}}]{Evoli:2019wwu}%
  \BibitemOpen
  \bibfield  {author} {\bibinfo {author} {\bibfnamefont {C.}~\bibnamefont
  {Evoli}}, \bibinfo {author} {\bibfnamefont {R.}~\bibnamefont {Aloisio}}, \
  and\ \bibinfo {author} {\bibfnamefont {P.}~\bibnamefont {Blasi}},\ }\href
  {\doibase 10.1103/PhysRevD.99.103023} {\bibfield  {journal} {\bibinfo
  {journal} {Phys. Rev. D}\ }\textbf {\bibinfo {volume} {99}},\ \bibinfo
  {pages} {103023} (\bibinfo {year} {2019})},\ \Eprint
  {http://arxiv.org/abs/1904.10220} {arXiv:1904.10220 [astro-ph.HE]}
  \BibitemShut {NoStop}%
\bibitem [{\citenamefont {Cirelli}\ \emph {et~al.}(2011)\citenamefont
  {Cirelli}, \citenamefont {Corcella}, \citenamefont {Hektor}, \citenamefont
  {Hutsi}, \citenamefont {Kadastik}, \citenamefont {Panci}, \citenamefont
  {Raidal}, \citenamefont {Sala},\ and\ \citenamefont
  {Strumia}}]{Cirelli:2010xx}%
  \BibitemOpen
  \bibfield  {author} {\bibinfo {author} {\bibfnamefont {M.}~\bibnamefont
  {Cirelli}}, \bibinfo {author} {\bibfnamefont {G.}~\bibnamefont {Corcella}},
  \bibinfo {author} {\bibfnamefont {A.}~\bibnamefont {Hektor}}, \bibinfo
  {author} {\bibfnamefont {G.}~\bibnamefont {Hutsi}}, \bibinfo {author}
  {\bibfnamefont {M.}~\bibnamefont {Kadastik}}, \bibinfo {author}
  {\bibfnamefont {P.}~\bibnamefont {Panci}}, \bibinfo {author} {\bibfnamefont
  {M.}~\bibnamefont {Raidal}}, \bibinfo {author} {\bibfnamefont
  {F.}~\bibnamefont {Sala}}, \ and\ \bibinfo {author} {\bibfnamefont
  {A.}~\bibnamefont {Strumia}},\ }\href {\doibase
  10.1088/1475-7516/2012/10/E01, 10.1088/1475-7516/2011/03/051} {\bibfield
  {journal} {\bibinfo  {journal} {JCAP}\ }\textbf {\bibinfo {volume} {1103}},\
  \bibinfo {pages} {051} (\bibinfo {year} {2011})},\ \bibinfo {note} {[Erratum:
  JCAP1210,E01(2012)]},\ \Eprint {http://arxiv.org/abs/1012.4515}
  {arXiv:1012.4515 [hep-ph]} \BibitemShut {NoStop}%
\bibitem [{\citenamefont {Maurin}\ \emph {et~al.}(2001)\citenamefont {Maurin},
  \citenamefont {Donato}, \citenamefont {Taillet},\ and\ \citenamefont
  {Salati}}]{Maurin:2001sj}%
  \BibitemOpen
  \bibfield  {author} {\bibinfo {author} {\bibfnamefont {D.}~\bibnamefont
  {Maurin}}, \bibinfo {author} {\bibfnamefont {F.}~\bibnamefont {Donato}},
  \bibinfo {author} {\bibfnamefont {R.}~\bibnamefont {Taillet}}, \ and\
  \bibinfo {author} {\bibfnamefont {P.}~\bibnamefont {Salati}},\ }\href
  {\doibase 10.1086/321496} {\bibfield  {journal} {\bibinfo  {journal}
  {Astrophys. J.}\ }\textbf {\bibinfo {volume} {555}},\ \bibinfo {pages} {585}
  (\bibinfo {year} {2001})},\ \Eprint {http://arxiv.org/abs/astro-ph/0101231}
  {arXiv:astro-ph/0101231} \BibitemShut {NoStop}%
\bibitem [{\citenamefont {Donato}\ \emph {et~al.}(2001)\citenamefont {Donato},
  \citenamefont {Maurin}, \citenamefont {Salati}, \citenamefont {Barrau},
  \citenamefont {Boudoul},\ and\ \citenamefont {Taillet}}]{Donato:2001ms}%
  \BibitemOpen
  \bibfield  {author} {\bibinfo {author} {\bibfnamefont {F.}~\bibnamefont
  {Donato}}, \bibinfo {author} {\bibfnamefont {D.}~\bibnamefont {Maurin}},
  \bibinfo {author} {\bibfnamefont {P.}~\bibnamefont {Salati}}, \bibinfo
  {author} {\bibfnamefont {A.}~\bibnamefont {Barrau}}, \bibinfo {author}
  {\bibfnamefont {G.}~\bibnamefont {Boudoul}}, \ and\ \bibinfo {author}
  {\bibfnamefont {R.}~\bibnamefont {Taillet}},\ }\href {\doibase
  10.1086/323684} {\bibfield  {journal} {\bibinfo  {journal} {Astrophys. J.}\
  }\textbf {\bibinfo {volume} {563}},\ \bibinfo {pages} {172} (\bibinfo {year}
  {2001})},\ \Eprint {http://arxiv.org/abs/astro-ph/0103150}
  {arXiv:astro-ph/0103150} \BibitemShut {NoStop}%
\bibitem [{\citenamefont {Maurin}\ \emph {et~al.}(2002)\citenamefont {Maurin},
  \citenamefont {Taillet}, \citenamefont {Donato}, \citenamefont {Salati},
  \citenamefont {Barrau},\ and\ \citenamefont {Boudoul}}]{Maurin:2002ua}%
  \BibitemOpen
  \bibfield  {author} {\bibinfo {author} {\bibfnamefont {D.}~\bibnamefont
  {Maurin}}, \bibinfo {author} {\bibfnamefont {R.}~\bibnamefont {Taillet}},
  \bibinfo {author} {\bibfnamefont {F.}~\bibnamefont {Donato}}, \bibinfo
  {author} {\bibfnamefont {P.}~\bibnamefont {Salati}}, \bibinfo {author}
  {\bibfnamefont {A.}~\bibnamefont {Barrau}}, \ and\ \bibinfo {author}
  {\bibfnamefont {G.}~\bibnamefont {Boudoul}},\ }\href@noop {} {\  (\bibinfo
  {year} {2002})},\ \Eprint {http://arxiv.org/abs/astro-ph/0212111}
  {arXiv:astro-ph/0212111} \BibitemShut {NoStop}%
\bibitem [{\citenamefont {Moskalenko}\ and\ \citenamefont
  {Strong}(1998)}]{Moskalenko:1997gh}%
  \BibitemOpen
  \bibfield  {author} {\bibinfo {author} {\bibfnamefont {I.~V.}\ \bibnamefont
  {Moskalenko}}\ and\ \bibinfo {author} {\bibfnamefont {A.~W.}\ \bibnamefont
  {Strong}},\ }\href {\doibase 10.1086/305152} {\bibfield  {journal} {\bibinfo
  {journal} {Astrophys. J.}\ }\textbf {\bibinfo {volume} {493}},\ \bibinfo
  {pages} {694} (\bibinfo {year} {1998})},\ \Eprint
  {http://arxiv.org/abs/astro-ph/9710124} {arXiv:astro-ph/9710124}
  \BibitemShut {NoStop}%
\bibitem [{\citenamefont {Strong}\ and\ \citenamefont
  {Moskalenko}(1998)}]{Strong:1998pw}%
  \BibitemOpen
  \bibfield  {author} {\bibinfo {author} {\bibfnamefont {A.~W.}\ \bibnamefont
  {Strong}}\ and\ \bibinfo {author} {\bibfnamefont {I.~V.}\ \bibnamefont
  {Moskalenko}},\ }\href {\doibase 10.1086/306470} {\bibfield  {journal}
  {\bibinfo  {journal} {Astrophys. J.}\ }\textbf {\bibinfo {volume} {509}},\
  \bibinfo {pages} {212} (\bibinfo {year} {1998})},\ \Eprint
  {http://arxiv.org/abs/astro-ph/9807150} {arXiv:astro-ph/9807150}
  \BibitemShut {NoStop}%
\bibitem [{\citenamefont {Strong}\ and\ \citenamefont
  {Moskalenko}(2001)}]{Strong:2001gh}%
  \BibitemOpen
  \bibfield  {author} {\bibinfo {author} {\bibfnamefont {A.~W.}\ \bibnamefont
  {Strong}}\ and\ \bibinfo {author} {\bibfnamefont {I.~V.}\ \bibnamefont
  {Moskalenko}},\ }in\ \href@noop {} {\emph {\bibinfo {booktitle} {{27th
  International Cosmic Ray Conference: Proceedings. p. 1942-1945}}}}\ (\bibinfo
  {year} {2001})\ pp.\ \bibinfo {pages} {1942--1945},\ \bibinfo {note}
  {[5,1942(2001)]},\ \Eprint {http://arxiv.org/abs/astro-ph/0106504}
  {arXiv:astro-ph/0106504} \BibitemShut {NoStop}%
\bibitem [{\citenamefont {Di~Bernardo}\ \emph {et~al.}(2010)\citenamefont
  {Di~Bernardo}, \citenamefont {Evoli}, \citenamefont {Gaggero}, \citenamefont
  {Grasso},\ and\ \citenamefont {Maccione}}]{DiBernardo:2009ku}%
  \BibitemOpen
  \bibfield  {author} {\bibinfo {author} {\bibfnamefont {G.}~\bibnamefont
  {Di~Bernardo}}, \bibinfo {author} {\bibfnamefont {C.}~\bibnamefont {Evoli}},
  \bibinfo {author} {\bibfnamefont {D.}~\bibnamefont {Gaggero}}, \bibinfo
  {author} {\bibfnamefont {D.}~\bibnamefont {Grasso}}, \ and\ \bibinfo {author}
  {\bibfnamefont {L.}~\bibnamefont {Maccione}},\ }\href {\doibase
  10.1016/j.astropartphys.2010.08.006} {\bibfield  {journal} {\bibinfo
  {journal} {Astropart. Phys.}\ }\textbf {\bibinfo {volume} {34}},\ \bibinfo
  {pages} {274} (\bibinfo {year} {2010})},\ \Eprint
  {http://arxiv.org/abs/0909.4548} {arXiv:0909.4548 [astro-ph.HE]} \BibitemShut
  {NoStop}%
\bibitem [{\citenamefont {Maurin}\ \emph {et~al.}(2010)\citenamefont {Maurin},
  \citenamefont {Putze},\ and\ \citenamefont {Derome}}]{Maurin:2010zp}%
  \BibitemOpen
  \bibfield  {author} {\bibinfo {author} {\bibfnamefont {D.}~\bibnamefont
  {Maurin}}, \bibinfo {author} {\bibfnamefont {A.}~\bibnamefont {Putze}}, \
  and\ \bibinfo {author} {\bibfnamefont {L.}~\bibnamefont {Derome}},\ }\href
  {\doibase 10.1051/0004-6361/201014011} {\bibfield  {journal} {\bibinfo
  {journal} {Astron. Astrophys.}\ }\textbf {\bibinfo {volume} {516}},\ \bibinfo
  {pages} {A67} (\bibinfo {year} {2010})},\ \Eprint
  {http://arxiv.org/abs/1001.0553} {arXiv:1001.0553 [astro-ph.HE]} \BibitemShut
  {NoStop}%
\bibitem [{\citenamefont {Génolini}\ \emph {et~al.}(2019)\citenamefont
  {Génolini} \emph {et~al.}}]{Genolini:2019ewc}%
  \BibitemOpen
  \bibfield  {author} {\bibinfo {author} {\bibfnamefont {Y.}~\bibnamefont
  {Génolini}} \emph {et~al.},\ }\href {\doibase 10.1103/PhysRevD.99.123028}
  {\bibfield  {journal} {\bibinfo  {journal} {Phys. Rev.}\ }\textbf {\bibinfo
  {volume} {D99}},\ \bibinfo {pages} {123028} (\bibinfo {year} {2019})},\
  \Eprint {http://arxiv.org/abs/1904.08917} {arXiv:1904.08917 [astro-ph.HE]}
  \BibitemShut {NoStop}%
\bibitem [{\citenamefont {Weinrich}\ \emph
  {et~al.}(2020{\natexlab{a}})\citenamefont {Weinrich}, \citenamefont
  {G{\'e}nolini}, \citenamefont {Boudaud}, \citenamefont {Derome},\ and\
  \citenamefont {Maurin}}]{Weinrich:2020cmw}%
  \BibitemOpen
  \bibfield  {author} {\bibinfo {author} {\bibfnamefont {N.}~\bibnamefont
  {Weinrich}}, \bibinfo {author} {\bibfnamefont {Y.}~\bibnamefont
  {G{\'e}nolini}}, \bibinfo {author} {\bibfnamefont {M.}~\bibnamefont
  {Boudaud}}, \bibinfo {author} {\bibfnamefont {L.}~\bibnamefont {Derome}}, \
  and\ \bibinfo {author} {\bibfnamefont {D.}~\bibnamefont {Maurin}},\ }\href
  {\doibase 10.1051/0004-6361/202037875} {\bibfield  {journal} {\bibinfo
  {journal} {Astron. Astrophys.}\ }\textbf {\bibinfo {volume} {639}},\ \bibinfo
  {pages} {A131} (\bibinfo {year} {2020}{\natexlab{a}})},\ \Eprint
  {http://arxiv.org/abs/2002.11406} {arXiv:2002.11406 [astro-ph.HE]}
  \BibitemShut {NoStop}%
\bibitem [{\citenamefont {Ptuskin}\ \emph {et~al.}(2006)\citenamefont
  {Ptuskin}, \citenamefont {Moskalenko}, \citenamefont {Jones}, \citenamefont
  {Strong},\ and\ \citenamefont {Zirakashvili}}]{Ptuskin:2005ax}%
  \BibitemOpen
  \bibfield  {author} {\bibinfo {author} {\bibfnamefont {V.~S.}\ \bibnamefont
  {Ptuskin}}, \bibinfo {author} {\bibfnamefont {I.~V.}\ \bibnamefont
  {Moskalenko}}, \bibinfo {author} {\bibfnamefont {F.~C.}\ \bibnamefont
  {Jones}}, \bibinfo {author} {\bibfnamefont {A.~W.}\ \bibnamefont {Strong}}, \
  and\ \bibinfo {author} {\bibfnamefont {V.~N.}\ \bibnamefont {Zirakashvili}},\
  }\href {\doibase 10.1086/501117} {\bibfield  {journal} {\bibinfo  {journal}
  {Astrophys. J.}\ }\textbf {\bibinfo {volume} {642}},\ \bibinfo {pages} {902}
  (\bibinfo {year} {2006})},\ \Eprint {http://arxiv.org/abs/astro-ph/0510335}
  {arXiv:astro-ph/0510335} \BibitemShut {NoStop}%
\bibitem [{\citenamefont {Génolini}\ \emph {et~al.}(2017)\citenamefont
  {Génolini} \emph {et~al.}}]{Genolini:2017dfb}%
  \BibitemOpen
  \bibfield  {author} {\bibinfo {author} {\bibfnamefont {Y.}~\bibnamefont
  {Génolini}} \emph {et~al.},\ }\href {\doibase
  10.1103/PhysRevLett.119.241101} {\bibfield  {journal} {\bibinfo  {journal}
  {Phys. Rev. Lett.}\ }\textbf {\bibinfo {volume} {119}},\ \bibinfo {pages}
  {241101} (\bibinfo {year} {2017})},\ \Eprint
  {http://arxiv.org/abs/1706.09812} {arXiv:1706.09812 [astro-ph.HE]}
  \BibitemShut {NoStop}%
\bibitem [{\citenamefont {Aguilar}\ \emph {et~al.}(2018)\citenamefont {Aguilar}
  \emph {et~al.}}]{Aguilar:2018njt}%
  \BibitemOpen
  \bibfield  {author} {\bibinfo {author} {\bibfnamefont {M.}~\bibnamefont
  {Aguilar}} \emph {et~al.} (\bibinfo {collaboration} {AMS collaboration}),\
  }\href {\doibase 10.1103/PhysRevLett.120.021101} {\bibfield  {journal}
  {\bibinfo  {journal} {Phys. Rev. Lett.}\ }\textbf {\bibinfo {volume} {120}},\
  \bibinfo {pages} {021101} (\bibinfo {year} {2018})}\BibitemShut {NoStop}%
\bibitem [{\citenamefont {Cholis}\ \emph {et~al.}(2016)\citenamefont {Cholis},
  \citenamefont {Hooper},\ and\ \citenamefont {Linden}}]{Cholis:2015gna}%
  \BibitemOpen
  \bibfield  {author} {\bibinfo {author} {\bibfnamefont {I.}~\bibnamefont
  {Cholis}}, \bibinfo {author} {\bibfnamefont {D.}~\bibnamefont {Hooper}}, \
  and\ \bibinfo {author} {\bibfnamefont {T.}~\bibnamefont {Linden}},\ }\href
  {\doibase 10.1103/PhysRevD.93.043016} {\bibfield  {journal} {\bibinfo
  {journal} {Phys. Rev.}\ }\textbf {\bibinfo {volume} {D93}},\ \bibinfo {pages}
  {043016} (\bibinfo {year} {2016})},\ \Eprint
  {http://arxiv.org/abs/1511.01507} {arXiv:1511.01507 [astro-ph.SR]}
  \BibitemShut {NoStop}%
\bibitem [{\citenamefont {Aguilar}\ \emph
  {et~al.}(2016{\natexlab{b}})\citenamefont {Aguilar} \emph
  {et~al.}}]{Aguilar:2016vqr}%
  \BibitemOpen
  \bibfield  {author} {\bibinfo {author} {\bibfnamefont {M.}~\bibnamefont
  {Aguilar}} \emph {et~al.} (\bibinfo {collaboration} {AMS collaboration}),\
  }\href {\doibase 10.1103/PhysRevLett.117.231102} {\bibfield  {journal}
  {\bibinfo  {journal} {Phys. Rev. Lett.}\ }\textbf {\bibinfo {volume} {117}},\
  \bibinfo {pages} {231102} (\bibinfo {year} {2016}{\natexlab{b}})}\BibitemShut
  {NoStop}%
\bibitem [{\citenamefont {Adriani}\ \emph {et~al.}(2013)\citenamefont {Adriani}
  \emph {et~al.}}]{Adriani:2012paa}%
  \BibitemOpen
  \bibfield  {author} {\bibinfo {author} {\bibfnamefont {O.}~\bibnamefont
  {Adriani}} \emph {et~al.},\ }\href {\doibase 10.1134/S002136401222002X}
  {\bibfield  {journal} {\bibinfo  {journal} {JETP Lett.}\ }\textbf {\bibinfo
  {volume} {96}},\ \bibinfo {pages} {621} (\bibinfo {year} {2013})}\BibitemShut
  {NoStop}%
\bibitem [{\citenamefont {Aguilar}\ \emph
  {et~al.}(2015{\natexlab{b}})\citenamefont {Aguilar} \emph
  {et~al.}}]{Aguilar:2015ctt}%
  \BibitemOpen
  \bibfield  {author} {\bibinfo {author} {\bibfnamefont {M.}~\bibnamefont
  {Aguilar}} \emph {et~al.} (\bibinfo {collaboration} {AMS collaboration}),\
  }\href {\doibase 10.1103/PhysRevLett.115.211101} {\bibfield  {journal}
  {\bibinfo  {journal} {Phys. Rev. Lett.}\ }\textbf {\bibinfo {volume} {115}},\
  \bibinfo {pages} {211101} (\bibinfo {year} {2015}{\natexlab{b}})}\BibitemShut
  {NoStop}%
\bibitem [{\citenamefont {Stone}\ \emph {et~al.}(2013)\citenamefont {Stone},
  \citenamefont {Cummings}, \citenamefont {McDonald}, \citenamefont {Heikkila},
  \citenamefont {Lal},\ and\ \citenamefont {Webber}}]{Stone150}%
  \BibitemOpen
  \bibfield  {author} {\bibinfo {author} {\bibfnamefont {E.~C.}\ \bibnamefont
  {Stone}}, \bibinfo {author} {\bibfnamefont {A.~C.}\ \bibnamefont {Cummings}},
  \bibinfo {author} {\bibfnamefont {F.~B.}\ \bibnamefont {McDonald}}, \bibinfo
  {author} {\bibfnamefont {B.~C.}\ \bibnamefont {Heikkila}}, \bibinfo {author}
  {\bibfnamefont {N.}~\bibnamefont {Lal}}, \ and\ \bibinfo {author}
  {\bibfnamefont {W.~R.}\ \bibnamefont {Webber}},\ }\href {\doibase
  10.1126/science.1236408} {\bibfield  {journal} {\bibinfo  {journal}
  {Science}\ }\textbf {\bibinfo {volume} {341}},\ \bibinfo {pages} {150}
  (\bibinfo {year} {2013})}\BibitemShut {NoStop}%
\bibitem [{\citenamefont {Davies}(1987)}]{Davies:1987zz}%
  \BibitemOpen
  \bibfield  {author} {\bibinfo {author} {\bibfnamefont {R.~B.}\ \bibnamefont
  {Davies}},\ }\href {\doibase 10.1093/biomet/74.1.33} {\bibfield  {journal}
  {\bibinfo  {journal} {Biometrika}\ }\textbf {\bibinfo {volume} {74}},\
  \bibinfo {pages} {33} (\bibinfo {year} {1987})}\BibitemShut {NoStop}%
\bibitem [{\citenamefont {Gross}\ and\ \citenamefont
  {Vitells}(2010)}]{Gross:2010qma}%
  \BibitemOpen
  \bibfield  {author} {\bibinfo {author} {\bibfnamefont {E.}~\bibnamefont
  {Gross}}\ and\ \bibinfo {author} {\bibfnamefont {O.}~\bibnamefont
  {Vitells}},\ }\href {\doibase 10.1140/epjc/s10052-010-1470-8} {\bibfield
  {journal} {\bibinfo  {journal} {Eur. Phys. J.}\ }\textbf {\bibinfo {volume}
  {C70}},\ \bibinfo {pages} {525} (\bibinfo {year} {2010})},\ \Eprint
  {http://arxiv.org/abs/1005.1891} {arXiv:1005.1891 [physics.data-an]}
  \BibitemShut {NoStop}%
\bibitem [{\citenamefont {{Yan}}\ \emph {et~al.}(2020)\citenamefont {{Yan}},
  \citenamefont {{Choutko}}, \citenamefont {{Oliva}},\ and\ \citenamefont
  {{Paniccia}}}]{2020NuPhA.99621712Y}%
  \BibitemOpen
  \bibfield  {author} {\bibinfo {author} {\bibfnamefont {Q.}~\bibnamefont
  {{Yan}}}, \bibinfo {author} {\bibfnamefont {V.}~\bibnamefont {{Choutko}}},
  \bibinfo {author} {\bibfnamefont {A.}~\bibnamefont {{Oliva}}}, \ and\
  \bibinfo {author} {\bibfnamefont {M.}~\bibnamefont {{Paniccia}}},\ }\href
  {\doibase 10.1016/j.nuclphysa.2020.121712} {\bibfield  {journal} {\bibinfo
  {journal} {{Nucl. Phys.}}\ }\textbf {\bibinfo {volume} {{A996}}},\ \bibinfo
  {pages} {{121712}} (\bibinfo {year} {2020})}\BibitemShut {NoStop}%
\bibitem [{\citenamefont {Weinrich}\ \emph
  {et~al.}(2020{\natexlab{b}})\citenamefont {Weinrich}, \citenamefont
  {Boudaud}, \citenamefont {Derome}, \citenamefont {Genolini}, \citenamefont
  {Lavalle}, \citenamefont {Maurin}, \citenamefont {Salati}, \citenamefont
  {Serpico},\ and\ \citenamefont {Weymann-Despres}}]{Weinrich:2020ftb}%
  \BibitemOpen
  \bibfield  {author} {\bibinfo {author} {\bibfnamefont {N.}~\bibnamefont
  {Weinrich}}, \bibinfo {author} {\bibfnamefont {M.}~\bibnamefont {Boudaud}},
  \bibinfo {author} {\bibfnamefont {L.}~\bibnamefont {Derome}}, \bibinfo
  {author} {\bibfnamefont {Y.}~\bibnamefont {Genolini}}, \bibinfo {author}
  {\bibfnamefont {J.}~\bibnamefont {Lavalle}}, \bibinfo {author} {\bibfnamefont
  {D.}~\bibnamefont {Maurin}}, \bibinfo {author} {\bibfnamefont
  {P.}~\bibnamefont {Salati}}, \bibinfo {author} {\bibfnamefont
  {P.}~\bibnamefont {Serpico}}, \ and\ \bibinfo {author} {\bibfnamefont
  {G.}~\bibnamefont {Weymann-Despres}},\ }\href {\doibase
  10.1051/0004-6361/202038064} {\bibfield  {journal} {\bibinfo  {journal}
  {Astron. Astrophys.}\ }\textbf {\bibinfo {volume} {639}},\ \bibinfo {pages}
  {A74} (\bibinfo {year} {2020}{\natexlab{b}})},\ \Eprint
  {http://arxiv.org/abs/2004.00441} {arXiv:2004.00441 [astro-ph.HE]}
  \BibitemShut {NoStop}%
\bibitem [{\citenamefont {Evoli}\ \emph {et~al.}(2020)\citenamefont {Evoli},
  \citenamefont {Morlino}, \citenamefont {Blasi},\ and\ \citenamefont
  {Aloisio}}]{Evoli:2019iih}%
  \BibitemOpen
  \bibfield  {author} {\bibinfo {author} {\bibfnamefont {C.}~\bibnamefont
  {Evoli}}, \bibinfo {author} {\bibfnamefont {G.}~\bibnamefont {Morlino}},
  \bibinfo {author} {\bibfnamefont {P.}~\bibnamefont {Blasi}}, \ and\ \bibinfo
  {author} {\bibfnamefont {R.}~\bibnamefont {Aloisio}},\ }\href {\doibase
  10.1103/PhysRevD.101.023013} {\bibfield  {journal} {\bibinfo  {journal}
  {Phys. Rev. D}\ }\textbf {\bibinfo {volume} {101}},\ \bibinfo {pages}
  {023013} (\bibinfo {year} {2020})},\ \Eprint
  {http://arxiv.org/abs/1910.04113} {arXiv:1910.04113 [astro-ph.HE]}
  \BibitemShut {NoStop}%
\end{thebibliography}%

\end{document}